\documentclass[twocolumn]{aastex631}

\DeclareRobustCommand{\vbar}{\allowbreak\textbar\allowbreak}

\usepackage{multirow}
\usepackage{placeins}
\usepackage{graphicx}
\DeclareGraphicsExtensions{.pdf,.png,.jpg}
\usepackage{longtable}   
\usepackage{rotating}    
\DeclareUnicodeCharacter{2212}{-}

\usepackage{array}
\usepackage{array}
\newcolumntype{J}[1]{>{\centering\arraybackslash}p{#1}}           
\newcolumntype{S}[1]{>{\raggedright\arraybackslash\hspace{0pt}\sloppy}p{#1}} 

\newcolumntype{J}[1]{>{\centering\arraybackslash}p{#1}} 
\newcolumntype{S}[1]{>{\raggedright\arraybackslash\hspace{0pt}\sloppy}p{#1}} 

\makeatletter
\@ifundefined{c@galrow}{\newcounter{galrow}}{}
\makeatother

\newcommand{\uvic}{Department of Physics and Astronomy, University of Victoria, Victoria, BC V8P 1A1, Canada}
\newcommand{\nrchaa}{National Research Council of Canada, Herzberg Astronomy \& Astrophysics Research Centre, Victoria, BC, V9E 2E7, Canada}
\newcommand{\csa}{Canadian Space Agency, 6767 Route de l’A\'eroport, Saint-Hubert, QC J3Y 8Y9, Canada}
\newcommand{\noirlab}{NSF’s NOIRLab, 950 N. Cherry Avenue, Tucson, AZ 85719, USA}
\def\statmorph{\texttt{statmorph} }

\begin{document}

\title{The Next Generation Virgo Cluster Survey (NGVS). XL. The Morphological Classification of Virgo Cluster Galaxies}

\correspondingauthor{Max Kurzner}
\email{mkurzner@uvic.ca}

\author[0000-0001-5208-4697]{Max M. Kurzner}
\affiliation{\uvic}

\author[0000-0003-1184-8114]{Patrick Côté}
\affiliation{\nrchaa}

\author[0000-0002-8224-11284]{Laura Ferrarese}
\affiliation{\nrchaa}

\author[0000-0003-4672-8497]{Kaixiang Wang}
\affiliation{Department of Astronomy, Peking University, Beijing, China
Kavli Institute for Astronomy and Astrophysics, Peking University, Beijing, China}

\author[0000-0002-2073-2781]{Eric W. Peng}
\affiliation{\noirlab}

\author[0000-0002-3303-4077]{Scott Wilkinson}
\affiliation{\uvic}

\author[0000-0002-0363-4266]{Joel C. Roediger}
\affiliation{\csa}

\author[0000-0002-1685-4284]{Chelsea Spengler}
\affiliation{\uvic}

\author[0000-0003-1845-0934]{Toby Brown}
\affiliation{\nrchaa}

\author[0000-0002-4718-3428]{Chengze Liu}
\affiliation{State Key Laboratory of Dark Matter Physics, Shanghai Key Laboratory for Particle Physics and Cosmology, School of Physics and Astronomy \& Tsung-Dao Lee Institute, Shanghai Jiao Tong University, Shanghai 200240, China}

\author[0000-0002-5049-4390]{Sungsoon Lim}
\affiliation{Division of Science Education, Kangwon National University, Chuncheon, Republic of Korea}

\author[0000-0003-4945-0056]{Rub\'en S\'anchez-Janssen}
\affiliation{UK Astronomy Technology Centre, Royal Observatory, Blackford Hill, Edinburgh EH9 3HJ, UK}

\author[0000-0001-6443-5570]{Elisa Toloba}
\affiliation{Department of Physics and Astronomy, University of the Pacific, Stockton, CA, USA}

\author[0000-0001-8867-4234]{Puragra Guhathakurta}
\affiliation{UCO/Lick Observatory, Department of Astronomy and Astrophysics, University of California Santa Cruz, Santa Cruz, CA, USA}

\author[0000-0002-5213-3548]{John P. Blakeslee}
\affiliation{\noirlab}

\author[0000-0001-9427-3373]{Patrick R. Durrell}
\affiliation{Department of Physics, Astronomy, GIS, Geology and Environmental Sciences, Youngstown State University, One Tressel Way, Youngstown, OH 44555 USA}

\author[0000-0002-7214-8296]{Ariane Lançon}
\affiliation{Université de Strasbourg, CNRS, Observatoire Astronomique de Strasbourg, UMR 7550, 11 rue de l’Université, F-67000 Strasbourg, France}

\author[0000-0002-7089-8616]{J. Christopher Mihos}
\affiliation{Department of Astronomy, Case Western Reserve University, Cleveland, OH 44106, USA}

\author[0000-0003-3009-4928]{Matthew A. Taylor}
\affiliation{University of Calgary, 2500 University Drive NW, Calgary Alberta T2N 1N4, Canada}

\author[0000-0003-1428-5775]{Tyrone E. Woods}
\affiliation{Department of Physics and Astronomy, Allen Building, 30A Sifton Road, University of Manitoba, Winnipeg, MB R3T 2N2, Canada}

\author[0009-0006-7485-7463]{Solveig Thompson}
\affiliation{University of Calgary, 2500 University Drive NW, Calgary Alberta T2N 1N4, Canada}

\author[0009-0003-5548-6773]{Lauren A. MacArthur}
\affiliation{Princeton University, Department of Astrophysical Sciences, Princeton, NJ 08544, USA}

\begin{abstract}
We present a study of  morphologies, based on deep $u^{*}g^{\prime}i^{\prime}z^{\prime}$ imaging of the Virgo Cluster from the Next Generation Virgo Cluster Survey (NGVS), for 3689 Virgo cluster members spanning a mass range of $\sim$$10^{11}M_{\odot}$ to $\sim$$10^5~M_{\odot}$. 
Our analysis introduces a new, two-component visual classification scheme developed to capture the morphological diversity of galaxies over more than six orders of magnitude in stellar mass.
Our morphological classifications use two parameters to describe the global structure and star formation activity of each galaxy. Structural sub-codes denote features such as spiral arms, bars, disks, shells, streams, while star formation sub-codes indicate the form and location of the current star formation activity (e.g., in cores, clumps, filaments, etc). These visual classifications rely on deep $g^\prime$-band images, supplemented by $u^{*}g^{\prime}i^{\prime}$ color images, as well as unsharp-masked images for a subset of objects. We compare our classifications to previous results for bright member galaxies that used more established schemes, finding good agreement. We also measure quantitative classification statistics (e.g., CASGM$_{20}$) for a subset of the brighter galaxies, and present catalogs for some galaxy types of special interest, including structurally compact galaxies, ultra-diffuse galaxies, candidate ultra-compact dwarf transition objects, as well as candidate post-merger systems. These morphological classifications may be useful as a training set in the application of machine learning tools to the next generation of wide-field imaging surveys. 
\end{abstract}


\section{Introduction} 
\label{sec:intro}

How is the general appearance of a galaxy connected to its evolutionary history and physical properties? This question has occupied astronomers for nearly a century \citep[see, e.g., the review of][]{conselice2014evolution}. A series of foundational papers by Edwin Hubble established what later came to be known as the Hubble Sequence --- a classification framework that dominated the field throughout much of the twentieth century \citep{Hubble1926,hubble1927classification,Hubble1936}. 

Numerous revisions and modifications to this system added 
the flexibility needed to capture the diversity of features observed in real galaxies, such as bars, rings, lenses, and spiral arm patterns
\citep[e.g.,][]{deVaucouleurs1959,vandenBergh1960a,vandenBergh1960b,Sandage1961,sersic1963influence, vandenBergh1976, Sandage1981, Elmegreen1987}.

Over time, researchers, motivated by a desire to  connect appearances to physical properties, began to use more quantitative methods to describe the light distribution within galaxies  \citep{Holmberg1958,Morgan1959,Roberts1963,Roberts1994,Conselice2006,Blakeslee2006,Allen2006,conselice2014evolution}. This approach has its roots in early analytic models of galaxy light profiles. The $r^{1/4}$ law for elliptical galaxies \citep{deVaucouleurs1948} and its generalization by \citet{sersic1963influence} laid the foundation for decades of quantitative classifications, in which de Vaucouleurs and Sérsic profiles were fitted to spiral bulges and elliptical galaxies \citep[e.g.,][]{Kent1985}. Such measurements helped establish empirical scaling relations that link structural parameters to internal kinematics, such as the Tully–Fisher relation for disk galaxies, the Faber–Jackson relation for ellipticals, and the Fundamental Plane \citep[e.g.,][]{Tully1977,faber1976velocity,Djorgovski1987}.

More recent efforts have used tools such as GALFIT \citep{Peng2002,Peng2010} to model galaxy light distributions using Sérsic, bulge–disk, and multi-component decompositions across large surveys \citep[e.g.,][]{Graham2005,Simard2011,Meert2015}. These quantitative approaches have become central to modern morphological analysis and continue to evolve through the development of new pipelines and deeper imaging datasets \citep[e.g.,][]{Gilhuly2018,Dimauro2018,Nogueira-Cavalcante2018,Casura2022}. Such tools are expected to play a central role in upcoming wide-field surveys such as Euclid \citep{Euclid,Euclid2023Morph}. Parametric fitting techniques have proven especially effective for describing the structure of massive, well-resolved galaxies, where features such as bulges and disks are prominent and can be robustly modeled. 
Nevertheless, their effectiveness may diminish for low-mass galaxies, where irregular morphologies, diffuse light profiles, and strong environmental processing challenge the assumptions of smooth, symmetric models \citep[e.g.,][]{Lisker2006,Boselli2006}.

At the same time, the introduction of ``non-parametric" structural parameters --- such as the Concentration, Asymmetry and Smoothness (CAS) system \citep{Abraham1996,Conselice2000,Bershady2000} --- allowed researchers to explore important physical processes, such as star formation and mergers, using measurements performed on digital images. Other parameters, such as the Gini coefficient \citep{Lotz2004} and the M$_{20}$ statistic \citep{Lotz2008}, gained popularity as diagnostics of mergers and star formation activity. Additional non-parametric methods and statistics continue to be developed, including those of \citet{Wen2014}, \citet{Curtis-Lake2016}, \citet{Wen2016} and \citet{Pawlik2016}, to name just a few.

Studies have utilized such non-parametric statistics to explore and characterize the morphologies of distinct and unique galaxy populations detected in deep and/or wide surveys, as well as galaxies generated in cosmological numerical simulations \citep[e.g.,][]{RodriguezGomez2019,Nelson2019, Pearson2019,RodriguezGomez2019,Roberts2020,Whitney2021,Wilkinson2022, Ferreira2022,Vega-Acevedo2022,Pandey2024}. Increasingly, machine learning and neural network techniques are being developed to classify and characterize galaxy morphologies, using both ``classical" \citep[e.g.,][]{Naim1995,Ball2006,lintott2008galaxy, Huertas-Company2008,Banerji2010,Huertas-Company2011} and more complex tools \citep{Dieleman2015,DominguezSanchez2018,Barchi2020,Hausen2020,Martin2020,Cheng2021a, Tohill2021, Walmsley2022}. These techniques are expected to see widespread use during the coming era of ultra-deep and ultra-wide-field imaging surveys, like Rubin Observatory's Legacy Survey of Space and Time (LSST; \citet{Ivezic2019} , Euclid \citep{Euclid} and Roman \citep{WFIRST}, where the analysis of huge galaxy samples will require fast, flexible and robust classification tools. At the same time, non-parametric and machine learning algorithms require training sets based on carefully selected samples of galaxies in the local universe that have been classified using traditional approaches. Such training sets will be especially important for intermediate- and low-mass galaxies; these populations have long been underrepresented in imaging programs but will be common in future surveys conducted with ground- and space-based telescopes.

A natural target for morphological studies of galaxies in the local universe is the Virgo cluster --- the nearest rich cluster of galaxies. At its distance of D $\approx$ 16.5 Mpc \citep{Mei2007,Blakeslee2009}, 1\arcsec~corresponds to $\simeq$ 80 pc so individual galaxies can be studied at a level of detail that will never be possible in more distant clusters, even those at modest redshifts.  Past studies of galaxies in Virgo, as well as other nearby groups and clusters \citep{dressler1980galaxy,Tinsley1980, Butcher1984,Moore1996,Moore1998,Kauffmann2004,Cappellari2006,Faber2007, Boselli2014, Buta2015}, revealed an impressive morphological diversity. Virgo itself is a complex, irregular system that, in some ways, mirrors the structure observed in some high-redshift clusters. It contains several thousand members, including virtually all known galaxy types, and spans a wide range in density and environment, from its periphery to dense core. Virgo and its surroundings thus provide a unique testbed for exploring how galaxy morphologies evolve --- before, during, and after cluster infall. 

The Virgo Cluster Catalog \citep[VCC;][]{Binggeli1985} --- a photographic survey of the Virgo cluster that identified 2096 confirmed or possible member galaxies within a 140 deg$^2$ region centered on the cluster --- established our baseline understanding of galaxy morphologies in a cluster environment. Indeed, the most enduring legacy of the VCC, aside from the galaxy catalog itself, may be its homogeneous database of  morphologies which continues to be utilized by astronomers to this day. A recent expansion of the VCC, based largely on the SDSS DR7 \citep{Abazaijan2009}, is the Extended Virgo Cluster Catalog (EVCC) of \citet{Kim2014} which surveyed a broader region than the original VCC to study galaxy morphologies in the outskirts of Virgo and beyond.

In this paper, we present a new study of galaxy morphologies in the Virgo cluster using multi-band imaging from the Next Generation Virgo Cluster Survey \citep[NGVS;][]{Ferrarese2012}. While numerous classification schemes have been developed to describe galaxy morphologies, most have been tailored to, and designed explicitly for, high-mass galaxies; thus, they rely on features such as well-defined spiral arms, bars, and bulges, which are less distinct or absent in lower-mass systems \citep[e.g.,][]{Binggeli1985, Lisker2006}. Dwarf galaxies, which dominate the low-mass end of the galaxy population, often exhibit more irregular structures, weak or absent disk features, and a greater susceptibility to environmental processes such as tidal interactions and ram pressure stripping \citep[e.g.,][]{Mayer2001, Boselli2006, Mayer2006,Smith2013}. Traditional classification schemes, including both visual and automated approaches, struggle to categorize robustly such galaxies, particularly when their structures are faint or diffuse. Our study addresses this gap by leveraging the depth, resolution, and multi-band coverage of the NGVS \citep{Ferrarese2012} to develop a morphological classification framework that is specifically designed to capture the diversity of low-mass cluster galaxies while being agnostic to labels and classifications based on stellar mass.

As we show in \S\ref{sec:observations}, the NGVS significantly improves on the VCC in terms of overall sample size (N$\sim3700$), limiting magnitude, angular resolution, and the availability of color information. The NGVS imaging allows us to characterize galaxy morphologies over roughly six orders of magnitude, from brightest cluster galaxies (M$_{*}$ $\gtrsim$ 10$^{11}$ M$_{\odot}$) down to the scale of dwarf galaxies, with stellar masses comparable to those of faint Milky Way satellites (M$_{*}$ $\sim$ 10$^{5}$ M$_{\odot}$). Indeed, low-mass galaxies dominate the NGVS sample, with $\sim$94\%  of the cataloged galaxies having stellar masses below 10$^9$ M$_{\odot}$.

Our analysis builds on a series of NGVS papers focused on galaxy evolution and the role of a cluster environment in shaping galaxy populations. Previous NGVS papers relevant to the present analysis include: 
the galaxy color-magnitude relation and red sequence relation \citep{Roediger2017}; 
the luminosity function and stellar populations of galaxies \citep{Grossauer2015,Ferrarese2016};
distance measurements using the method of surface brightness fluctuations \citep{Cantiello2018,Cantiello2024};
the properties of structurally compact \citep{Liu2015,Zhang2018,Liu2020} or diffuse galaxies \citep{Lim2020,Junais2022,Toloba2018,Toloba2023}, and the evolutionary processes that may link them \citep{Guérou2015,Wang2023};
galaxy shapes and nucleation properties \citep{Spengler2017,Sanchez-Janssen2019,Sanchez-JanssenMNRAS2019}; plus targeted analyses of unusual or noteworthy galaxies and groups \citep{Arrigoni2012,Paudel2013,Paudel2017}.

This paper is structured as follows. In \S\ref{sec:observations}, we describe the NGVS imaging and data products used in this study. Our discussion of galaxy morphologies, and a comparison to previous results for Virgo cluster galaxies, is presented in \S\ref{sec:methodology} while \S\ref{sec:results} presents an analysis of our morphological classifications. A discussion of our findings, including an overview of some morphologically or structurally interesting galaxies, is presented in \S\ref{sec:discussion}. We conclude and summarize in \S\ref{sec:conclusions}.


\section{Data and Observations} 
\label{sec:observations}

\subsection{Survey Design and Observations}\label{subsec:design}

We use images and data products from the NGVS \citep{Ferrarese2012} to visually classify 3689 confirmed, probable or possible members of the Virgo Cluster --- roughly 2200 of which represent new discoveries (for details, see Ferrarese et~al. 2026, in preparation). We introduce a customized classification system designed to avoid some of limitations of past schemes, most notably the separate treatment of high- and low-mass galaxies (i.e., so-called ``giant" and  ``dwarf" galaxies). In this section, we briefly describe the survey itself, including the imaging products and catalog measurements used in our study.

The NGVS is a Large Program that was carried out with the MegaPrime instrument on the 3.6m Canada France Hawaii Telescope (CFHT). The NGVS was performed in the $u^{*}g^{\prime}r^{\prime}i^{\prime}z^{\prime}$ bands, covering Virgo's two major sub-clusters \citep[A and B;][]{Binggeli1987} out to their virial radii. The survey footprint, covering 104 deg$^{2}$, was tiled with 117 distinct MegaCam pointings overlapping by 3\arcmin, each covering an area of roughly 1$^{\circ}{\times}1^{\circ}$ (after stacking of multiple, dithered exposures). 

Each of these fields was observed in four MegaCam filters, $u^{*}g^{\prime}i^{\prime}z^{\prime}$, whose bandpasses, with the exception of $u^{*}$, closely resemble those of their SDSS counterparts. The original NGVS survey strategy called for coverage in the $r^{\prime}$-band as well, but $r^{\prime}$-band exposures were only obtained for the core of the cluster plus a few additional fields, due to time lost to weather and mechanical problems, so this coverage is largely ignored in this work. Two separate image stacks were obtained for each field: a long exposure, reaching a 10$\sigma$ point-source limit of 25.9 mag in the g$^{\prime}$-band, and a short exposure designed to recover the centers of galaxies that saturate in the long exposures, as shown in Table 2 in \cite{Ferrarese2012}. For all long exposures, the observing strategy followed a ``step-dither" pattern described in \cite{Ferrarese2020}, whereby the dithered exposures that would normally be acquired in a continuous sequence for each individual field and each filter are applied, instead, to a {\it block} of fields. 

The NGVS images reach 5$\sigma$ limiting magnitudes, for point sources, of 26.3, 26.6, 25.8, and 24.8 mag in the stacked ${u}^{* }$, $g^{\prime}$, i$^{\prime}$, and z$^{\prime}$ images, respectively. For extended sources, the NGVS reaches a surface brightness limit of $\mu_{g^{\prime}}$ $\sim$ 29 mag arcsec$^{-2}$, enabling the detection of galaxies at the faint end of the luminosity function. The NGVS thus supersedes previous optical surveys of Virgo in terms of depth, number of passbands and areal coverage, including the landmark study of \citet{Binggeli1985}. 

For reference, the 5$\sigma$ point-source limits of the stacked NGVS images differ from the expected 10-year depths of LSST by +0.7, $-$0.3, $-$0.5 and $-$0.8 mag in the ${u}^{*}, g, i$ and $z$ passbands, respectively \citep{Ivezic2019}. These differences are in the sense that the NGVS depth exceeds that of LSST in the $u^*$ band, and is slightly shallower in the remaining bands. 

\subsection{Image Analysis and Data Products}\label{subsec:ImageAnalysis}

The NGVS observing strategy adopted a method commonly used in near-infrared observations, as outlined in \cite{Erben2005}, which allowed for the application of the {\it Elixir-LSB} reduction package \citep{Ferrarese2012}. {\it Elixir-LSB} is important for NGVS observations, especially for the low-surface brightness galaxies observed in Virgo as it creates an accurate map of the spatially varying component of the sky and scattered light background by median averaging all fields observed in a continuous sequence for a given filter. This map is then subtracted from each individual exposure within the sequence before stacking. Typical residuals in the {\it Elixir-LSB}-processed, scattered light-subtracted images are 0.2\% of the sky background in all filters, corresponding to a surface brightness of $\sim$29 mag arcsec$^{-2}$ (see \citealt{Duc2015} for a detailed discussion). As discussed there, some faint background variations are still visible after the {\it Elixir-LSB} processing. Nevertheless, the stacked, {\it Elixir-LSB} images are useful for morphological classication given their sensitivity to extended, LSB features like shells, halos and stellar streams.

The intrachip regions (measuring $\sim$13$^{\prime\prime}$ wide between columns and $\sim$80$^{\prime\prime}$ wide between rows) are recovered by the dithering strategy but do not reach the full exposure time and therefore have lower S/N. Moreover, bright, saturated foreground stars can be surrounded by faint (surface brightness $\geq$ 26.5 mag arcsec$^{-2}$  in the $g^{\prime}$-band) extended stellar halos while scattered light from Galactic cirrus also affects the images at g-band surface brightnesses $\leq$ 27 mag arcsec$^{-2}$. We do not correct for any of these effects, and as shown by the simulations employed in \citet{Ferrarese2020}, about 10\% of artificial galaxies injected in the frames are not recovered by the automated detection algorithm precisely because they fall in a region of variable background, due to either high-surface brightness or low S/N. However, these objects are ultimately recovered based on visual inspection, leading to the detection limit for low-mass galaxies being set by noise, rather than residual background variations. 

To visually classify each galaxy, and to measure non-parametric morphological parameters, we use fitted two-dimensional model images and residuals available from the NGVS. The model images were created with either the IRAF ELLIPSE task \citep{Jedrzejewski1987} or through fitting of two-dimensional Sérsic models using the GALFIT code \citep{Peng2002,Peng2010}, as detailed in \citet{Ferrarese2020}. Residual images generated by subtracting these models from individual galaxy images were also useful for identifying, with SAOImage DS9 \citep{DS9}, certain morphological features, such as spiral arms, bars, central nuclear star clusters, and irregular or twisted isophotes. For brighter galaxies with higher-S/N imaging, unsharp-masked images \citep{MalinUnsharp} were also generated; by sharpening the contrast, these images proved useful for classifying certain faint, extended features.

Our classifications also made use of color ``thumbnail" images created from the stacked $i^{\prime}g^{\prime}u^{*}$ NGVS images using a customized python code. Such images are especially useful when examining the stellar populations of individual galaxies and their sub-components.

\subsection{Supplementary Information}\label{subsec:supplementary_info}

Our classifications also make use of some other NGVS data products and catalogs that provide information on the structure and stellar content of individual galaxies and their surroundings. These include catalogs of candidate globular clusters selected from the NGVS on the basis of magnitude, colors and inverse concentrations \citep{2014ApJ...794..103D, 2024arXiv240309926L, Lim2025} and catalogs of candidate ultra-compact dwarf galaxies \citep{Liu2015,Liu2020,Wang2023} as well as cataloged LSB features from the Burrell Schmidt survey of \citet{Mihos2005,Mihos2015, Mihos2017}, including tidal tails, shells, plumes and diffuse halos.

Our sample is defined using NGVS membership probabilities from \citet{Ferrarese2020} that distinguish Virgo galaxies from contaminants, particularly when redshift data are unavailable. These probabilities rely on multi-dimensional color-magnitude space, trained on objects with known spectroscopic redshifts, and incorporate structural parameters—such as surface brightness, Sérsic index, and effective radius—as well as spatial clustering (given that Virgo members are more centrally concentrated than background field galaxies). Pre-processing techniques like object masking and ring filtering \citep{Secker1995} were utilized to help suppress bright sources that could obscure low-surface-brightness galaxies. Candidate Virgo members were then selected based on structural and photometric properties, with diagnostic plots (e.g., isophotal area vs. surface brightness) aiding separation from contaminants. Two-dimensional Sérsic profile fitting with GALFIT \citep{Peng2010} refined these measurements, and spectroscopic redshifts supersede photometric classifications when available. Even at the faint end of the luminosity function, where spectroscopic completeness is low, combining photometric and structural priors enabled a robust probabilistic classification. Ultimately, membership probabilities are assigned by weighting multiple independent criteria to maximize agreement with a training set of known cluster members, ensuring reliable Virgo galaxy selection while minimizing contamination.

We also make use of the NGVS stellar masses (Roediger et al. 2026, in preparation). The derivation of stellar masses for NGVS galaxies follows a methodology designed to ensure consistency across the full mass range of the sample. Stellar masses are estimated by fitting multi-band photometry to stellar population synthesis models. The inclusion of NGVS $u^*g'i'z'$ photometry allows for robust constraints on age and dust attenuation, minimizing systematic uncertainties in mass estimates. For galaxies with spectroscopic data, the photometric masses are cross-validated against spectroscopically derived mass-to-light ratios. This approach ensures reliable mass determinations, even for the low-mass dwarf population, where single-band mass-to-light ratio assumptions can introduce significant biases.


\section{Methodology}
\label{sec:methodology}

\subsection{Classification Scheme}
\label{subsec:MorphScheme}

\begin{figure*}[htbp]
\centering
\includegraphics[width=\textwidth]{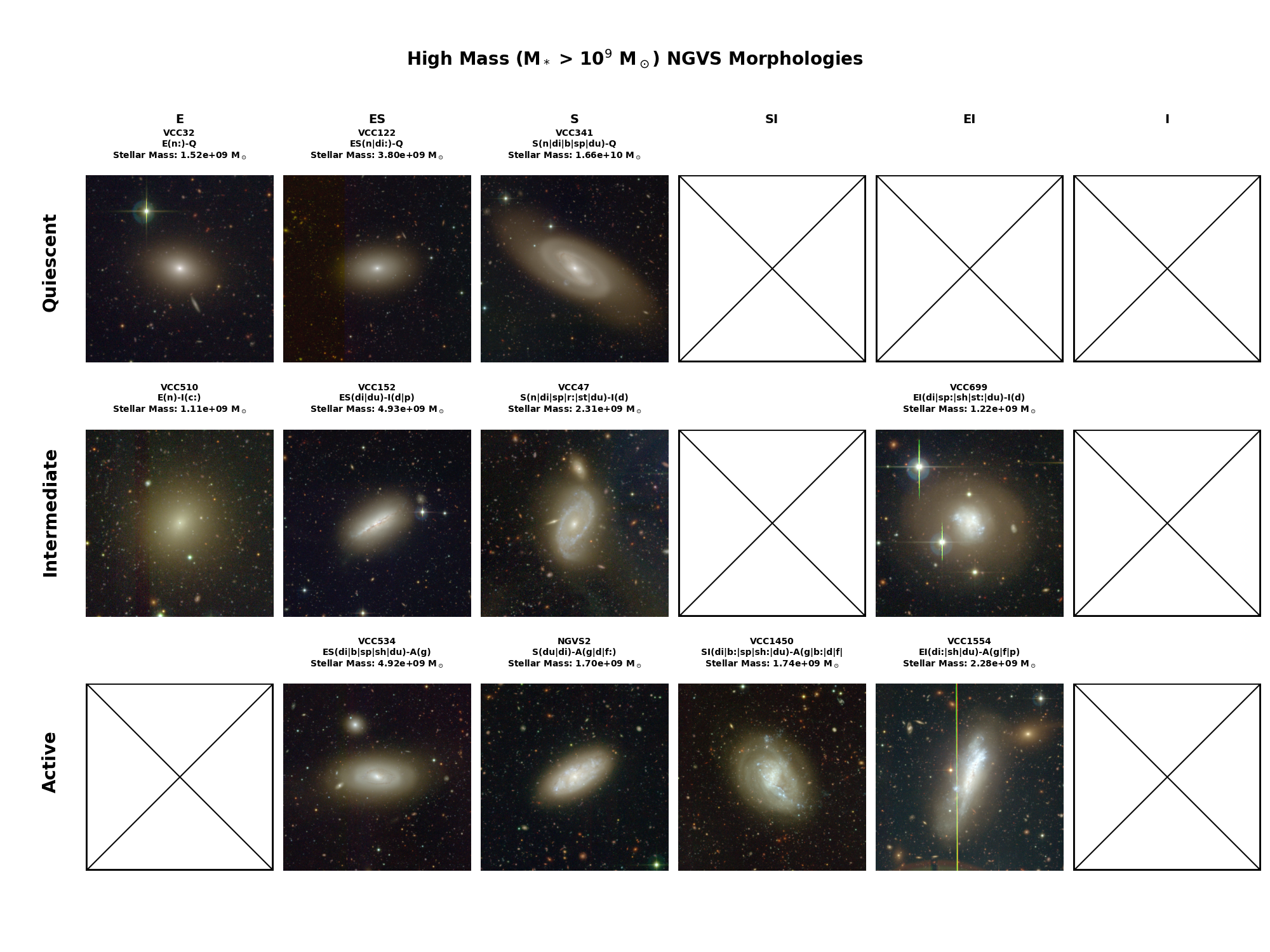}
\caption{A schematic illustration of the NGVS morphological classification scheme for a selection of high-mass (M$_{*} \ge $ 10$^{9}$M$_{\odot}$) galaxies. The six possible structural codes are arranged roughly from early to late types (E $\rightarrow$ S $\rightarrow$ I), including ``transitional" categories (ES, SI, EI). The three rows correspond to the three possible star formation codes: Quiescent (Q), Intermediate (I) and Active (A). Galaxies are labeled by their NGVS {\it identification number} --- or VCC number if the galaxy was previously cataloged in the VCC \citep{Binggeli1985} --- {\it morphology} and {\it stellar mass}. Empty entries in this matrix, denoted by an 'X', are either galaxies that do not exist in nature, or plausibly exist but are not found in the NGVS, as explained in the text.
\label{fig:NGVS_HighMass_Morphologies}}
\end{figure*}

\begin{figure*}[htbp]
\centering
\includegraphics[width=\textwidth]{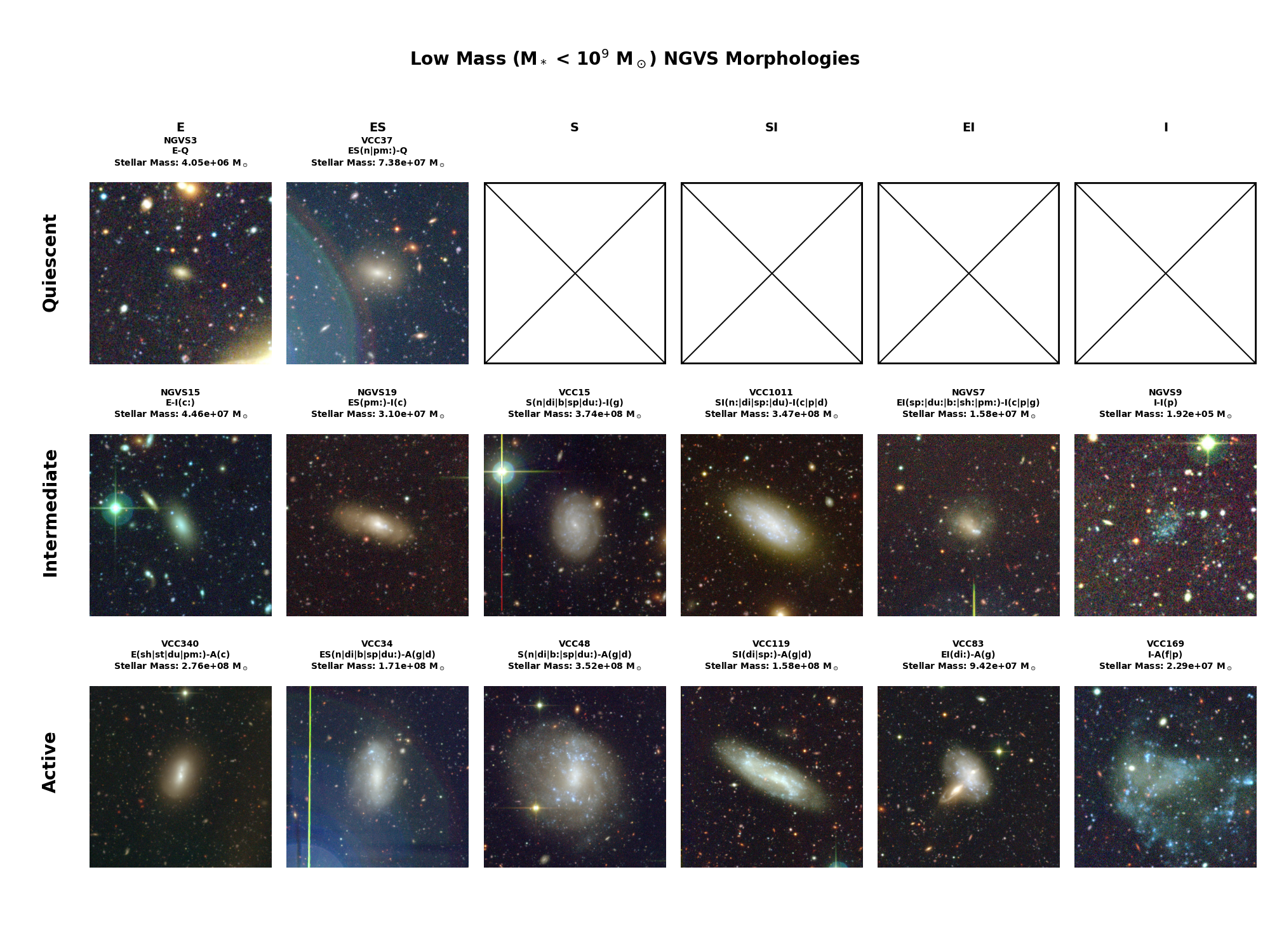}
\caption{Same as the previous figure, except for the case of low-mass (M$_{*} \leq$ 10$^{9}$M$_{\odot}$) galaxies. The classification categories in terms of structural and star formation properties are applicable across the full range of stellar mass probed by the NGVS. 
\label{fig:NGVS_LowMass_Morphologies}}
\end{figure*}

\begin{figure*}[t!]
  \centering
  \includegraphics[width=\textwidth]{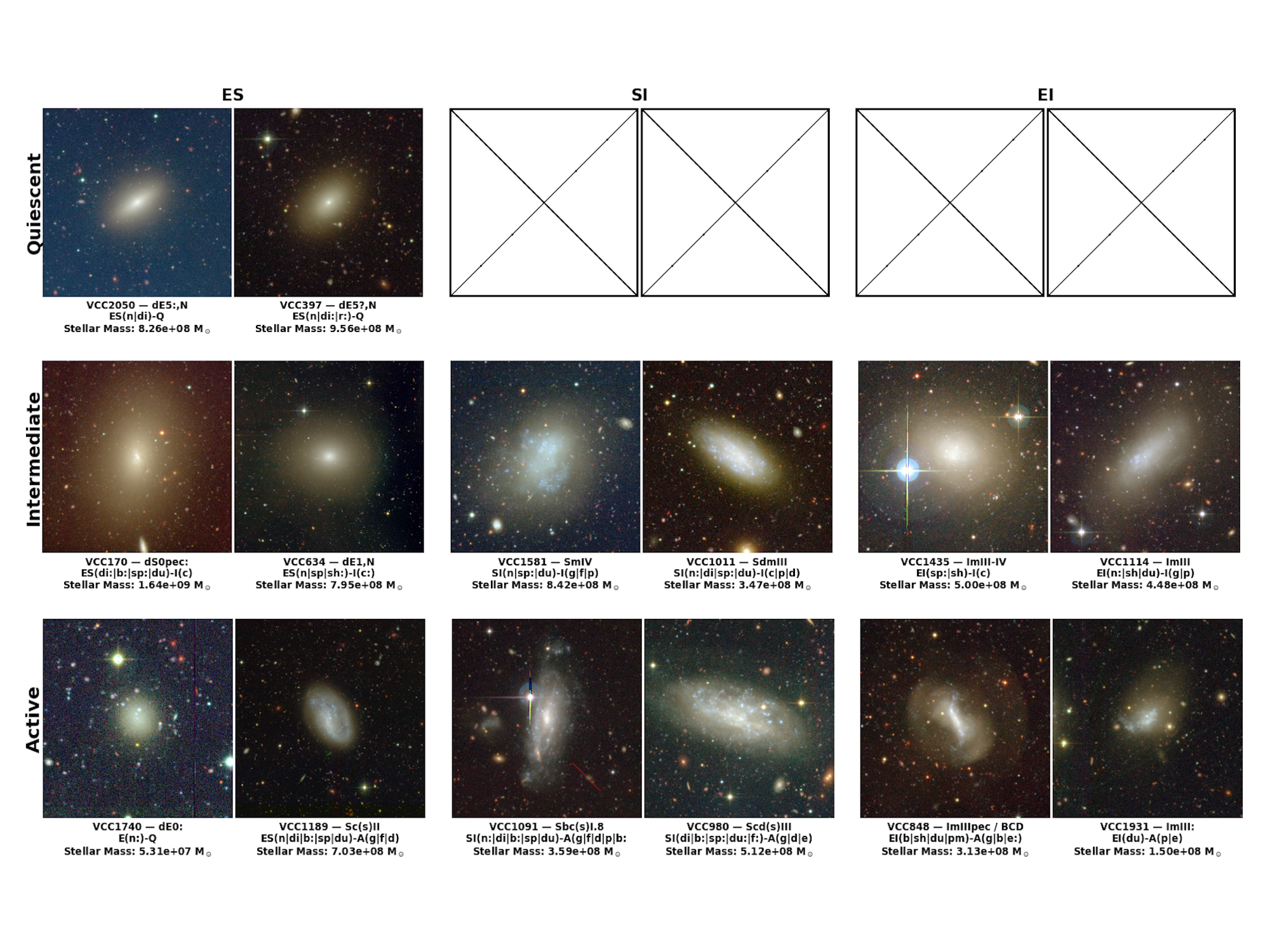}
  \caption{As in Figs.~\ref{fig:NGVS_HighMass_Morphologies} and
  \ref{fig:NGVS_LowMass_Morphologies}, but highlighting \emph{transitional}
  NGVS structural codes ({\tt ES}, {\tt SI}, and {\tt EI}). Thumbnails are
  annotated (top to bottom) with the VCC identifier and VCC type from
  \citet{Binggeli1985}, the compact NGVS code, and the stellar mass.}
  \label{fig:rare_morphologies}
\end{figure*}

The overall usefulness of morphological classifications for studies of galaxy formation and evolution can be a matter of debate. On one hand, visual classifications are known to have drawbacks: e.g., being subjective in nature, they can vary between classifiers; the classifications can depend on the source materials (such as the wavelength, depth, and resolution of the imaging used in the analysis); and some popular classification schemes are focused heavily --- sometimes exclusively --- on high-mass galaxies, whereas low-mass systems dominate galaxy populations by number.

On the other hand, morphologies have played a prominent role in many past investigations, and nearly every previous imaging survey of Virgo galaxies has provided classifications of some sort (e.g., \citealt{Reaves1983, Binggeli1985, Trentham2002, Lisker2007, Kim2014}). Moreover, history has shown that morphological classifications can enjoy usage long after the original imaging materials have been superseded: i.e., the Hubble tuning fork system \citep{Hubble1926} endures to this day, and the classifications of \citet{Binggeli1985} are still widely used four decades after the publication of the original VCC catalog. 

We have therefore devised a visual classification scheme that builds on previous morphological classifications while taking advantage of the depth, angular resolution and multi-wavelength coverage of the NGVS. Each member galaxy is assigned two parameters that broadly describe its global structure and current level of star formation activity. For the first parameter, galaxies are divided into six broad categories, with the following defining characteristics: 

\small
\begin{enumerate}
\item[{\tt E}:] Smooth and regular in appearance, having {\tt elliptical} isophotes.
\item[{\tt ES}:] Intermediate in properties between {\tt E} and {\tt S}. ES galaxies often exhibit a largely spheroidal outer light distribution consistent with elliptical galaxies but contain inner features such as weak or embedded disks, bars, or shell-like structures that indicate a dynamically colder component. Some edge-on systems are also placed in this category when flattened light profiles or dust lanes suggest hidden disk components. These galaxies often lie near the boundary between classic ellipticals and lenticular (S0) galaxies. For examples, see VCC122, VCC534 and VCC34 in Figures~\ref{fig:NGVS_HighMass_Morphologies} and \ref{fig:NGVS_LowMass_Morphologies}.
\item[{\tt S}:] A {\tt spiral} galaxy morphology, with features that may include bars, spiral arms or flattened disks.
\item[{\tt SI}:] Intermediate in properties between {\tt S} and {\tt I}. SI galaxies often show a global disk-like structure but are more irregular, with patchy, asymmetric, or clumpy features replacing coherent spiral arms. Many show active star formation in irregular patterns or dust lanes that distort the underlying morphology. See VCC1450 and VCC119 in Figures~\ref{fig:NGVS_HighMass_Morphologies} and \ref{fig:NGVS_LowMass_Morphologies}.
\item[{\tt I}:] An {\tt irregular}, patchy, or filamentary appearance lacking an obvious spiral or ellipsoidal structure.
\item[{\tt EI}:] Intermediate in properties between {\tt I} and {\tt E}. EI galaxies have broadly elliptical envelopes but show significant irregularities in their inner structure, including off-centered, star-forming knots, clumps, or mild asymmetries. These features suggest either a disturbed morphology or residual star formation that disrupts the appearance of a smooth spheroid. Examples include VCC1554 and VCC83 in Figures~\ref{fig:NGVS_HighMass_Morphologies} and \ref{fig:NGVS_LowMass_Morphologies}.
\end{enumerate}
\normalsize

\normalsize
\noindent The three intermediate structural classes---{\tt ES}, {\tt SI}, and {\tt EI}---are used for galaxies that do not fit cleanly into a single structural type, often exhibiting transitional morphologies or complex structures. These types are especially useful for classifying systems with intermediate masses or galaxies impacted by evolutionary processes that appear to have altered or reshaped their morphologies.

\noindent The {\tt ES} class in our scheme includes most of the high-mass galaxies that, in the canonical Hubble system, would likely be classified as S0 galaxies or, less frequently, early-type spirals. Rather than enforce a strict division between ellipticals and spirals, our scheme provides a more flexible description that reflects the observed continuity in structural properties, particularly among galaxies that show both spheroidal envelopes and disk-like substructure. In this way, {\tt ES} types overlap substantially with the traditional S0 class, but also accommodates edge-on systems and shell-bearing galaxies, as well as low-mass galaxies with disk-like structures, that do not fit cleanly within the Hubble tuning fork.

\noindent For the second parameter in our visual classification scheme, galaxies are divided into three categories on the basis of their apparent state of star formation:

\small
\begin{enumerate}
\item[{\tt Q}:] {\tt Quiescent}, Galaxies with negligible or no recent star formation, characterized by red optical colors with little or no $u^{*}$ band emission.
\item[{\tt I}:] {\tt Intermediate}, Galaxies exhibiting sporadic or low-level star formation, either localized (e.g., in specific knots or regions) or fading globally, often associated with weak u* band emission and moderate optical colors.
\item[{\tt A}:] {\tt Active}, Galaxies with vigorous star formation across significant portions of the disk, often visible in blue optical colors and structured features like star-forming clumps or filaments.
\end{enumerate}
\normalsize

The NGVS classifications rely primarily on stacked $g^\prime$-band images. Our decision to use the $g^{\prime}$-band images as our primary classification resource is a compromise between a desire to trace the underlying stellar mass in each galaxy and the need for sensitivity to recent star formation; it also allows a direct connection to earlier photographic classifications carried out using blue-sensitive plates. Our two-parameter classification system differs from most previous schemes (such as the ``extended Hubble” scheme employed in the VCC; see \citep{Sandage1981} and \citep{Sandage1984} but has the advantage that it can be applied equally well to high- and low-mass galaxies, thus avoiding the introduction of conceptual “dichotomies” due to the scheme itself (e.g., ``giants” vs. ``dwarfs”; \citealt{Kormendy2009}). 

This flexibility is illustrated in Figures~\ref{fig:NGVS_HighMass_Morphologies} and \ref{fig:NGVS_LowMass_Morphologies} which show the application of this scheme to a representative sample of high- and low-mass galaxies, respectively. Note that some spaces in this matrix are unpopulated because no examples exist in nature: i.e., there are no examples of {\tt Active Ellipticals}, or {\tt Quiescent Irregulars}. Other elements of the matrix might exist, at least in principle, but no such galaxies appear in our current database for the Virgo cluster.

In Figure~\ref{fig:rare_morphologies}, we show some more examples of transitional types --- {\tt ES}, {\tt SI} and {\tt EI} galaxies with intermediate masses. Galaxies are labeled with both the VCC and NGVS classifications. While such transitional systems comprise a modest fraction  of the NGVS sample, they exhibit properties that are not easily captured within the simple {\tt E}, {\tt S} and {\tt I} classes.

\subsection{Internal Comparisons and Consistency Checks
\label{subsec:comparisons}}

In this section, we describe the classification sub-samples and results, which provides insights into overall repeatability and possible biases. These issues are common to all visual classification schemes, including the one used here, which are inherently subjective in nature. It is thus important to verify that the classifiers are coming to the same general conclusions when attempting to connect classifications to the physics of galaxy formation and evolution.

Initially, classifications according to the scheme that forms the basis of our analysis were carried out for roughly a thousand galaxies by one of the authors (PC, hereafter Classifier 1 = C1). Following a joint review by C1 and MMK (hereafter C2)  of a training set large enough to include all galaxy types, including rare systems, classifications then performed independently by~C2.

We consider two different sub-samples for our internal comparison. First, a  sub-sample of 100 galaxies was randomly selected from the full catalog for classification by C2. These galaxies span the full range in magnitude of Virgo cluster members and include many of the brightest and most structurally complex objects. We find very good agreement between the independent classifications for this sub-sample, the results of which are shown in the third column of Table~\ref{tab:ngvs_combined_comparison}.

For the 100 galaxies in this sample, 91\% are classified with the same structural code and 93\% are assigned the same star formation code. Moreover, in the few cases where the assigned codes differ, the discrepancy nearly always amounts to a difference of one sub-code (i.e., {\tt E} vs. {\tt ES}, or {\tt A} vs {\tt I}). This high level of agreement demonstrates that the classifiers are broadly identifying the same structural or star formation patterns when making their classifications.  

As a second comparison, two independent sets of classifications were carried out for 1275 galaxies with magnitudes $18 \lesssim g^{\prime} \lesssim 24.4$. These galaxies represent the faintest third of the sample that lacked prior classifications and account for precisely half of all newly classified galaxies in this magnitude range. Results in this case are shown in the fourth column of Table~\ref{tab:ngvs_combined_comparison}.
 
Once again, the two sets of classifications are in very good agreement, with 97.1\% and 95.5\% of galaxies having the same structural and star formation codes, respectively. Note that this sub-sample is heavily weighted towards faint, low-mass galaxies which are dominated by apparently compact, low-surface brightness galaxies typically assigned E-Q types.

Student's t-tests \citep{Student1908} were performed on the independent classifications to see if the observed breakdowns in structural and star formation codes are consistent with their belonging to the same distribution. We find that all labeled structure codes and star formation codes are similar, with mean classifications within one standard deviation in code vs. absolute magnitude space with p-values $\ge$ 0.05 indicating a high level of internal agreement. These p-values provide strong evidence that the samples are drawn from the same underlying parent distribution. As a final step in the calibration and standardization of the classifications, a third classifier (LF = C3) performed an independent review of all classifications, identifying a small fraction of systems with discrepant structural or star formation codes, which were then resolved jointly to ensure maximum internal consistency across the full dataset. 

We conclude on the basis of these tests and comparisons that there is very good internal consistency for the NGVS morphologies. With this established, we now examine how our morphological classifications compare to previous results for the Virgo cluster.

\subsection{Comparison to the Virgo Cluster Catalog
\label{subsec:VCCComp}}

The survey of \citet{Binggeli1985} provides an obvious comparison point for our classifications. Their survey consisted of confirmed or probable cluster members that were identified using photographic $B$-band imaging taken with the duPont 2.5m telescope at Las Campanas Observatory. From a visual inspection of these blue-sensitive plates, they placed each galaxy within the Hubble classification system, including its extension to late-type spirals and irregular galaxies. They also presented an extension into the regime of dwarf galaxies, for both early- and late-type systems. Full details on the classification system are presented in \citet{RSA1981} and \citet{Sandage1984}, with the individual classifications tabulated in \citet{Binggeli1985}.

Of the 2096 galaxies in the VCC, 1466 lie within the NGVS footprint and have been classified by the NGVS as confirmed, probable or possible members of the Virgo cluster. This sample forms the basis of the comparison below.  A wide diversity is available within this sample --- from massive quenched elliptical galaxies, to grand-design spirals, to star-forming irregulars, and  to many different types of dwarfs scattered throughout the cluster. Like their high-mass counterparts, these dwarfs show a range in star formation activity; in some cases, this activity was likely triggered by interactions with other cluster galaxies, or with the cluster itself. 

\begin{figure*}[ht!]
\centering
\includegraphics[width=1.0\textwidth]{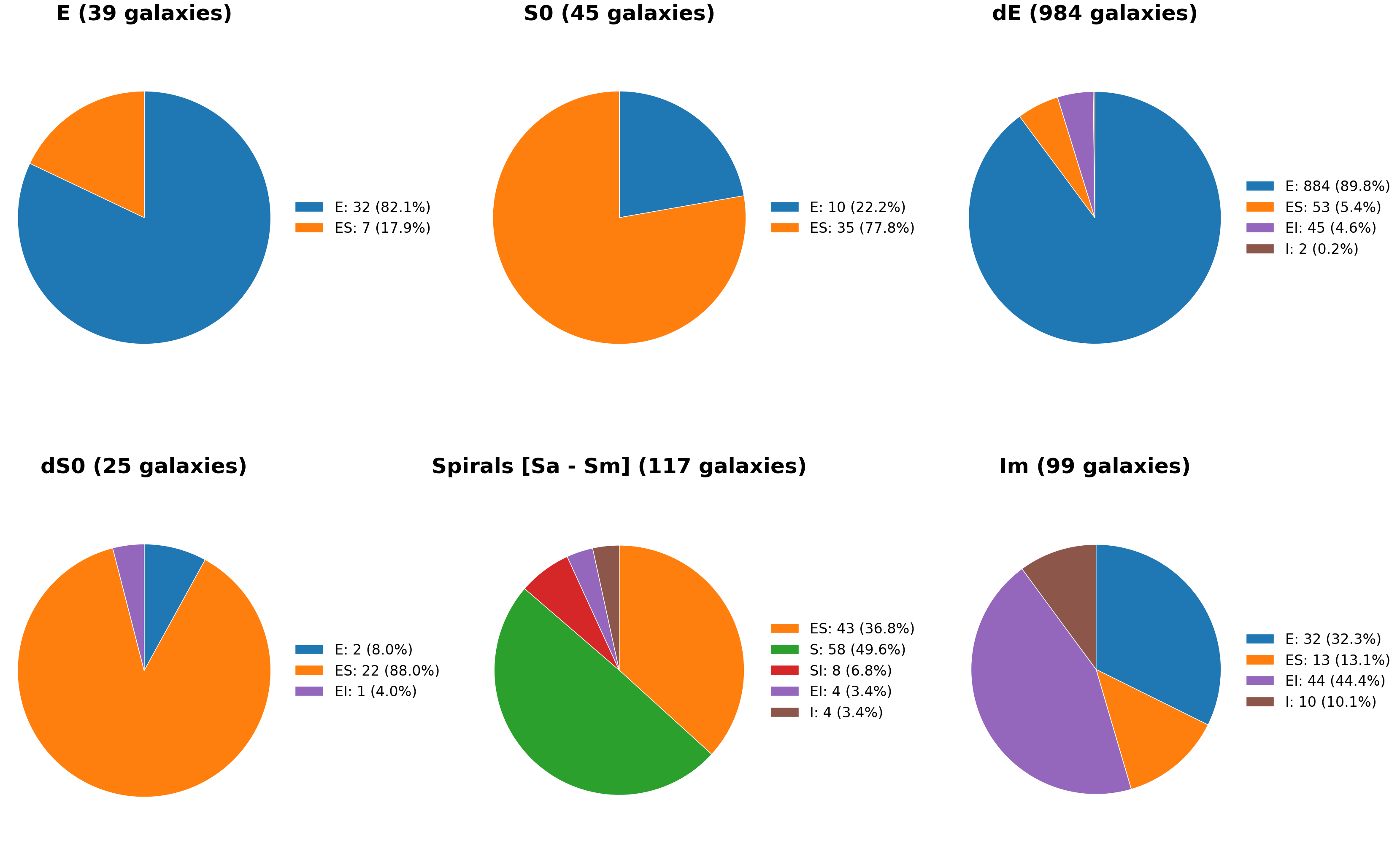}
\caption{Comparison of our morphological classifications to those from the Virgo Cluster Catalog of \citet{Binggeli1985}. The pie charts show the breakdown of our NGVS classifications for each of six major morphological categories from the VCC: Es; S0s; dEs/dE,Ns; dS0s/dS0,Ns; spirals (Sa -- Sm); and irregulars (Im).
\label{fig:VCC_Pie_Chart}}
\end{figure*}

Figure \ref{fig:VCC_Pie_Chart} compares the VCC and NGVS classifications in the form of pie charts. Each chart shows the division of NGVS types for six of the most common galaxy types in the VCC. From left to right, these are {\tt E}  galaxies, {\tt S0} (lenticular) galaxies, {\tt dE} galaxies, {\tt dS0} galaxies, spiral galaxies (types {\tt Sa} through {\tt Sm}) and {\tt Im} galaxies. Each pie chart shows the division of structural codes according to the NGVS classifications.\footnote{Another 174 galaxies in the overlap sample have uncertain classifications (with catalog entries of {\tt ?}) or represent very rare types in the VCC; as a result, they are not included in this comparison.}

\begin{figure*}[ht!]
\centering
\includegraphics[width=0.85\textwidth]{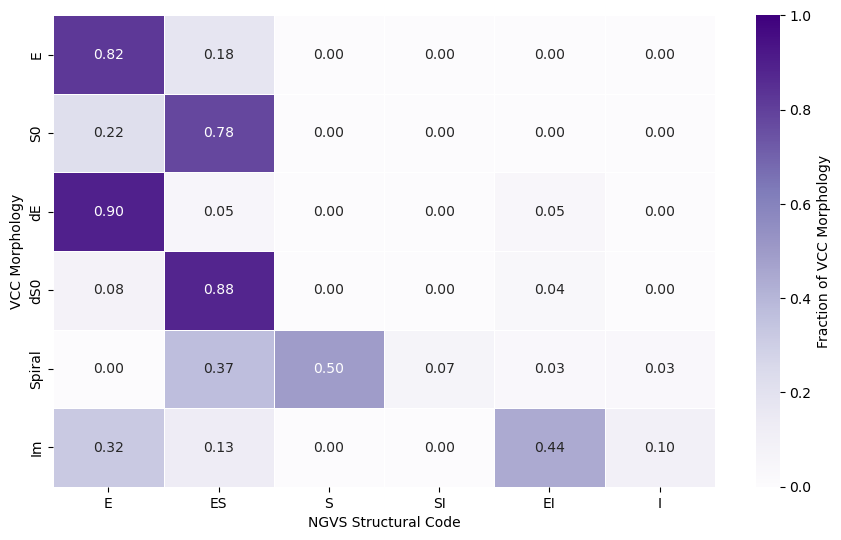}
\caption{Fractional confusion matrix comparing VCC and NGVS structural classifications. Each row corresponds to a VCC morphological class, and the colour shading indicates the fraction of galaxies assigned to each NGVS structural category.}
\label{fig:VCC_ConfusionMatrix}
\end{figure*}

To complement these pie charts, we present in Figure~\ref{fig:VCC_ConfusionMatrix} a confusion matrix that shows the fractional breakdown of NGVS structural classifications for each of the six major VCC classes. This heatmap provides a more quantitative view of the classification overlap between the two catalogs. As expected, early-type classes (e.g., {\tt E}, {\tt S0}, {\tt dE}, and {\tt dS0}) show very good alignment with NGVS {\tt E} and {\tt ES} classifications. Later-type systems exhibit more diversity, with VCC spirals often classified as {\tt ES} or, less frequently, other types ({\tt SI, EI, I}) in the NGVS. VCC irregulars show a broad spread, including a surprising number of galaxies with regular elliptical morphologies in their low-surface brightness outskirts. These findings underscore the importance of deep imaging when analyzing galaxy structural properties.

For the sample of galaxies classified as {\tt E}s in the VCC, we find good agreement, with 82.1\% of these objects similarly classified in the NGVS; the remaining 17.9\% are classified as ES galaxies, meaning that we see some evidence for disk-like or intermediate properties in the CFHT imaging. Similarly, 77.8\% of {\tt S0} galaxies in the VCC are classified as {\tt ES} types, which is expected given that this NGVS-defined class corresponds most closely to the traditional {\tt S0} class. The remaining 22.2\% of {\tt S0} galaxies in the VCC are assigned {\tt E} or {\tt S} classifications in the NGVS, which is again understandable given the transitional nature of the {\tt S0} class as originally defined.

The {\tt dE} class dominates this comparison sample with 984 galaxies classified as such in the VCC. The vast majority (89.8\%) of these objects are similarly classified as {\tt E} types in the NGVS, with the small number of exceptions classified as either {\tt ES}, {\tt EI}, or {\tt I} galaxies. The {\tt dS0} class is much less common in the VCC, with just 25 examples in their catalog. As expected, these galaxies are usually classified as {\tt ES} galaxies in the NGVS (88\%), with the remaining objects assigned {\tt E} or {\tt EI}  types, meaning that a disky nature was not obvious in the deep CFHT imaging. Overall, though, we find good agreement with the early-type classes in the VCC (i.e., the {\tt E} and {\tt S0} galaxies as well as their low-mass counterparts).

There is a greater diversity in the classifications for the two later-type classes in the VCC. Roughly half (49.6\%) of the galaxies classified as spirals in the VCC are similarly classified in the NGVS, but fully a third (36.8\%) are assigned {\tt ES} types due to the clear presence of diffuse, extended and roughly elliptical components in which the star-forming regions are embedded. The remaining systems (13.6\%) are divided among the {\tt SI}, {\tt EI} and {\tt I} types (roughly 5\% in each case). There is even greater diversity in the case of the 99 galaxies classified as {\tt Im} types in the VCC. In only 10.1\% of cases do we assign an {\tt I} type structural code. In 32.3\% of the cases, we find a predominant {\tt E} type morphology: our deeper imaging reveals more regular elliptical isophotes in these galaxies, resulting in fewer pure {\tt I} classifications and instead favoring more transitional morphologies. This shows that many of these systems, while not entirely devoid of star formation activity or substructures, appear more regular and elliptical than suggested by the VCC Im designations. The remaining {\tt Im} galaxies from the VCC have {\tt ES} or {\tt EI} structural codes (in 13.1\% and 44.4\% of the case, respectively). Again, the deep CFHT imaging points to more transitional morphologies, rather than pure {\tt I} systems.

\subsection{Comparison to the Extended Virgo Cluster Catalog\label{subsec:EVCCComp}}

A second important comparison sample is provided by the Extended Virgo Cluster Catalog (EVCC) of \citet{Kim2014}. The EVCC expanded on the original VCC by using SDSS DR7 data \citep{Abazaijan2009} to carry out a census of galaxies within a 725 deg$^{2}$ region centered on the cluster --- an area 5.2 times larger than the VCC and reaching to a projected distance of $\sim3.5$ virial radii. The EVCC utilized both SDSS imaging and spectroscopic data, defining a {\it primary morphology} (based on monochromatic $g$-, $r$- and $i$-band images as well as combined $gri$ color images) as well as a {\it secondary morphology}, determined from the  SED. 

The EVCC contains a total of 1589 galaxies, 676 of which do not appear in the VCC. Cataloged galaxies were selected spectroscopically, with a requirement that radial velocities be less than 3000 km s$^{-1}$. A total of 1324 galaxies were selected on the basis of SDSS redshifts, while an additional 265 galaxies without SDSS spectra were identified using redshifts collected from the NASA Extragalactic Database (NED).\footnote{The NASA/IPAC Extragalactic Database (NED)
is operated by the Jet Propulsion Laboratory, California Institute of Technology,
under contract with the National Aeronautics and Space Administration.} The EVCC covers a much larger area than the NGVS but our comparison is naturally limited to galaxies in common between the two surveys, which amounts to just 766 galaxies that can be matched directly to the NGVS --- all in the inner region of the EVCC. Still, as we show below, this matched sample exhibits broad morphological diversity and thus provides a useful point of comparison.  

The EVCC primary classifications are based on optical CCD imaging from the SDSS \citep{York2000}. \citet{Kim2014} employed a classification scheme that differs from the one used here, dividing galaxies into four categories: {\tt ellipticals}, {\tt disk galaxies}, {\tt irregulars} and {\tt early-type dwarfs}. These classes are assigned respective digital codes of 100, 200, 300 and 400 in the EVCC catalog, sub-classes excluded. The overlap between the two studies is limited by the depth of the SDSS photometry and spectroscopy, and includes only galaxies brighter than g$^{\prime}\sim$ 20. 

\begin{figure*}[ht!]
\centering
\includegraphics[width=0.75\textwidth]{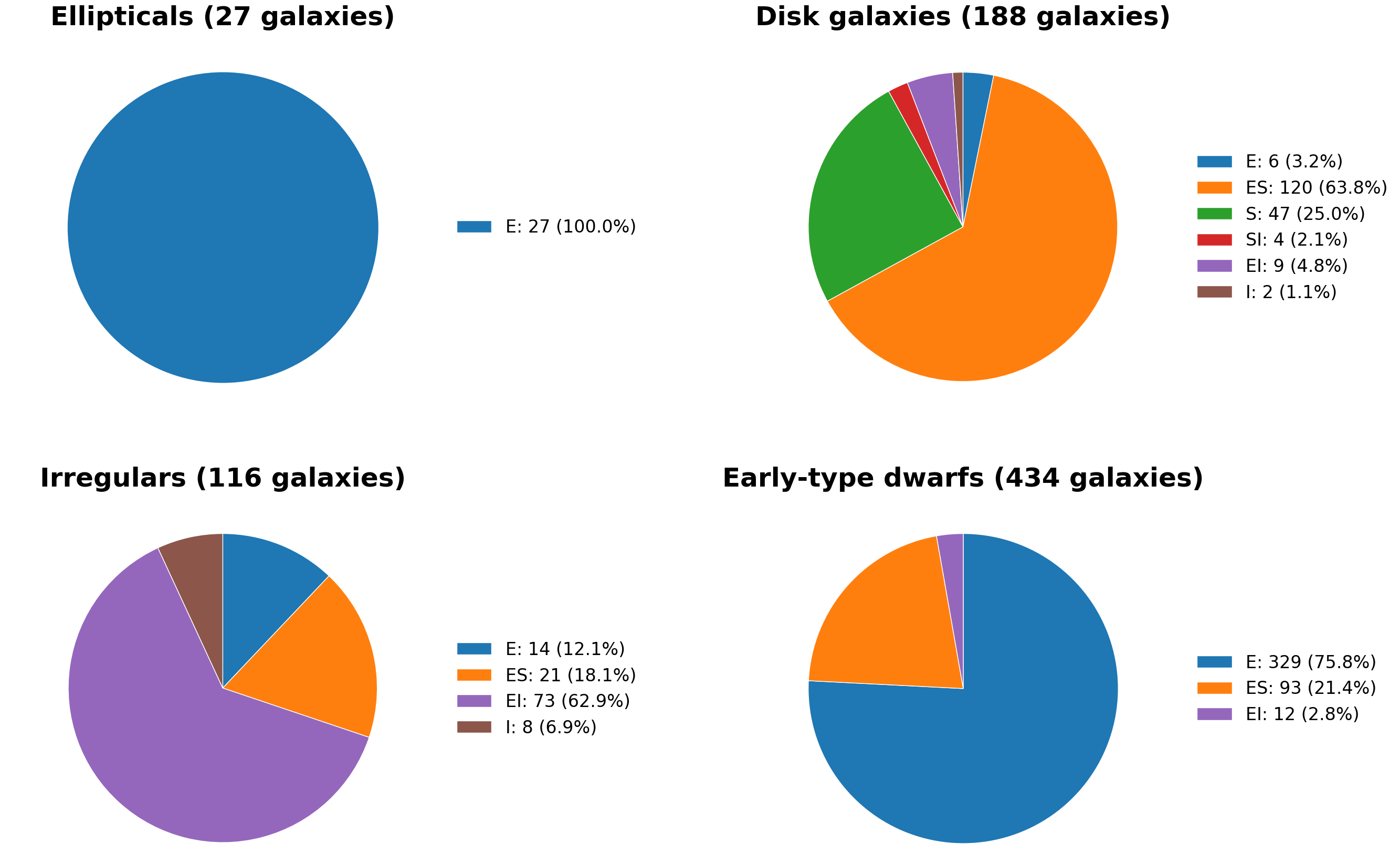}
\caption{Comparison of our morphological classifications to those from the Extended Virgo Cluster Catalog (EVCC) of \citet{Kim2014}. The pie charts show the breakdown of our NGVS classifications for each of four primary morphological categories in the EVCC: ellipticals; disk galaxies; irregulars; and early-type dwarfs.
\label{fig:EVCC_Pie_Chart}}
\end{figure*}

Figure \ref{fig:EVCC_Pie_Chart} shows a comparison of the NGVS  morphologies to those from the EVCC. The four panels correspond to the main EVCC morphological classes, as labeled in the figure. The number of matched galaxies in each EVCC category is also labeled in the figure. The pie charts show the breakdown of NGVS structural codes within these four EVCC classes. 

\begin{figure*}[ht!]
\centering
\includegraphics[width=0.85\textwidth]{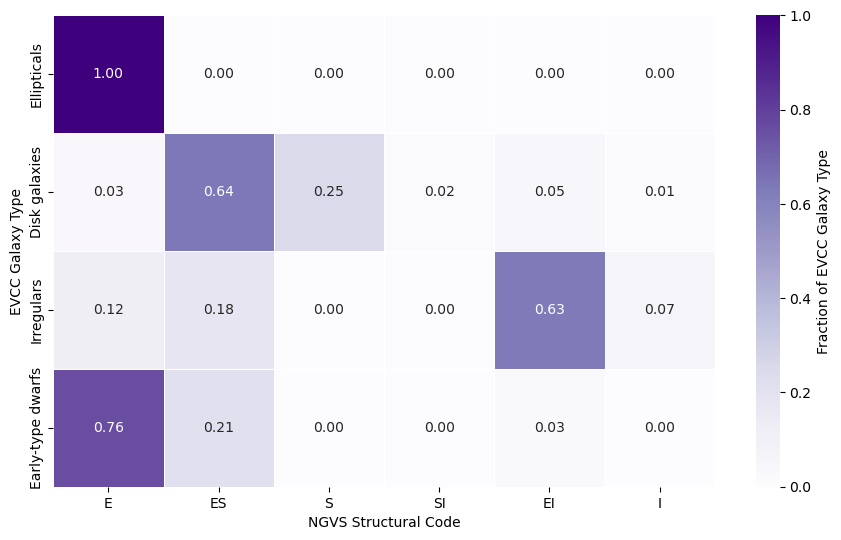}
\caption{Same as Figure \ref{fig:VCC_ConfusionMatrix} for comparing the primary morphological classifications in the EVCC with NGVS structural codes.}
\label{fig:EVCC_ConfusionMatrix}
\end{figure*}

An alternate comparison between the EVCC and NGVS structural classifications is shown in Figure~\ref{fig:EVCC_ConfusionMatrix}. Here we display a fractional confusion matrix, akin to Figure~\ref{fig:VCC_ConfusionMatrix}, where now each row corresponds to one of the four primary EVCC morphological classes. The figure highlights the strong consistency for early-type dwarfs and ellipticals, with every matched EVCC ellipticals being classified as {\tt E} in the NGVS, and 97\% of early-type dwarfs being assigned either {\tt E} or {\tt ES} types. Disk galaxies show a broader spread but still largely fall within disk-like NGVS classifications (i.e.,  91\% are classified as having {\tt ES-}, {\tt S-} or {\tt SI-}type morphologies). Irregular galaxies display the widest distribution of structural types in the NGVS. Only 7\% are classified as {\tt I}, while 63\% are labeled as {\tt EI}, meaning that a significant fraction of Irregular EVCC galaxies appear more elliptical-like than in their original EVCC designation, as was the case for {\tt Im} galaxies in the VCC.

Several conclusions can be drawn from this comparison. The sample is, unsurprisingly, dominated by galaxies belonging to the EVCC's early-type dwarf class, which comprises 57\% (434/766) of the entire sample. 422 of these 434 galaxies are assigned either {\tt E} or {\tt ES} types in the NGVS; the remaining twelve are classified as {\tt EI}s: i.e., transitional objects that show some signs of a later morphological type. Among the 188 objects classified as disk galaxies in the EVCC, 167 are classified as either {\tt ES}, {\tt S}  or {\tt SI} in the NGVS, meaning they show some disk-like features in the deeper CFHT imaging; the remaining 21 objects are classified as {\tt E}, {\tt EI} or {\tt I}. Among the 116 galaxies classified as irregular in the EVCC, nearly 70\%  are found to have {\tt EI} or {\tt I} types in the NGVS. In fact, a sizable number of galaxies in this EVCC class are assigned earlier-type morphologies in the NGVS (i.e., 14 and 21 objects are classified as {\tt E} or {\tt ES}, respectively) which we interpret as the result of more regular, diffuse features appearing in the deeper CFHT imaging. Finally, we find excellent agreement among the 27 (high-mass) galaxies classified as {\tt E}s in the EVCC with every object similarly classified as {\tt E} in the NGVS.
Despite the differing classification schemes, the agreement between the two studies is very good, especially for EVCC ellipticals, disk galaxies and early-type dwarfs. 

\begin{figure*}[ht!]
\centering
\includegraphics[width=0.75\textwidth]{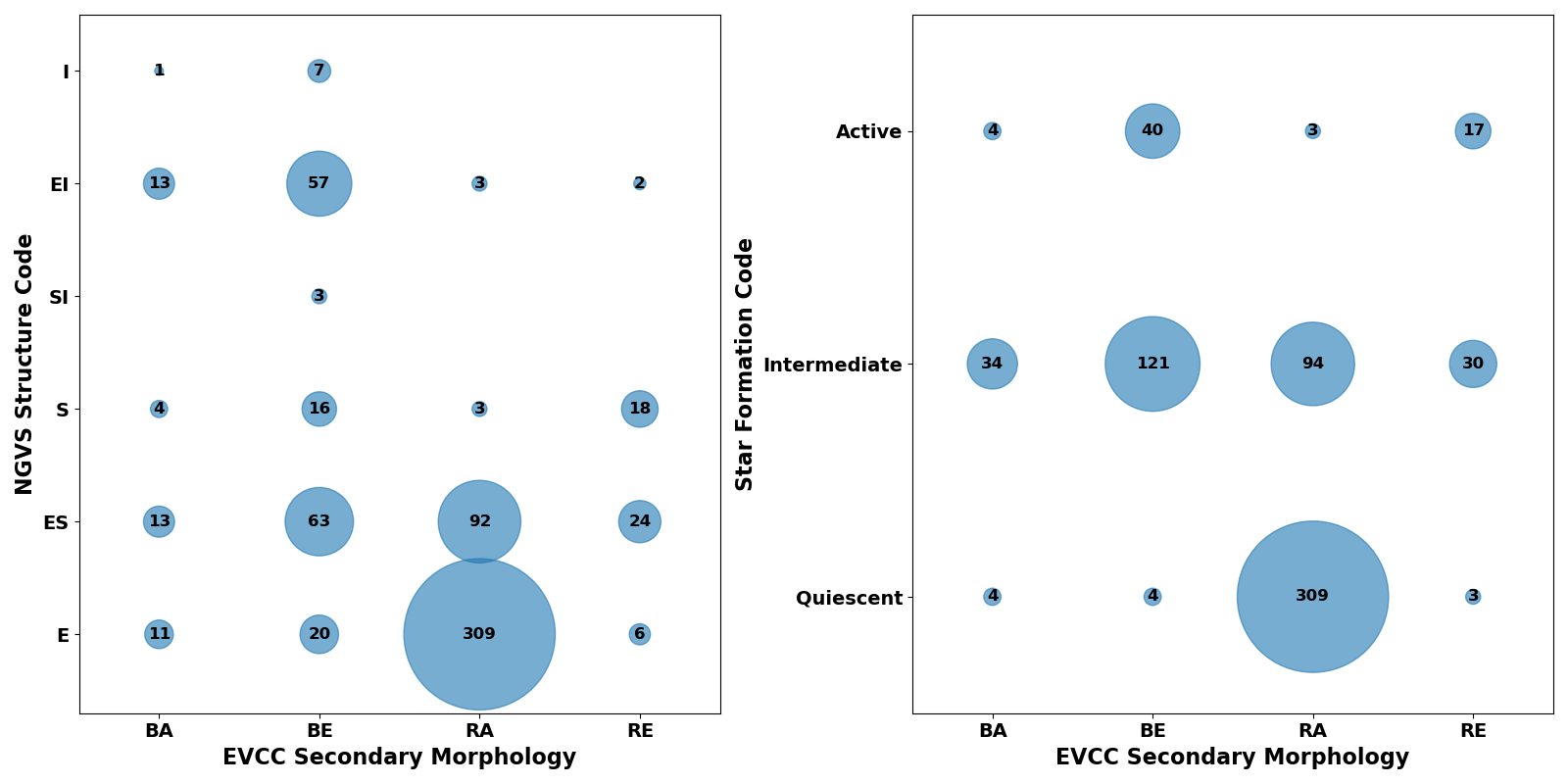}
\caption{Secondary morphologies from the EVCC compared to our structural and star formation codes (left and right panels, respectively). The symbol size is proportional to the number of galaxies at each point on this grid, with the number of galaxies labeled in each case. Note that most of the 661 galaxies in this comparison have either {\tt RA} or {\tt BE} morphologies in the EVCC.
\label{fig:EVCC_Comparisons}}
\end{figure*}

Figure \ref{fig:EVCC_Comparisons} compares our classifications to the {\it secondary} morphologies from the EVCC. As explained in \citet{Kim2014}, there are four categories for these secondary morphologies, which are based on a combination of SDSS optical spectroscopy and broadband SEDs: {\tt Red Absorption (RA)}, { \tt Red Emission (RE)}, { \tt Blue Absorption (BA)} or { \tt Blue Emission (BE)}. Galaxies are classified as ``red" if their spectrum matches that of a model for an early-type galaxy and ``blue" if their spectrum matches that of a blue, star-forming galaxy; the absorption or emission criteria are determined from the appearance of H$\alpha$ in either absorption or emission. 

Most galaxies with secondary morphologies (74.8\% of the matched sample) fall into the {\tt BE} and {\tt RA} categories, as would be expected from the broad separation of galaxies into star-forming and quiescent populations \citep[e.g.,][]{Strateva2001,Kauffmann2003,Baldry2004, Schawinski2014}. This general picture is supported by the NGVS classifications: 98.5\% of the {\tt RA} galaxies have {\tt E} or {\tt ES} structural codes, whereas 88\% of {\tt BE} galaxies have some level of morphological structure, and are not smooth {\tt E} galaxies (mainly {\tt ES}, {\tt EI} or {\tt S}). In terms of star formation codes, 97.5\% of {\tt BE} galaxies are classified as {\tt A} or {\tt I} types, while 76.1\% of {\tt RA} galaxies are classified as {\tt Q}.

Of course, it is worth noting that the SDSS spectral classifications utilized in the EVCC are based on spectra taken through a 3\arcsec fiber, meaning they reflect conditions in the central region of the galaxy rather than its global properties. In contrast, our classification account for the entire galaxy so some discrepancies between the two methods may be expected -- for instance, a galaxy that appears quiescent overall but hosts a central starburst or AGN could be classified differently, as could a star-forming disk with a central old bulge. 

The rarer, and more anomalous, secondary types --- the RE and BA classes --- provide a glimpse into the often complex evolutionary processes operating in the cluster. Visual inspection of the 50 galaxies belonging to the RE class (6.5\% of the matched sample) reveals these objects to have predominantly late-type morphologies and low levels of star formation, or disk-like emission embedded with the elliptical isophotes typical of older, redder stellar populations (in some cases with appreciable dust content). This class also includes a small number of E galaxies that appear quiescent but evidently exhibit nuclear emission from their centers, likely the signature of centrally concentrated star formation or emission from an active nucleus.

The 42 BA galaxies (5.5\% of the sample) appear to trace similar populations but are generally comprised either of objects with more elliptical morphologies and embedded/distributed star formation, or galaxies with transition-type morphologies having red central stellar populations and offset star formation. These complexities are well captured by the classification scheme introduced here, which includes several transitional, or mixed, structural and star formation categories.

Taken together, the comparisons to the VCC and EVCC reveal some consistent trends. In both cases, we find excellent agreement for massive early-type galaxies and early-type dwarfs, which are overwhelmingly classified as {\tt E} or {\tt ES} in the NGVS. Differences are most apparent among late-type systems: galaxies originally classified as spirals or irregulars are often assigned intermediate types such as {\tt ES}, {\tt EI}, or even {\tt E} in the NGVS, likely reflecting features revealed thanks to the greater depth and spatial resolution of the CFHT imaging. For example, the CFHT imaging allows us to detect diffuse outer envelopes or regular isophotes that may have been missed in earlier, shallower surveys. While the EVCC includes a spectroscopically motivated secondary classification scheme, mismatches with NGVS star formation classes may be largely attributable to differences in spatial coverage as EVCC spectroscopy is confined to the central 3\arcsec, while NGVS classifications are based on the galaxy as a whole. 

Finally, we briefly examined classifications from Galaxy Zoo 2 (GZ2; \cite{Willet2013}), which provided morphological votes for $\sim$300,000 SDSS galaxies with $m_{r} < 17$. Although Virgo systems are underrepresented in GZ2, owing to their low redshift, large angular sizes, and the tendency of the SDSS pipeline to mis-handle nearby galaxies through shredding, deblending, or crowding effects (\citealt{Blanton2005, Lintott2011, Willet2013}) we identified 117 NGVS galaxies with one-to-one positional matches (within 3 arcseconds). All matched galaxies were classified as “smooth” in the primary GZ2 task, broadly consistent with their predominantly early-type morphologies. However, this agreement is largely superficial: 62.3\% were flagged as potential mergers, and 57.6\% had bar vote fractions $P_{\mathrm{bar}} > 0.5$, suggesting structural features not reflected in the dominant vote. Additionally, all matched galaxies were also classified as having “no bulge” in the GZ2 bulge task. These inconsistencies highlight the limitations of crowd-sourced classifications for faint, compact, or low-surface-brightness galaxies in SDSS imaging, and underscore the importance of high-resolution visual classifications from the NGVS.

\section{Results} 
\label{sec:results}

\subsection{Breakdown of Morphological Types} \label{sec:distribution}
\label{sec:breakdown}

We consider the implications of our classifications for the evolution of the cluster and its member galaxies, including the interpretation of the color-magnitude relation, the morphology density relation, and  other galaxy scaling relations. Before doing so, however, we examine the breakdown of morphological sub-types within the cluster. This global breakdown is reported in Table~\ref{tab:full_ngvs_morphologies} which summarizes the distribution of galaxies by both structural and star formation code. 

For the cluster as a whole, the vast majority of galaxies are classified as quiescent ({\tt Q}) in star formation properties (82.3 $\pm$ 1.5\%) and {\tt E}-like in structure (84.0 $\pm$ 1.5\%). This preponderance is unsurprising given that galaxies in clusters are known to be quenched when they enter the cluster potential \citep{Fabjan2010,Taranu2014,Paccagnella2016}, after pre-processing at the group scale \citep{Sanchez-Jannsen2008,Haines2015, Jaffe2018,Cortese2021,Boselli2022}. This is especially true for the low-mass galaxies ($\lesssim 10^9 M_{\odot}$) that dominate the NGVS sample: i.e., 3431 out of the 3689 galaxies in the NGVS sample are in the low-mass regime (93.0\%). Most low-mass galaxies are found to be quiescent (85.1 $\pm$ 1.6\%) and have an E-like structure (88.4 $\pm$ 1.6\%). The rest of the low-mass systems predominantly exhibit transitional structures and intermediate levels of star formation. 

Although they comprise only a small fraction of the sample overall, galaxies with irregular, transition-type morphologies are more common within the low-mass sample. The low-mass sample, outside of elliptical galaxies, is dominated by {\tt ES} (4.5 $\pm$ 0.4\%) and {\tt EI} (5.4 $\pm$ 0.4\%) structure codes, but they are rarely seen to be actively forming stars (2.3 $\pm$ 0.3\%); more often, they are classified as having intermediate levels of star formation activity (12.5 $\pm$ 0.6\%). This may be the result of the short quenching times for dwarf galaxies in the cluster; alternatively, it may point to the onset of incompleteness for very faint and blue galaxies in the NGVS sample \citep[see the discussion in][]{Ferrarese2020}. Environmental processes such as ram pressure stripping and tidal interactions likely accelerate quenching, rapidly transforming star-forming dwarfs into quiescent systems before they can be detected \citep{Boselli2006,Boselli2014, Ferrarese2020}. Additionally, the detection of diffuse, low-surface-brightness galaxies may be biased due to incomplete recovery of their surface brightness profiles \citep{Davies2016,Ferrarese2020}. It is also true that, for the faintest and most compact galaxies in our sample, the limited S/N and relatively small number of spatial resolution elements may make it difficult to discern subtle structural peculiarities or clearly identify low levels of star formation activity.

To estimate the extent of incompleteness in our sample, we make use of the simulations presented in Section 4 of \citep{Ferrarese2020}, to which we refer the reader for full details. Briefly, the NGVS catalog is 50\% complete at $M_g = -9.1$, and we detect 403 galaxies fainter than this limit, including 28 systems with $(u - i) < 1.0$, indicative of blue stellar populations. Assuming 50\% completeness, we infer that an additional $\sim$ 400 faint galaxies, including $\sim$ 30 faint, blue systems could be undetected. This hypothetical population of ``missing" galaxies could partially explain the apparent paucity of blue, actively star-forming dwarfs at the faintest magnitudes.

At the high-mass end ($M_{*} \ge 10^{9} M{_\odot}$), late-type morphologies such as {\tt S} (26.3 $\pm$ 3.3\%) and {\tt ES} (48.7 $\pm$ 4.5\%) systems are at their peak abundance, with intermediate ({\tt I}) (31.8 $\pm$ 3.7\%) and active ({\tt A}) star-forming galaxies (23.3 $\pm$ 3.1\%) combining to form a slight majority of the high-mass sample. This abundance of star-forming systems is less pronounced at the highest stellar masses: among the 236 high-mass galaxies, 52 are quenched ellipticals ({\tt E} and {\tt Q}), corresponding to 22.0\% $\pm$ 3.1\%. Massive, quenched ellipticals are usually located in the densest cluster regions and are responsible for much of the quenching of lower-mass satellite galaxies \citep{dressler1980galaxy,Thomas2005,Boselli2006,vanderWel2010,Peng2010,McConnell2013,Ma2014,Gabor2015,Cappellari2016,Boselli2022}. It is thus expected that, compared to field galaxy populations, Virgo will contain a richer diversity of galaxy morphologies. This should be true even at the highest masses given the frequency of mergers, galaxy interactions and feedback processes, including bursts of star formation triggered in a dense cluster environment \citep{Gunn1972,Press1974,White1978,dressler1980galaxy,Smith2012,Kauffmann2004,Wetzel2013}. Notably, pure {\tt I} structure codes are not found at the high-mass end, as these galaxies tend to be classified more frequently as transition-types (e.g., {\tt ES}, {\tt EI}) due to their underlying structural complexity. A purely irregular classification requires significant morphological disturbance without any discernible ordered structure, a criterion rarely met by high-mass galaxies. This follows the trend identified in \citet{Sandage1985a}, where no Im galaxies were found with $ B_{T} \lesssim 14$, with more massive irregulars typically exhibiting features characteristic of more complicated transition-type morphologies Our classification similarly reflects this trend, reinforcing the expectation that truly irregular, high-mass galaxies should be exceedingly rare.

The morphological complexity of galaxies in Virgo is captured by our classification scheme which allows us to broadly map galaxy properties to observables from the NGVS catalog. In S\ref{sec:special_types}, we will use our morphological classifications, catalog measurements, and non-parametric morphologies to highlight the full diversity of galaxies in the Virgo cluster, and identify members of rare, unusual or otherwise noteworthy galaxy classes.

\begin{figure*}
\begin{center}
\begin{tabular}{c}
\includegraphics[height=9.0cm]{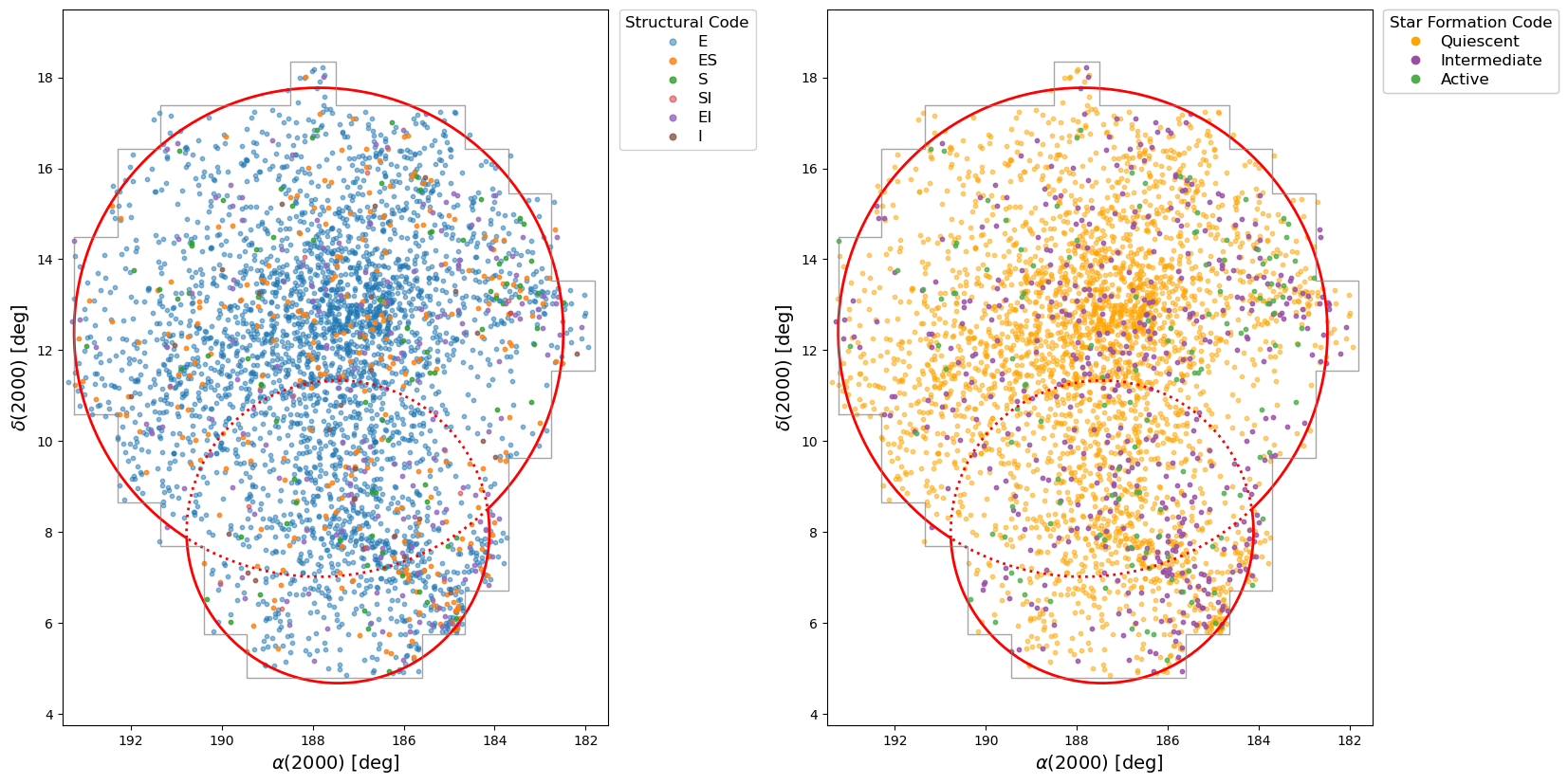}
\end{tabular}
\end{center}
\caption 
{(Left panel) Distribution on the sky of 3689 confirmed or probable Virgo cluster members from the NGVS. Symbol colors indicate different morphological structural codes, as indicated in the legend. For reference, the large red circles indicate the virial radii of Virgo's two major subclusters, centered on M87 (A) and M49 (B). The footprint of the NGVS is shown in light gray. (Right panel) Same as the previous panel, except now with symbols showing different star formation codes.} 
\label{fig:NGVS_Both}
\end{figure*} 

\begin{figure*}[t]
\centering
\includegraphics[width=0.9\textwidth]{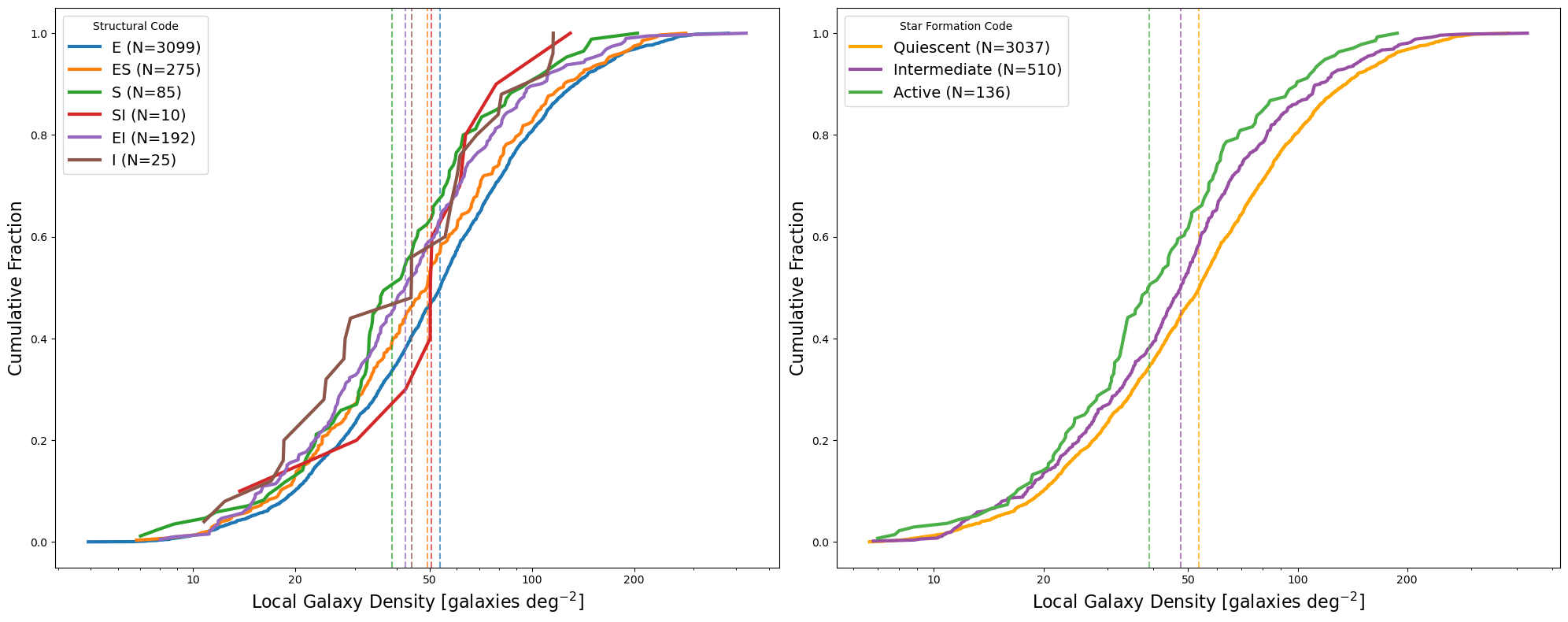}
\caption{Cumulative distribution functions (CDFs) of local surface density for 3675 NGVS galaxies with measured surface density (out of the 3689 total in the sample). The left panel shows the distribution by structural code, while the right panel presents the distribution by star formation code. Dotted lines indicate the median density for each classification.}
\label{fig:CDF}
\end{figure*}

\begin{figure*}
\begin{center}
\begin{tabular}{c}
\includegraphics[height=9.0cm]{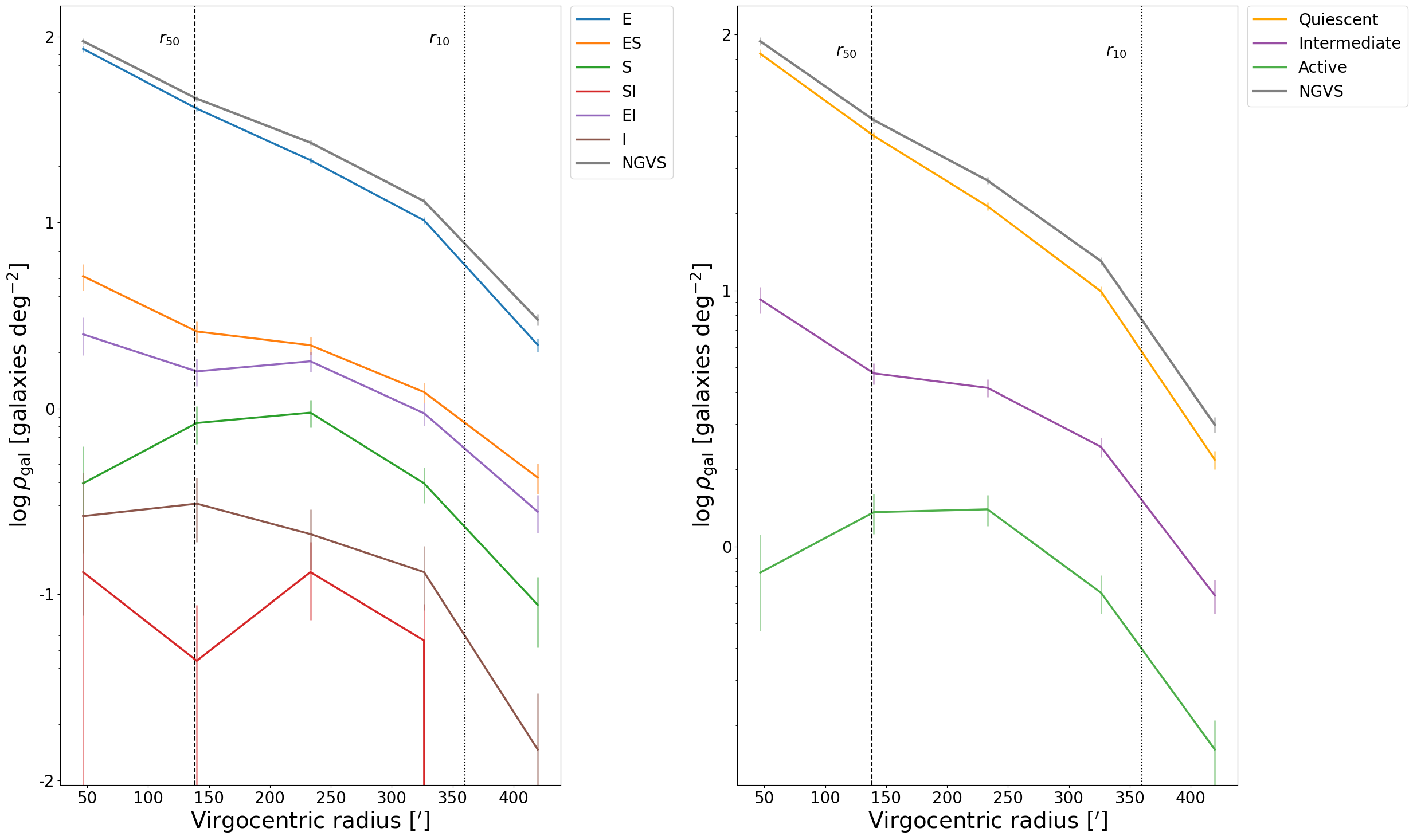}
\end{tabular}
\end{center}
\caption 
{Surface density profiles for Virgo cluster galaxies. Profiles are measured in circular annuli centered on M87. The left and right panels show  profiles found for subsets of the galaxy population after dividing by structural code and star formation code, respectively. The vertical dashed lines show the radii at which the total (i.e., full sample) surface density profile falls to one-half ($r_{50}$) and one-tenth ($r_{10}$) of its peak value.}

\label{fig:GalaxyDensity}
\end{figure*} 

\begin{figure*}
\begin{center}
\begin{tabular}{c}
\includegraphics[height=7.0cm]{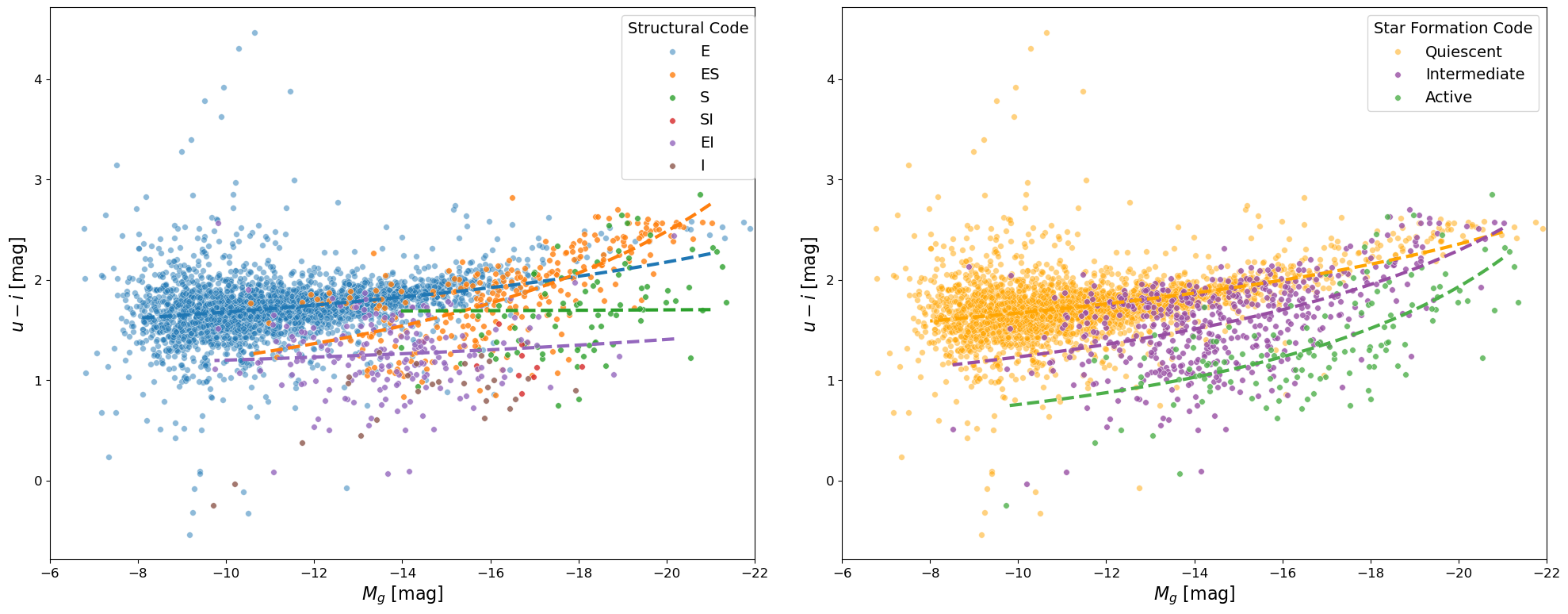}
\end{tabular}
\end{center}
\caption 
{(Left panel) Color-magnitude diagrams for 3689 members of the Virgo cluster from the NGVS. The  distribution of galaxies in the $g-(u$-$i)$ plane is shown with symbols color coded according to NGVS structural code, as indicated in the legend. Note that power-law fits for the SI and I structural codes are not shown to the sparsity of data. (Right panel) Same as the previous panel, except with color coding based on the NGVS star formation code.} 
\label{fig:combined_CMDs}
\end{figure*}

\subsection{Ensemble Properties}
\label{sec:ensemble}

How are the different morphological subpopulations distributed within cluster and across the color-magnitude relation? Figure~\ref{fig:NGVS_Both} shows a map for the full sample of NGVS galaxies inside the survey footprint; individual galaxies are labeled by structural and star formation code in the left and right panels, respectively. The dominance of galaxies with early-type morphologies and quiescent star formation codes is apparent in both panels. In fact, it is this large population of mainly quiescent, E-like galaxies that most clearly defines the main subclusters and filaments within the cluster. The highest-density regions are marked principally by these early-type populations, consistent with previous findings \citep[e.g.,][]{Binggeli1987,Binggeli1991,Schindler1999}.

A useful way to explore possible trends with environment is to connect the visual morphologies to local density. The left and right panels of Figure~\ref{fig:CDF} show cumulative distribution functions for local surface density as a function of structural and star formation code, respectively, with the median density for each code overlaid. The surface densities used in this analysis were calculated directly from the NGVS catalog using the areas inscribed within the radial distance to the 10th nearest neighbor. In both panels, the distributions for the subpopulations show considerable overlap, presumably due to projection effects. Galaxies with late-type structural codes ({\tt S}) and disturbed morphologies ({\tt EI, I}) or elevated star formation rates ({\tt I, A}) reach their peaks at low densities, as would be expected if these objects preferentially occupy the outer cluster regions where infall is underway. Nevertheless, we see signs that some of these morphologically complex or active galaxies are found in denser regions. The peak of {\tt SI} galaxies in slightly denser regions suggests that these transitional, late-type galaxies may be associated with galaxy interactions and mergers. This likely points to episodes of star formation triggered by recent interactions or close passes with companion galaxies. It is also possible that some of these galaxies appear in denser environments because of projection effects rather than physical associations. Given the lack of clear merger signatures, projection may, in some cases, be a more plausible explanation.

Building on these observed qualitative trends, we conducted a more rigorous statistical analysis of the local density distributions across star formation and morphological codes in the NGVS sample using both Kolmogorov–Smirnov (KS; \citealt{Kolmogorov1933, Smirnov1948}) and Anderson–Darling (AD; \citealt{Anderson1952, Scholz1987}) tests. A KS test confirmed a strong correlation between star formation activity and environment, with {\tt Q} and {\tt A} galaxies exhibiting a highly significant difference (p $\approx$ 0.001). 
{\tt I} and {\tt A} galaxies also differ (KS p $\approx$ 0.041; AD p $\approx$ 0.001), indicating that star formation activity is environmentally regulated. {\tt Q} and {\tt I} galaxies likewise show a significant separation (p $\approx$ 0.001 in both tests). The AD test consistently yields higher significance than the KS test, reflecting its greater sensitivity to differences in the distribution tails. 
Together, these results reinforce the conclusion that local density significantly influences star formation activity.

For morphological codes, both the KS and AD tests identify strong environmental differences between {\tt E} and {\tt S}, with p $\approx$ 0.001 in both tests. 
This finding supports the well-established trend that ellipticals preferentially reside in denser environments, while spirals are more common in lower-density regions \citep[e.g.,][]{dressler1980galaxy, Postman1984, Goto2003}. 
{\tt ES} and {\tt S} show only marginal separation in the KS test (p $\approx$ 0.068), but the AD test indicates a highly significant difference (p $\approx$ 0.001). 
This discrepancy suggests that the median densities of these two populations are similar, but their extremes differ in a statistically meaningful way, which is captured by the AD test. 
Ellipticals and transition ellipticals ({\tt E} vs.\ {\tt EI}) also differ significantly (KS p $\approx$ 0.002; AD p $\approx$ 0.001), with both tests agreeing on the environmental distinction.

Beyond these key contrasts, many comparisons among intermediate categories (e.g., {\tt EI} vs.\ {\tt I}, {\tt S} vs.\ {\tt SI}) yield weak or non-significant results in the KS test, yet the AD test often reports differences at the $\sim$0.003 level. 
This pattern underscores the complementarity of the two tests: the KS test is most sensitive to differences near the median of the distributions, while the AD test is more sensitive to differences in the tails. For highly significant results where p-values approach numerical limits, we report them as p $\approx$ 0.001, though they may in fact be smaller. 
Overall, the KS and AD tests converge on the broad trends—especially in distinguishing {\tt E} from {\tt S} and separating {\tt Q} from {\tt A}, but the AD test highlights additional differences that are less apparent in the KS test. 
This emphasizes the importance of applying both methods together to fully capture the environmental dependence of galaxy morphology and star formation activity.

To further highlight trends with position, Figure~\ref{fig:GalaxyDensity} shows density profiles for the different morphological populations, with structural code and star formation code shown in the left and right panels, respectively. Profiles are calculated using five radial bins in Virgocentric radius. We see the expected (albeit gentle) morphology-density relation wherein {\tt S} and other late-type structure codes are found at preferentially larger distances from the center of the cluster, while early-type, quiescent galaxies are especially dominant at smaller radii. The overall density profile for cluster members (shown as the gray curves in each panel) is dominated by E-like, quiescent galaxies with quiescent star formation, as noted above. We summarize these trends more qualitatively in Table~\ref{tab:profile_results}. From left to right, this table records, for each morphological sub-population, the peak density and the projected radii where the profiles fall to one-half ($r_{\rm 50}$) and one-tenth ($r_{\rm 10}$) these peak values.

Figure~\ref{fig:combined_CMDs} shows how the different morphological sub-populations are distributed across the color-magnitude diagram (CMD). Results are shown after dividing by structural code (on the left) and star formation code (on the right). We parameterize the overall behavior by fitting power-laws to each sub-population having enough data points to fit reliably. The various loci are most clearly delineated in terms of star formation, where there is a clear connection between integrated color and star formation code: i.e., bluer galaxies tending to be actively forming stars, while red galaxies are overwhelmingly classified as quiescent. 

Dividing by structural code results in greater overlap among the power-law fits, making it more difficult to meaningfully distinguish between different structural codes in color-magnitude space. This is understandable given the high degree of structural complexity captured by each code and the fact that the most common morphological type ($\tt{E}$) extends from the highest-mass ellipticals down to some of the faintest dwarfs in the cluster. These caveats aside, the CMDs demonstrate that the NGVS classifications are reflecting --- independent of integrated color --- expectations that the Red Sequence is composed of mostly quenched, passively evolving galaxies having smooth, elliptical-like structures, while the Blue Cloud is populated by later-types exhibiting active or intermediate star formation \citep{Roediger2017}.

Our conclusions are consistent with previous findings, based on significantly smaller samples, that the Virgo cluster --- despite its relatively unrelaxed state --- exhibits a clear morphology-density relation, with early-type (Red Sequence) systems dominating its high-density regions (e.g., \citep{Binggeli1987, Phillips1998, Trentham2002, Lieder2012}. This observation is usually taken as evidence that these low-mass galaxies correspond to subhalos that have experienced strong mass loss, or spent extended periods inside massive halos \citep{Lisker2009}. 

\begin{figure*}[t]
  \centering
  \includegraphics[width=\textwidth]{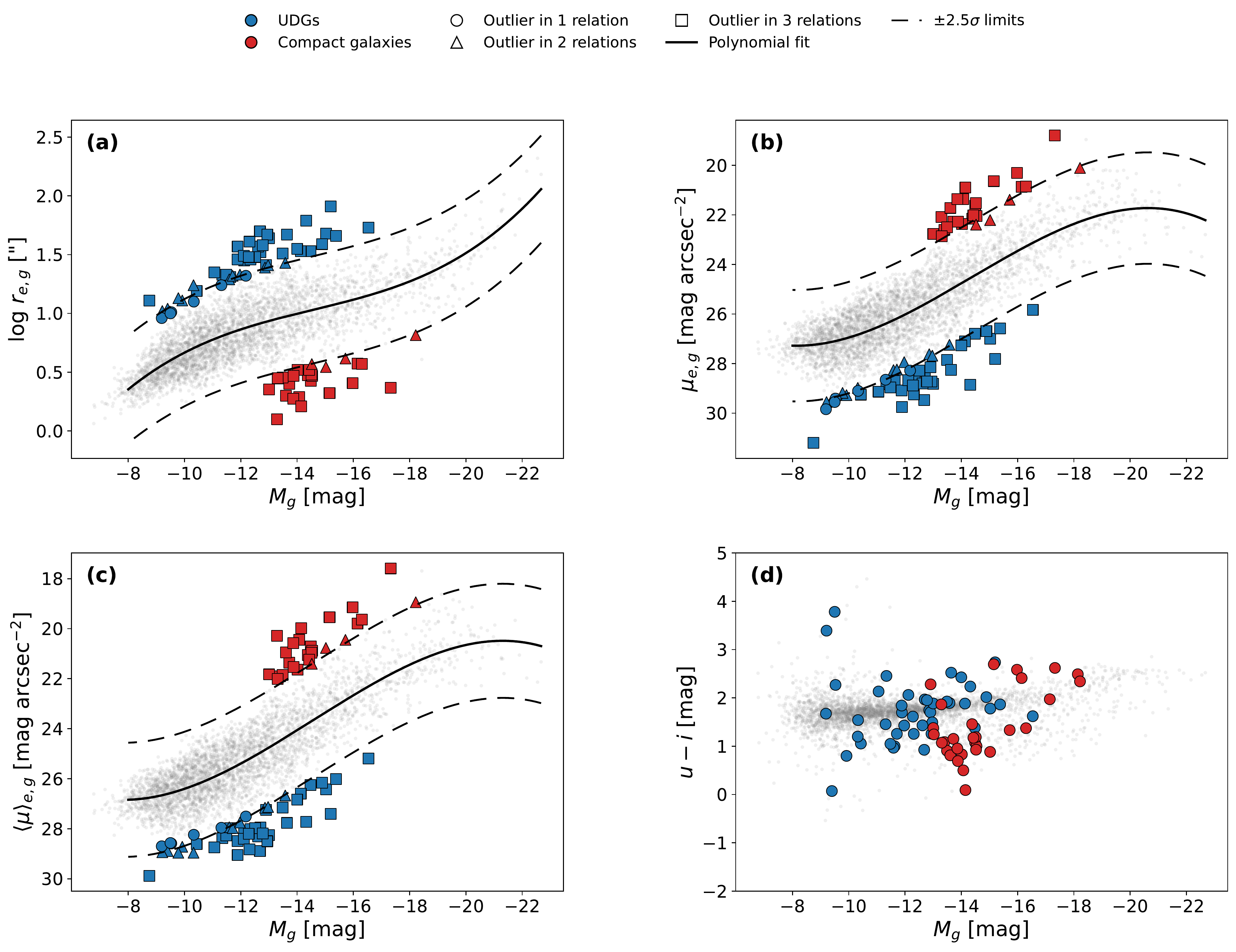}
  \caption{Scaling relations for NGVS galaxies with $\log M_\star \le 10$. 
  Panels: (a) size--magnitude ($\log r_{e,g}$ vs.\ $M_g$), 
  (b) effective surface brightness ($\mu_{e,g}$ vs.\ $M_g$), 
  (c) mean surface brightness within one effective radius ($\langle\mu\rangle_{e,g}$ vs.\ $M_g$), and 
  (d) color--magnitude ($u-i$ vs.\ $M_g$). 
  Black curves show polynomial fits to the NGVS population; long-dashed black lines mark the $\pm2.5\sigma$ envelopes. 
  Ultra-diffuse galaxy candidates (blue) and compact galaxies (red) are highlighted. 
  Marker shape encodes how many relations the object is an outlier in: circles $=$ 1, triangles $=$ 2, squares $=$ 3. 
  Outliers are defined as points lying outside the $\pm2.5\sigma$ envelope of the corresponding fit.}
  \label{fig:size-magnitude}
\end{figure*}


\section{Discussion}
\label{sec:discussion}

\subsection{Galaxy Types of Special Interest} 
\label{sec:special_types}

Previous studies of the Virgo (\citealt{Binggeli1985, Lisker2006, Ferrarese2006, Kim2014}) and Fornax clusters (\citealt{Ferguson1989,Venhola2019}) have compiled catalogs of galaxies with unusual or notable morphologies. We do the same in this section, focusing on galaxy types that may be of particular interest for future study. Tables~\ref{tab:_types_ce}-\ref{tab:_types_mergers} provide basic data for these noteworthy objects, including morphological classifications and brief notes; similar data for the complete sample of NGVS galaxies will be presented in the upcoming NGVS galaxy catalog (Ferrarese et al. 2026, in preparation).

\subsubsection{Compact Galaxies}
\label{sec:types_ce}

In cluster environments, quiescent galaxies with small sizes are generally thought to be byproducts of interactions with massive companions or the mean cluster gravitational field. Such interactions are known to strip the lower-mass galaxy of much of its star forming disk, leaving behind a dense core with little or no star-forming gas \citep{Bekki2001,Bekki2003,Meza2003,Ferrarese2016} although other formation channels are certainly possible \citep[see, e.g.,][and references therein]{Du2019}. Such ``compact elliptical galaxies" (cEs) are similar in structure and morphology to the archetypal Local Group galaxy, M32, which is thought to have been heavily influenced by tidal interaction with M31 \citep{Bekki2001, Ibata2001,Choi2002, DSouza2018}. This class of galaxy was therefore referred to as ``M32 type" systems by \cite{Binggeli1985} who identified a handful of compact, intermediate-luminosity, high-surface brightness ellipticals in Virgo (see their Table~XIII).

Virgo is also known to contain a number of Blue Compact Dwarf (BCD) galaxies --- compact, metal-poor galaxies that exhibit bursts of star formation: i.e.,  \cite{Searle1973,Gerola1980,Thuan1981,Sandage1984,Thuan1991, Amrutha2024}. Table~XII of \cite{Binggeli1985} gives a catalog of BCD candidates selected from the~VCC. Once thought to be young galaxies undergoing their first bursts of star formation \citep{Sargent1970,Lequeux1980, Kunth1988}, the prevailing view now holds that BCDs are subject to repeated bursts of star formation caused by interactions with massive galaxies and/or the cluster potential \citep{Thuan1991,Papaderos1996,Cairos2001a,Cairos2001b,dePaz2003,Hunter2004,Cairos2010}. As a class, BCDs are sometimes referred to as dwarf irregulars and may be subject to the same extreme tidal forces that give rise to M32-type galaxies \citep{Bekki2008,Kado-Fong2020,Zhang2020a,Chhatkuli2023}. 

Since the exact definition of what constitutes a BCD or cE is still a topic of debate \citep{Ekta2010,Yin2011,Lelli2012,Shen2014, Hsyu2018}, we proceed by restricting the sample selection to structurally compact galaxies: i.e., galaxies located at the extreme end of NGVS galaxy size distribution. Following \citet[hereafter L20]{Lim2020}, we use several scaling relations to identify structurally extreme sizes: compact galaxies and ultra diffuse galaxies (UDGs; discussed in \S\ref{sec:types_udg}). From left to right, the first three panels of Figure~\ref{fig:size-magnitude} show the relationship between absolute $g$-band magnitude, $M_g$, and effective radius, $r_e$, effective surface brightness, $\mu_{e,g}$, and mean effective surface brightness, $\langle{\mu_{e,g}}\rangle$, respectively. All 3601 confirmed, probable and possible cluster members with log M$_{*}$/M$_{\odot}$ $\le$ 10 are plotted as gray points, where the high mass galaxies are excluded in order to reduce the potential bias in the trend introduced by the highest mass galaxies which are best treated as a separate class of galaxies entirely. The fourth panel of this figure shows the color-magnitude diagram for this same sample, with $(u-i)$ as the color index.

For this analysis, we use similar scaling relations as L20 who identified UDGs and studied their globular cluster systems. We define compact galaxies as those objects that deviate by at least 2.5$\sigma$ from the mean scaling relations, though in the opposite direction of UDGs (i.e., we select outliers with small size and high-surface brightness). Table~\ref{tab:_types_ce} lists properties for the 30 compact galaxies that are outliers in at least two of the three scaling relations in Figure~~\ref{fig:size-magnitude}. From left to right, this table records the NGVS ID, NGVS ``nickname", VCC number (if available), right ascension, declination, $g$-band magnitude and morphological classification. The final column gives brief notes from the visual classifications.

Rather than separating M32-like galaxies and BCDs, once we exclude the most massive galaxies we simply select galaxies that fall at least 2.5$\sigma$ below a polynomial curve in log $r_{e,g}$--$M_g$ space and at least 2.5$\sigma$ above the respective surface brightness curves. These are the regions where we expect to find the most concentrated objects. In fitting each scaling relation, we found that a cubic polynomial was able to capture the overall trends in the data while imposing few additional restrictions on the sample. Note that the polynomials used in L20 are fourth-order polynomials determined by maximum likelihood fitting of the observed scaling relations in the Virgo core region and differ slightly from those used here. 

Inherent in this method is the assumption of uniform scatter in the standard deviation at all magnitudes, instead of assigning magnitude-dependent confidence intervals based on the scatter within magnitude bins. Our approach is more robust given the relative sparsity of objects in the tails of our distributions (i.e., we wish to avoid biases that might be introduced by using a more complicated functional form). 

\begin{figure*}[h]
\centering
\includegraphics[width=1.0\textwidth]{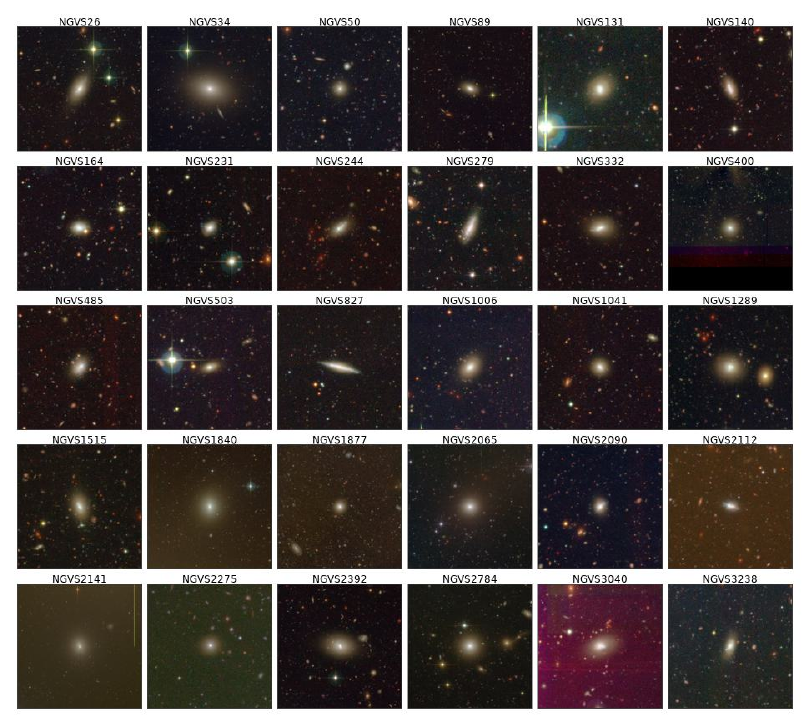}
\caption{A mosaic of the 30 compact galaxies listed in Table~\ref{tab:_types_ce}. The sample includes both M32-like objects, Blue Compact Dwarf (BCD) galaxies, as well as intermediate systems. Each panel measures 1.5\arcmin$\times$1.5\arcmin~ with North at the top and East to the left. The sample includes both M32-like objects (e.g., NGVS50 and NGVS400) and BCD galaxies (e.g., NGVS1515 and NGVS231).
\label{fig:compact_mosaic}}
\end{figure*}
As is evident from Table~\ref{tab:_types_ce}, the compact galaxies are predominantly early-type in structure, with {\tt E} systems comprising 16/30 (53\%), alongside 10 {\tt EI} (33\%) and four {\tt ES} (13\%) galaxies. Consistent with the thumbnails in Figure~\ref{fig:compact_mosaic}, this indicates modest morphological diversity within an overall early-type population. The star formation codes likewise show some diversity in the properties of these galaxies: 21/30 (70\%) are classified as {\tt I}, while the remaining 9/30 (30\%) are quiescent {\tt Q}).
The varied star formation types are additional evidence that a sample of compact galaxies selected in size-magnitude space is populated by both star-forming and quiescent compact galaxies belonging to distinct sub-populations. This diversity is also apparent in the right panel of Figure \ref{fig:size-magnitude}. The CMD in the figure shows that the compact galaxy sample includes both very red and very blue objects.

This color spread hints at the different evolutionary pathways within for compact galaxies. The redder compact galaxies may be the remnants of once more massive systems that have undergone significant stripping, while the bluer objects could be more akin to BCDs which have retained or recently reignited star formation. Altogether, the sample thus provides a snapshot of the evolutionary channels that drive morphological changes within the cluster.
The varied star formation modes provide additional evidence that a sample of compact galaxies selected in size-magnitude space includes both star-forming and quiescent sub-populations. This apparent bimodality follows with the agreed upon picture that compact galaxies can follow different evolutionary pathways, with some retaining or reigniting star formation while others remain quenched.

Interestingly, only eight galaxies in this sample were classified as either M32-like or BCD systems by \citet{Binggeli1985}. Of these eight galaxies, five were classified as cEs and three as BCDs. We attribute this modest overlap due to the superior depth and resolution provided by the NGVS, bearing in mind that some discrepancies will be inevitable given differing definitions of ``compact". Our selection criteria have the advantage of being based on quantitative and uniform cuts in the size-magnitude surface brightness-magnitude diagrams but some caution is advisable since a few galaxies usually considered to cE-type systems (e.g., VCC1192, VCC1440, VCC1545, etc) just fail to satisfy our selection criteria.

\subsubsection{Ultra-Diffuse Galaxies} 
\label{sec:types_udg}

Although large and diffuse low-mass galaxies had been known for some time \citep{Sandage1984,Binggeli1985,Impey1988,Impey1996,Dalcanton1997,Blanton2005}, the Ultra Diffuse Galaxy (UDG) class was introduced by \citet{vanDokkum2015} for Coma cluster galaxies with large effective radii (r$_{e,g} \geq 1.5$ kpc) which were also exceptionally faint and diffuse ($\mu_g \geq 24$).\footnote{Table~XIV of \cite{Binggeli1985} gives a compilation of ``dwarfs of very large size and low-surface brightness".} 

Recent work on UDGs has revealed some strange and seemingly contradictory properties. Some UDGs without dark matter  have been reported \citep{vanDokkum2018,vanDokkum2019, Trujillo2021, Toloba2023}, posing a potential challenge to $\Lambda$CDM \citep{Bullock2017}. At the same time, other UDGs have been found to be dark-matter-dominated in their central regions \citep{Peng2016,Beasley2016,vanDokkum2017, Toloba2018, Carleton2019, Toloba2023}. More work is needed to understand these seemingly contradictory results and ascertain their homogeneity as a class. 

Using the same methodology as in \S\ref{sec:types_ce}, we identify 50 UDGs as diffuse outlier galaxies in our scaling relations. These galaxies are labeled in Figure~\ref{fig:size-magnitude} and listed in Table~\ref{tab:_types_udg}. A natural comparison for our analysis is the catalog of UDGs presented in L20. These authors defined a sample of 44 UDGs, 26 in a ``primary" sample of objects that are at least 2.5$\sigma$ outliers in each of three scaling relations (i.e., magnitude versus effective radius, effective surface brightness and mean effective surface brightness). They further defined a ``secondary" sample of 18 galaxies that are 2.5$\sigma$ outliers in at least one of these scaling relations, bringing their sample to 44 galaxies.

For our UDG sample selection, we opt to utilize the same 2.5$\sigma$ outlier selection used in \S\ref{sec:types_ce} while also adopting the approach of defining primary and secondary samples. We focus on the galaxies that lie at least 2.5$\sigma$ above the best-fit curve derived from our polynomial fits, rather than those in L20, in each of the scaling relations to present a uniform and consistent sample of UDGs across all mass regimes. By design, we are less stringent with our selection criteria than L20 who limited their sample only to ``certain" and ``probable" NGVS members (membership classes 0 and 1, respectively) whereas; by contrast, we include ``possible" NGVS members (class 2) in our sample in order to compile a more complete sample of UDG candidates. In total, 33 of our galaxies fall into our primary sample while another 17 objects belong to our secondary sample.

Some galaxies are absent from either our sample or that of L20 due to slight differences in how we define UDGs. In total, 30 galaxies in our sample are also present in L20, while 14 galaxies are exclusive to L20, and 21 galaxies are unique to our catalog.
For images of some of these galaxies, we refer the reader to L20, where they present visual representations of their UDG sample as part of their analysis. The 14 galaxies found only in the L20 sample fall just outside our outlier regions; however, we still consider them viable UDG candidates that narrowly missed our slightly different selection criteria.

An additional method to study the evolutionary properties of UDGs in cluster environments is to look for signs of nucleation, i.e. the presence of a compact nuclear star cluster (NSC). NSCs are dense stellar systems, typically only a few parsecs in size but reaching masses up to $10^{7-8}M_\odot$, and are among the most common central components of low- and intermediate-mass galaxies \citep{Boker2002, Côté2006, Neumayer2020} Their incidence depends on both galaxy mass and environment, and nucleation fractions provide important clues to the formation and survival of galaxies in clusters \citep{Georgiev2016, Antonini2015, Leaman2022}. Within our sample, 44 galaxies show no visual evidence for a nucleus, four are classified as potentially nucleated, and only two exhibit a clear nucleus. We further investigated these systems using structural decompositions with GALFIT \citep{Peng2010}, following the approach of L20. In that analysis, 48 galaxies were found to be non-nucleated, one was deemed likely non-nucleated, and two were confirmed to host a nucleus. Both approaches therefore converge on the conclusion that nuclear star clusters are rare in UDGs, with an occupation fraction of only (2–6)/50 $ = $ 4–12\%. This is markedly lower than the typical nucleation fractions of 20–60\% found in galaxies of comparable luminosity in Coma and Virgo \citep{Sanchez-Janssen2019, Lim2018, Lim2020}. Although the precise mass ranges of the two UDG samples differ, both analyses consistently point to a suppressed nucleation fraction in UDGs relative to normal galaxies at fixed stellar mass.

Table \ref{tab:_types_udg} shows that most of our UDG candidates are quiescent (38/42 $\simeq 90$\%) and roughly elliptical in structure (40/42 $\simeq$ 95$\%$). This is consistent with past work on UDGs \citep{vanderBurg2016,Rong2017, Rong2020}. However, even though we expect the majority of the UDG population to be faint, diffuse and quiescent, we are most likely missing some information on UDG structure and star formation due to the challenging nature of these objects for in-depth study. For example, there is a chance that some of our Class 2 objects may actually be very faint image artifacts \citep[see the discussion in][]{Ferrarese2020}.


Unlike compact galaxies, which span both quiescent and star-forming regimes, UDGs in our sample are almost exclusively quiescent. Of the 50 UDGs, 46 ($\simeq$ 92$\%$) are classified as quiescent, while only four ($=$ 8$\%$) show intermediate or irregular star formation signatures. This reinforces the picture of UDGs as diffuse, red systems that largely follow passive evolutionary tracks, with star-forming cases being the exception rather than the rule.

A handful of morphologically interesting systems warrant mention. NGVS227 (VCC197) is a quiescent elliptical that nevertheless hosts an embedded nucleus, likely the result of a tidal stripping event. NGVS421 (VCC360) is a diffuse EI galaxy with irregular structure and evidence of residual star formation throughout its body. NGVS1160 (VCC811) is an extended LSB system possibly interacting with a nearby faint LSB companion. NGVS1528 (VCC1017) is an EI system with multiple shell-like features suggesting recent interactions. NGVS1593 (VCC1052) shows a disturbed morphology and hints of low-level star formation. Finally, NGVS3404, although only a possible cluster member, exhibits EI structure with filaments or streams.

Taken together, these few exceptions highlight that most UDGs in Virgo appear structurally simple and quiescent, but a minority do retain signatures of star formation or environmental disturbance. The full degree of morphological complexity exhibited by these objects, however, will be better assessed using resolved stellar populations based on future space-based studies.

\subsubsection{UCD Transition Objects}
\label{sec:types_ucdtrans}

Ultra-compact dwarf (UCD) galaxies are dense, compact stellar systems with luminosities comparable to faint dwarf galaxies but sizes roughly an order of magnitude smaller ($10 \lesssim r_e \lesssim 50$~pc). Located between globular clusters and classical dwarf galaxies in the size-magnitude relation, the origin of these objects has been debated since their discovery more than two decades ago \citep{Hilker1999,Drinkwater2000, Phillipps2001}. It was quickly recognized that these systems did not fit neatly into the context of dwarf galaxy formation \citep{Moore1996, Bekki2001}, which led to suggestions they arise from tidal ``threshing" of nucleated dwarf galaxies \citep{Bekki2003, Mieske2008}. Early studies established that some UCDs do indeed show evidence for tidal distortion \citep{Drinkwater2003,Hasegan2005}, supporting of this interpretation. However, other formation mechanisms remain feasible, at least in some cases, including the coalescence of super stellar clusters formed in gas-rich mergers \citep{Fellhauer2002,Fellhauer2005} or formation from small-scale peaks in the primordial dark matter power spectrum \citep{Drinkwater2004}.

Recent studies \citep[e.g.,][]{Zhang2015, Fahrion2020, Saifollahi2021, Mayes2021, Mahani2021, Napolitano2022} have tried to discern the most likely formation channels for UCDs. In the Virgo cluster, \citet{Liu2015, Liu2020} used NGVS imaging to assemble a sample 612 UCD confirmed or candidate UCDs distributed across the entire cluster. They showed that these UCDs are concentrated towards the most massive galaxies and noted an apparent connection between local density and envelope size for UCDs and nucleated dwarf galaxies (dE,N), suggestive of an evolutionary link. This connection was examined in more detail by \citet{Wang2023} who presented a quantitative link between location within the cluster and the prominence of the diffuse component surrounding both UCDs and the nuclei of dE,N galaxies. 

Our visual inspection of NGVS galaxies bears directly on a possible evolutionary link between UCDs and nucleated dwarfs. According to the threshing scenario, some of these transition-type galaxies may be the progenitors of UCDs. As a byproduct of our morphological analysis, we identified an initial sample of galaxies as candidate UCD ``transition objects": i.e., galaxies that appear to contain an especially prominent nucleus, or a small, diffuse or elongated envelope that could be the sign of advanced tidal stripping. 

To refine this initial sample, we impose additional criteria to select galaxies with prominent nuclei and faint, compact, or elongated envelopes.  

Our final sample of 115 candidates is presented in Table~\ref{tab:_types_ucdtrans}.

A key question is how our sample compares with that of \citet{Wang2023}, who used surface-brightness profile fitting to study the connection between UCDs and nucleated dwarfs. Cross-matching their 48 ``strongly nucleated'' systems to our UCD-transition catalog yields 31 galaxies in common ($\approx$65\% of their sample and $\approx$27\% of our sample). This difference reflects complementary selection criteria: our morphology-driven approach, focusing on asymmetric/irregular, low-surface-brightness envelopes, naturally admits many systems at intermediate stripping phases, whereas the profile-fitting strategy of \citet{Wang2023} favors objects with smooth, well-behaved diffuse components. Taken together, the two samples bracket different points along the stripping sequence: our non-overlap objects plausibly represent earlier, more disturbed phases, while their non-overlap systems tend to be more regular and may be further along in the transformation.

Our identified UCD transition candidates display a wide range of morphologies, from elongated, asymmetric systems to compact galaxies dominated by bright nuclei. Several candidates (e.g., NGVS308, NGVS1032, NGVS1719, NGVS1849, NGVS2024, NGVS2244, NGVS2681, NGVS2785, NGVS3099) exhibit elongated or irregular low-surface-brightness envelopes, consistent with ongoing tidal interactions. Others (e.g., NGVS149, NGVS787, NGVS1032, etc) show bright nuclei embedded in only small, faint envelopes, suggestive of systems in advanced stages of stripping where most of the diffuse component has already been lost \citep{Wang2023}. In addition, NGVS317 stands out as having a possible offset or double nucleus, while NGVS935 is classified as both a candidate UDG and a UCD transition system, reflecting the extreme tidal effects shaping its morphology. This morphological diversity highlights the transitional status of these objects and their potential role in bridging nucleated dwarfs and bona fide UCDs.

Our sample of candidate transition objects provides an opportunity to investigate the physical processes driving envelope disruption during an apparently short-lived evolutionary phase. These objects may be attractive targets for future spectroscopic and photometric studies aimed at understanding the role of tidal stripping in UCD growth.  In particular, a quantitative comparison of envelope sizes and luminosities between our sample and those of \cite{Wang2023} could further clarify whether these differences arise due to selection effects or reflect intrinsic variations in the stripping process.

By characterizing the structural properties of these objects in detail and comparing them to both intact nucleated dwarfs and fully stripped UCDs, it may be possible to refine models of tidal stripping and nuclear survival. Future work will explore the implications of these findings for the structural evolution of low-mass galaxies in cluster environments (e.g., Wang et al. 2026, in preparation), while a broader investigation of nuclear star clusters in Virgo galaxies will be presented in a future paper in this series.



\begin{figure*}[h]
\centering

\includegraphics[width=\textwidth,clip,trim=2pt 2pt 2pt 2pt]{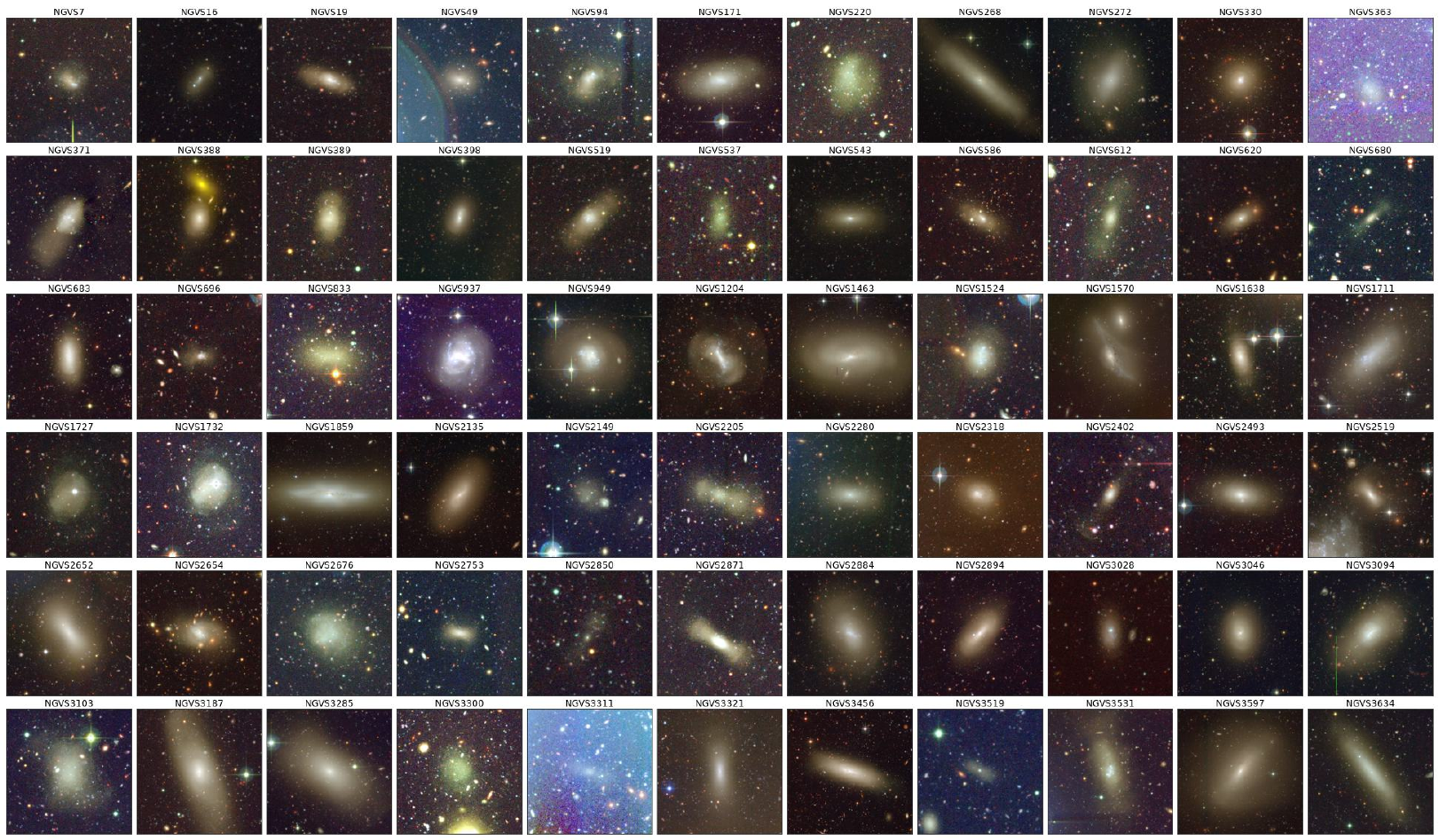}
\caption{A mosaic of 66 candidate merger or post-merger galaxies from the NGVS. Each panel measures 3\arcmin$\times$3\arcmin~with North at the top and East to the left.
\label{fig:post_merger_mosaic}}
\end{figure*}

\subsubsection{Candidate Merger and Post-Merger Galaxies}
\label{sec:types_mergers} 

Mergers, both major and minor, are recognized as important drivers of galaxy evolution \citep{Efstathiou1983,Blumenthal1984,Silk1993} since they effect the growth and morphological transformation of galaxies. Their importance as a function of environment and galaxy mass is a topic of active study \citep{Thomas2010,Jogee2009,Darg2010,Kocevski2012,Ellison2019,Vulcani2021,Driver2022}. This is particularly true in cluster environments \citep[e.g.,][]{Sheen2012,Oh2016,Oh2018,Oh2019,Piraino-Cerda2024,Kim2024} and specific examples of merger and post-merger systems in Virgo may bear on the structure of the cluster itself \citep{Bohringer1994, Lee2021}. 

Our visual analysis reveals a number of Virgo galaxies that show signs of interactions or tidal encounters --- stellar streams, tails, plumes, shells or disturbed isophotes. While the NGVS structure and star formation codes record such features when present, some galaxies appear sufficiently disturbed to suggest an ongoing or recent merger. In this section, we discuss such systems, noting that they are best described as {\it candidate} merger or post-merger galaxies given that our classifications have been made on the basis of visual appearance alone, without the benefit of the kinematic observations or numerical modeling that is usually needed to identify such systems unambiguously.

A sample of 66 candidate merger or post-merger galaxies is presented in Table~\ref{tab:_types_mergers}.
As noted above, these candidates were identified purely from their visual appearance: i.e., based on the presence of prominent shells, filaments, twisted isophotes, misaligned spiral arms and distorted morphologies that are indicative of ongoing or recent mergers. Many galaxies in the cluster show such features, so only the most prominent cases are listed in Table~\ref{tab:_types_mergers}. The sample is thus somewhat subjective in nature and should not be considered complete.

Figure~\ref{fig:post_merger_mosaic} presents thumbnail images for these 66 candidates. Each panel measures $3\arcmin\times3\arcmin$ (i.e., $14.4$ kpc on a side at the distance of Virgo). The telltale signs of ongoing or recent mergers are readily apparent in most cases. As expected, the galaxies span a broad range in mass ($6.47 \le \log(M_{*}/M_{\odot}) \le 10.81$) and are distributed across the cluster ($182.087^{\circ} \le \alpha \le 192.415^{\circ}$, $5.181^{\circ} \le \delta \le 15.679^{\circ}$), encompassing a wide range of structural and star formation codes.

This compilation may be useful for future studies of mergers in cluster environments, particularly among intermediate- and low-mass galaxies. The median galaxy mass is log M${*}$/M$_{\odot}$ = 7.89, placing the majority of our sample in the dwarf galaxy regime. Additionally, 53/66 (80\%) of our post merger galaxies have log M$_{*}$/M$_{\odot}$ $\leq$ 9, indicating that our sample is mostly comprised of dwarfs. 

We also note the inclusion of three dwarf mergers previously identified as early-type dwarfs with shell-like features by \citet{Paudel2017}: VCC1361 (NGVS2205), VCC1447 (NGVS2402), and VCC1668 (NGVS2871). Although the faint and extended nature of these shells makes them difficult to discern in NGVS imaging, they remain indicative of recent merger activity that has perturbed the morphologies of these galaxies. Our sample additionally includes more striking and strongly disturbed dwarf merger systems such as VCC848 and VCC479, known dwarf–dwarf mergers whose NGVS imaging has been used in the respective studies of \citet{Zhang2020a,Zhang2020b} and \citet{Sun2025}. While our visually identified sample of post-merger systems in Virgo is necessarily incomplete due to the subjective nature of classification, we nonetheless observe a significant presence of merger activity. The diversity of morphological signatures within our sample, from subtle tidal features to well-defined merger remnants, underscores the prevalence of interactions among low-mass galaxies and highlights the broad range of dynamical states they exhibit within the cluster.

\begin{figure*}[h]
\centering
\includegraphics[width=1.0\textwidth]{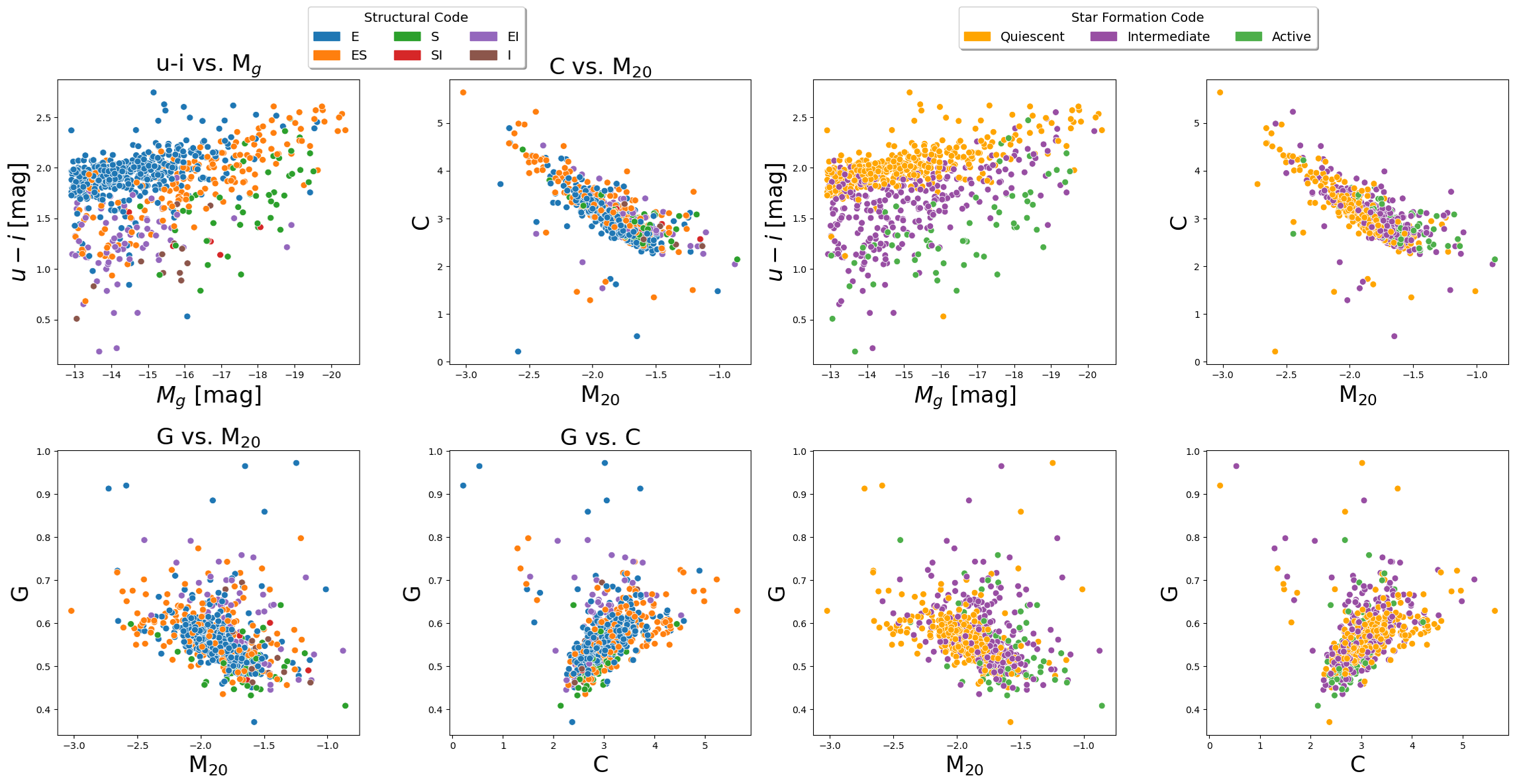}
\caption{The color-magnitude diagram for NGVS galaxies along with several scatter plots using non-parametric statistics. Galaxies are labeled by their structural code (left) and star formation code (right) for the sample of 762 galaxies having \statmorph measurements with M$_{g}$  $\lesssim $ -13 as discussed in the text. 
\label{fig:full_sample_cas_combined}}
\end{figure*}

\begin{figure*}[htbp]
\centering
\includegraphics[width=1.0\textwidth]{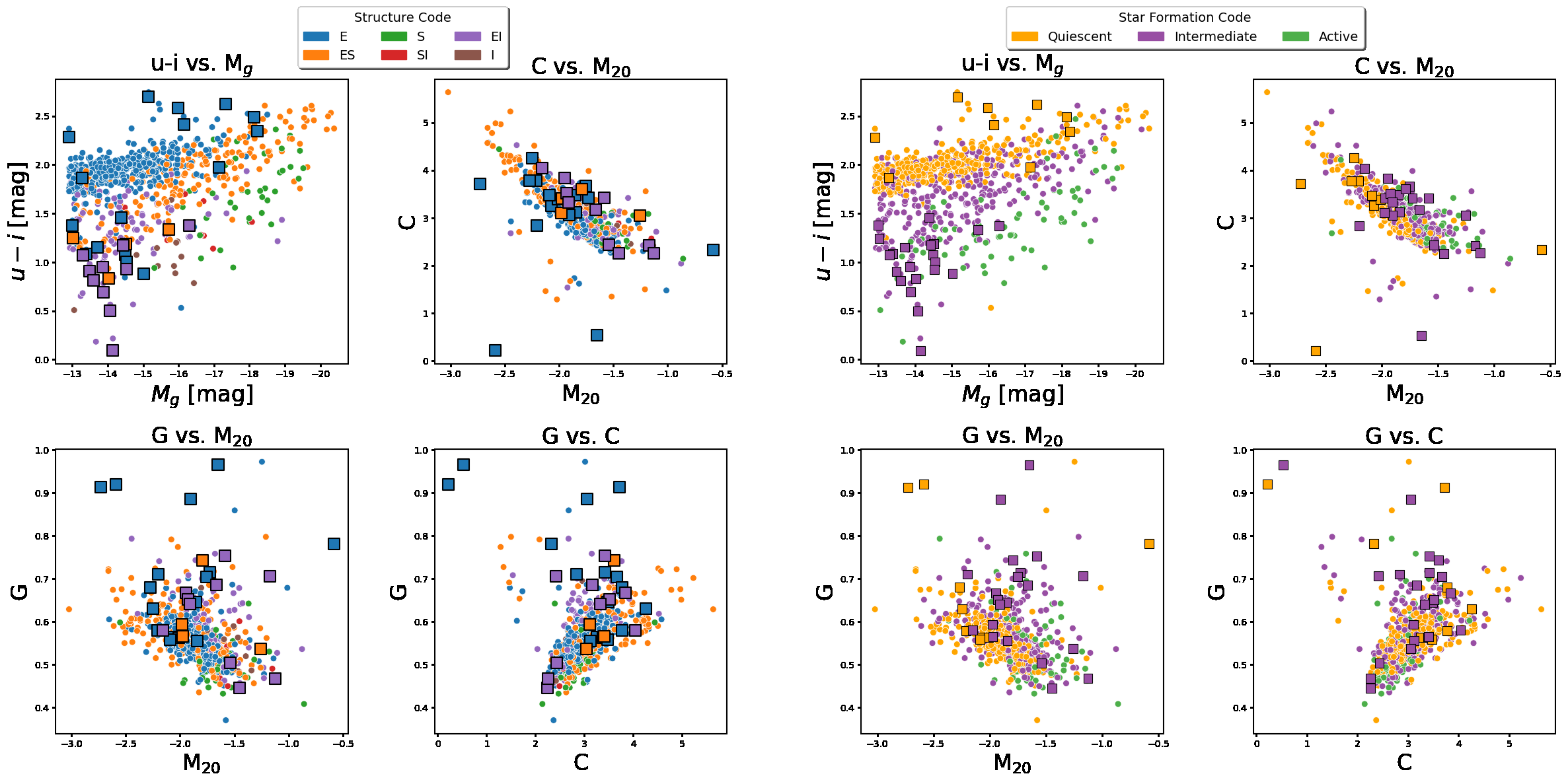}

\caption{Same as Figure~\ref{fig:full_sample_cas_combined} except with compact galaxies from \S\ref{sec:types_ce} denoted by large squares. We note that compact galaxies bifurcate in the the CMD as expected and that these galaxies are generally categorized as being more morphologically complex with higher values of C, G and M$_{20}$.
\label{fig:cas_combined}}
\end{figure*}

\subsection{Non-Parametric Measurements}
\label{sec:CAS_measurements}

\subsubsection{\statmorph measurements of the NGVS}
\label{sec:statmorph}

Non-parametric morphological statistics can be used to supplement traditional galaxy classifications. Such statistics have usually been used to classify galaxies automatically with little or no human intervention, and have never been applied to a large cluster sample dominated by low-mass, faint galaxies as is the case here.  We have run the \statmorph code of \cite{RodriguezGomez2019} on our NGVS sample to measure non-parametric morphological statistics for as many Virgo galaxies as possible.
Before describing our selection criteria, we briefly review the utility of non-parametric morphological statistics. For more thorough reviews, the reader is directed to the deeper dives on non-parametric morphological statistics in \citet{Wilkinson2022},  \citet{RodriguezGomez2019} and the section on non-parametric morphologies in the review on galaxy structure formation by \cite{conselice2014evolution}.

We focus on some of the most commonly applied non-parametric morphology statistics. These statistics are the CAS system \citep{Conselice2003} and Gini-M$_{20}$ statistics \citep{Lotz2004}. Concentration (C) \citep{Bershady2000, Conselice2003}, Asymmetry (A) \citep{Schade1995, Abraham1996, Conselice2000} and Smoothness (S) \citet{Conselice2003} define the CAS system, which quantifies different aspects of a galaxy's structure: concentration measures the fraction of light contained within a central aperture, asymmetry captures the difference between a galaxy and its image rotated by 180$^{\circ}$ and smoothness assesses the presence of small-scale substructures by comparing the original image to a smoothed version. Together, these metrics provide insight into the morphological complexity of a galaxy. This system has been successfully applied to both simulated and observed galaxy populations (e.g., \citealt{Hambleton2011,Whitney2021,Wilkinson2022, Ferreira2023,Wilkinson2024}) to explore properties such as star formation and degree of morphological transformation from galaxy interactions and mergers.

The CAS system can, as we show below, be unreliable without some restriction of our sample. These statistics were developed specifically for use on bright, high-mass galaxies observed with space-based telescopes. Since the majority of our sample are faint, diffuse galaxies, the A and S statistics in particular will suffer from the inclusion of near-zero pixel values \citep{Sazonva2024}. Very smooth and diffuse galaxies are smoothed further, due to their faint nature, in the case of S, which can erase the little morphological information available in the images. These factors will limit the utility of CAS measurements for our sample, restricting their application to only high- and intermediate-luminosity galaxies. The concentration parameter (C) is more robust to the effects of low signal-to-noise imaging and can be measured reliably for a larger number of objects.

The G-M$_{20}$ system \citep{Lotz2004} uses the Gini coefficient (G) to measure the concentration of light in an image, irrespective of where the light is concentrated. This makes it particularly sensitive to off-center knots of star formation, providing discrimination between centrally concentrated early-type galaxies and star-forming, late-type galaxies or morphologically disturbed systems. By contrast, the M$_{20}$ statistic measures the brightest 20$\%$ of galaxy flux normalized by the total second moment of light of the galaxy. It provides spatial information on the separation between the brightest regions of a galaxy, making it sensitive to disturbed morphologies and merger signatures.

With these considerations in mind, we used the \statmorph code \cite{RodriguezGomez2019} to measure non-parametric morphologies for Virgo galaxies based on $g'$-band images from the NGVS. For the reasons described above, we restricted our analysis  to galaxies with reliable measurements following the recommended procedures from \cite{RodriguezGomez2019}.  Galaxies with low-quality or unreliable \statmorph measurements were rejected, as denoted by an output \texttt{Flag} $\ge 2$. We also rejected galaxies with images having low mean S/N per pixel, $\langle$S/N$\rangle$ $\le 2.5$ \citep{Lotz2006}, since \statmorph fits are known to be unreliable in this noise regime. We applied one final cut by removing all galaxies with M$_{g}$ $\geq -13$ as galaxies fainter than this often had suspect \statmorph output diagnostic plots and measured non-parametric morphological statistics.   Applying these cuts reduced the size of the NGVS sample to 762 galaxies with trustworthy \statmorph morphologies (i.e., just the brightest $\sim$ one fifth of the catalog). 

We examined a variety of CAS diagnostic plots using the available parameters (C, A, S, M$_{20}$ and G) in order to find the most instructive combinations. Figure \ref{fig:full_sample_cas_combined} shows some general trends exhibited by our cleaned sample of galaxies. Results are shown after color-coding the galaxies by both structural code (left panels) and star formation code (right panels). In each case, the first panel shows the distribution of galaxies in the $M_g$-$(u-i)$ CMD. The remaining three panels show different combinations of CASGM$_{20}$ parameters: (1) C vs. M$_{20}$; (2) G vs. M$_{20}$; and (3) C vs. G. 

Figure \ref{fig:full_sample_cas_combined} suggests that elevated rates of star formation are correlated with higher M$_{20}$ at fixed C and G; this is more obvious in the right hand panels which are color-coded by star formation code. This trend is expected, as highly star-forming galaxies exhibit bluer colors. They also tend to appear more concentrated at a fixed M$_{20}$ and have higher G values at the same M$_{20}$. The panels on the right side of the figure show the general trend but with galaxies color-coded by structural code. Outside of the expected separation seen in color-magnitude space between early- and late-type morphologies, spanned by transition-type structure codes the distinctions between populations are more subtle in non-parametric morphology space. E-type, quiescent galaxies tend to show lower concentrations at constant M$_{20}$ values compared to later-type objects with $\tt A$ or $\tt I$ star formation codes. ES galaxies tend to define the tail of the main galaxy population in the C-M$_{20}$ plan, to high C values and more negative M$_{20}$ values.

\subsubsection{CASGM\texorpdfstring{$_{20}$}{ 20 } Measurements for Compact Galaxies}
\label{sec:casgm20_compact}

To illustrate the connections between our visual morphologies, non-parametric CASGM$_{20}$ measurements and NGVS catalog parameters, we consider the population of compact galaxies discussed in \S\ref{sec:types_ce}. These objects are attractive targets for such a comparison because they are bright enough to have reliable \statmorph measurements and may be undergoing morphological changes within the cluster environment.

Figure~\ref{fig:cas_combined} shows the same scatter plots as Figure \ref{fig:full_sample_cas_combined} but now with the sample of compact galaxies from \S\ref{sec:types_ce} highlighted as large squares. As before, the structural and star formation codes for individual galaxies (compact objects included) are given at the top of the left and right panels, respectively.

The compact galaxies in Figure~\ref{fig:cas_combined} appear to show two main components in the CMDs --- a blue, star forming population and a redder, quiescent population. This separation is most easily seen in the right panels where galaxies are labeled by their star formation code. In the right panels, where galaxies are labeled by their structural code, the picture is less clear but we still see that galaxies having higher levels of morphological complexity are generally found to have higher values of C, G and M$_{20}$, indicative of ongoing star formation. 

Within the CASGM$_{20}$ scatter plots, most of the compact galaxies fall inside the main galaxy distributions in the M$_{20}$-C, M$_{20}$-G and C-G planes. There is, however, a slight separation within these clouds according both structural code and star formation code: i.e., earlier morphologies and quiescent types have, on average, slightly lower values of M$_{20}$ at constant C. A small number of extreme outliers fall well beyond the main cloud of points in each of these relations. These outliers a mixture of structural codes and star formation codes but share the common feature of having a small number of bright pixels in their inner regions, either the result of a steeply rising brightness profile, in the case of M32-like systems, or the presence of compact star-forming regions, in the case of BCD-like systems.

\subsection{Application to Machine Learning Tools} 
\label{sec:ML}

Ongoing and imminent imaging surveys --- such as the Ultraviolet Near Infrared Optical Northern Survey \citep[UNIONS;][]{Unions,Gwyn2025}, Euclid \citep{Euclid} and LSST \citep{Ivezic2019} --- will deliver huge datasets containing billions of galaxies. Such data volumes far exceed the capacity for human classification, so machine learning and deep learning techniques will be essential for efficiently analyzing these data  \citep[e.g.,][]{Lahav1995,Dieleman2015,Cheng2021a,DominguezSanchez2019,walmsley2020galaxy}. It is hoped that such tools will be able to efficiently identify and classify galaxies of all morphological types, including objects that have often eluded detection and classification, such as low-surface-brightness galaxies and/or low-mass dwarfs.

To assess the compatibility of our classification system with past machine learning approaches, we cross-matched our NGVS sample with the $\sim$700{,}000 galaxies in \citet{Huertas-Company2011} (HC2011). That early, pioneering study used a support vector machine (SVM) trained on a visually classified subset of SDSS galaxies to assign probabilistic morphological types: E, S0, Sab, and Scd. Each galaxy in their catalog is associated with a probability vector over these four types, reflecting the SVM’s learned decision boundaries from the training set.

We identified 462 unambiguous matches (within 1 arcsecond) between the NGVS and HC2011. These galaxies lie exclusively at the bright end of the NGVS sample ($g \lesssim 18$ mag), where SDSS detections are reliable and morphologies are well constrained. Within this matched subset, 57\% are classified as {\tt E}, 26\% as {\tt ES}, and 10\% as {\tt S}, {\tt SI}, or {\tt I} in the NGVS morphology 
  scheme.

Among the NGVS sample of {\tt E} galaxies, 48\% are classified as Sab and 42\% as Scd by HC2011, while NGVS {\tt ES} galaxies are classified as Sab and Scd in 43\% and 50\% of cases, respectively. These results indicate poor agreement between the schemes for early-type systems, with most NGVS ellipticals and compact spheroidals being misclassified as late-type spirals by the HC2011 model. This almost certainly reflects the limited number of compact or low-surface-brightness (LSB) galaxies in their training set, which biases the classifier toward later types when applied to diffuse systems. For NGVS spirals and transition-types (e.g., {\tt S}, {\tt SI}, {\tt I}), 76\% are classified as Scd and 24\% as Sab, more consistent with expectations. These findings highlight the potential of the NGVS catalog as a training set for future morphological classification models, including learning methods which aim to infer galaxy morphologies directly from imaging data \cite{Banerji2010,DominguezSanchez2018, Bottrell2019b, bickley2021convolutional, Walmsley2022decals, Walmsley2023}.


\section{Summary and Conclusions}
\label{sec:conclusions}

The Next Generation Virgo Cluster Survey used the MegaCam instrument on Canada-France-Hawaii Telescope to perform deep, multi-band ($u^*giz$) optical imaging of the Virgo Cluster from its central core out to its virial radius. This region --- covering an area of $\sim$100~deg$^2$ centered on the largest overdensity in the local universe --- contains a vast reservoir of galaxies that span nearly all known morphological types.

As described in \citet{Ferrarese2016,Ferrarese2020}, the NGVS images have been carefully searched for Virgo cluster members using a variety of techniques, resulting in a final catalog of 3689 confirmed, probable and possible cluster members (Ferrarese~et~al. 2026, in preparation). While the catalog includes more than 100 distinct photometric and structural parameters measured for each cluster member, future studies based on this sample and survey would benefit from the addition of morphological information on the sample. Although visual morphologies have some well known limitations (e.g., they are inherently subjective, to some degree), such classification continue to see widespread use within the community, especially in the selection of targets and sub-samples for follow-up studies. 

In this paper, we have introduced a customized classification framework for NGVS galaxies. This system builds on previous morphological classification schemes while taking advantage of the depth, angular resolution and multi-wavelength coverage of the NGVS imaging, with the flexibility to describe high- and low-mass galaxies within a single framework. Each member galaxy was assigned two parameters that broadly describe its global structure and current level of star formation activity. In terms of structure, galaxies are divided into six broad categories, including transitional types: i.e., {\tt E, ES, S, SI, EI} and {\tt I}. {The morphology catalog will be made publicly available upon the publication of the full NGVS catalog (Ferrarese~et~al. 2026, in preparation).}

Our morphological system is shown to be self-consistent among classifiers, with good internal agreement. Comparisons with previous Virgo surveys, such as the VCC \citep{Binggeli1985} and EVCC \citep{Kim2014}, also show a high degree of consistency while revealing additional structural details due to the fainter point-source and surface brightness limits of the CFHT imaging. We confirm a gentle morphology-density relation within Virgo, consistent with prior studies of the cluster, but with an expanded sample.

Examining our morphological classifications, we highlight several galaxy types of interest for future study, including compact galaxies, ultra diffuse galaxies, possible examples of ultra-compact galaxies forming through the disruption of nucleated dwarfs, and candidate merger and post-merger systems. These objects may potentially serve as testbeds for understanding the physical mechanisms shaping galaxy structure and star formation histories, offering constraints before the era of big-data-driven astronomical surveys makes it nearly impossible  to visually classify galaxies and manually verify derived models and outputs. Non-parametric morphological measures, which are reliable for the brightness of $\sim$ 760 galaxies in our sample, provide insight into the underlying physical processes. 

This study bridges classical classification methods with modern, data-driven approaches, providing a new perspective on galaxy formation and evolution, in a cluster environment, and spanning roughly seven decades in stellar mass. The next generation of wide-field imaging surveys will push the limits of spatial resolution and depth, and the NGVS may serve as a benchmark for calibrating and interpreting automated classification frameworks. By combining visual and non-parametric techniques, this work lays the foundation for deeper insights into the structural evolution of galaxies, particularly those in dense environments.

 \begin{acknowledgments}
The authors thank Hong-Xin Zhang for insightful discussions on the selection of candidate merger and post-merger systems. The authors acknowledge the use of the Canadian Advanced Network for Astronomy Research (CANFAR) Science Platform. Our work used the facilities of the Canadian Astronomy Data Center, operated by the National Research Council of Canada with the support of the Canadian Space Agency, and CANFAR, a consortium that serves the data-intensive storage, access, and processing needs of university groups and centers engaged in astronomy research \citep{gaudet10}. This research has also made use of the NASA/IPAC Extragalactic Database (NED), which is operated by the Jet Propulsion Laboratory, California Institute of Technology, under contract with the National Aeronautics and Space Administration. This work was supported by the Natural Sciences and Engineering Research Council of Canada (NSERC).  S.L. acknowledges the support from the Sejong Science Fellowship Program by the National Research Foundation of Korea (NRF) grant funded by the Korea government (MSIT) (No. NRF-2021R1C1C2006790). PG acknowledges support from NSF grant AST-2206328. E.T. is thankful for the support from NSF-AST- 2206498 grant. CL acknowledges support from the National Natural Science Foundation of China (NSFC, Grant No. 12173025), National Key R\&D Program of China(2023YFA1607800, 2023YFA1607804), 111 project (No. B20019), and Key Laboratory for Particle Physics, Astrophysics and Cosmology, Ministry of Education. M.A.T. and S.T. acknowledge the support of the NSERC, 2023-03298.

\end{acknowledgments}

%

\vspace{5mm}
\facilities MegaCam(CFHT)


\software{Astropy \citep{2013A&A...558A..33A,2018AJ....156..123A}, Source Extractor \citep{1996A&AS..117..393B}
          }
matplotlib \citep{Matplotlib}, numpy \citep{numpy}, pandas \citep{McKinney2010,Reback2020}, photutils
\citep{photutils}, 
seaborn \citep{seaborn}

\bibliographystyle{aasjournal}
\bibliography{MorphPaper._resubmit}

\begin{thebibliography}{}
\expandafter\ifx\csname natexlab\endcsname\relax\def\natexlab#1{#1}\fi
\providecommand{\url}[1]{\href{#1}{#1}}
\providecommand{\dodoi}[1]{doi:~\href{http://doi.org/#1}{\nolinkurl{#1}}}
\providecommand{\doeprint}[1]{\href{http://ascl.net/#1}{\nolinkurl{http://ascl.net/#1}}}
\providecommand{\doarXiv}[1]{\href{https://arxiv.org/abs/#1}{\nolinkurl{https://arxiv.org/abs/#1}}}

\bibitem[{{Abazajian} {et~al.}(2009){Abazajian}, {Adelman-McCarthy}, {Ag{\"u}eros}, {Allam}, {Allende Prieto}, {An}, {Anderson}, {Anderson}, {Annis}, {Bahcall}, {Bailer-Jones}, {Barentine}, {Bassett}, {Becker}, {Beers}, {Bell}, {Belokurov}, {Berlind}, {Berman}, {Bernardi}, {Bickerton}, {Bizyaev}, {Blakeslee}, {Blanton}, {Bochanski}, {Boroski}, {Brewington}, {Brinchmann}, {Brinkmann}, {Brunner}, {Budav{\'a}ri}, {Carey}, {Carliles}, {Carr}, {Castander}, {Cinabro}, {Connolly}, {Csabai}, {Cunha}, {Czarapata}, {Davenport}, {de Haas}, {Dilday}, {Doi}, {Eisenstein}, {Evans}, {Evans}, {Fan}, {Friedman}, {Frieman}, {Fukugita}, {G{\"a}nsicke}, {Gates}, {Gillespie}, {Gilmore}, {Gonzalez}, {Gonzalez}, {Grebel}, {Gunn}, {Gy{\"o}ry}, {Hall}, {Harding}, {Harris}, {Harvanek}, {Hawley}, {Hayes}, {Heckman}, {Hendry}, {Hennessy}, {Hindsley}, {Hoblitt}, {Hogan}, {Hogg}, {Holtzman}, {Hyde}, {Ichikawa}, {Ichikawa}, {Im}, {Ivezi{\'c}}, {Jester}, {Jiang}, {Johnson}, {Jorgensen}, {Juri{\'c}}, {Kent}, {Kessler}, {Kleinman}, {Knapp},
  {Konishi}, {Kron}, {Krzesinski}, {Kuropatkin}, {Lampeitl}, {Lebedeva}, {Lee}, {Lee}, {French Leger}, {L{\'e}pine}, {Li}, {Lima}, {Lin}, {Long}, {Loomis}, {Loveday}, {Lupton}, {Magnier}, {Malanushenko}, {Malanushenko}, {Mandelbaum}, {Margon}, {Marriner}, {Mart{\'\i}nez-Delgado}, {Matsubara}, {McGehee}, {McKay}, {Meiksin}, {Morrison}, {Mullally}, {Munn}, {Murphy}, {Nash}, {Nebot}, {Neilsen}, {Newberg}, {Newman}, {Nichol}, {Nicinski}, {Nieto-Santisteban}, {Nitta}, {Okamura}, {Oravetz}, {Ostriker}, {Owen}, {Padmanabhan}, {Pan}, {Park}, {Pauls}, {Peoples}, {Percival}, {Pier}, {Pope}, {Pourbaix}, {Price}, {Purger}, {Quinn}, {Raddick}, {Re Fiorentin}, {Richards}, {Richmond}, {Riess}, {Rix}, {Rockosi}, {Sako}, {Schlegel}, {Schneider}, {Scholz}, {Schreiber}, {Schwope}, {Seljak}, {Sesar}, {Sheldon}, {Shimasaku}, {Sibley}, {Simmons}, {Sivarani}, {Allyn Smith}, {Smith}, {Smol{\v{c}}i{\'c}}, {Snedden}, {Stebbins}, {Steinmetz}, {Stoughton}, {Strauss}, {SubbaRao}, {Suto}, {Szalay}, {Szapudi}, {Szkody}, {Tanaka},
  {Tegmark}, {Teodoro}, {Thakar}, {Tremonti}, {Tucker}, {Uomoto}, {Vanden Berk}, {Vandenberg}, {Vidrih}, {Vogeley}, {Voges}, {Vogt}, {Wadadekar}, {Watters}, {Weinberg}, {West}, {White}, {Wilhite}, {Wonders}, {Yanny}, {Yocum}, {York}, {Zehavi}, {Zibetti}, \& {Zucker}}]{Abazaijan2009}
{Abazajian}, K.~N., {Adelman-McCarthy}, J.~K., {Ag{\"u}eros}, M.~A., {et~al.} 2009, \apjs, 182, 543, \dodoi{10.1088/0067-0049/182/2/543}

\bibitem[{{Abraham} {et~al.}(1996){Abraham}, {Tanvir}, {Santiago}, {Ellis}, {Glazebrook}, \& {van den Bergh}}]{Abraham1996}
{Abraham}, R.~G., {Tanvir}, N.~R., {Santiago}, B.~X., {et~al.} 1996, \mnras, 279, L47, \dodoi{10.1093/mnras/279.3.L47}

\bibitem[{{Allen} {et~al.}(2006){Allen}, {Driver}, {Graham}, {Cameron}, {Liske}, \& {de Propris}}]{Allen2006}
{Allen}, P.~D., {Driver}, S.~P., {Graham}, A.~W., {et~al.} 2006, \mnras, 371, 2, \dodoi{10.1111/j.1365-2966.2006.10586.x}

\bibitem[{{Amrutha} {et~al.}(2024){Amrutha}, {Das}, \& {Yadav}}]{Amrutha2024}
{Amrutha}, S., {Das}, M., \& {Yadav}, J. 2024, \mnras, 530, 2199, \dodoi{10.1093/mnras/stae907}

\bibitem[{Anderson \& Darling(1952)}]{Anderson1952}
Anderson, T.~W., \& Darling, D.~A. 1952, The Annals of Mathematical Statistics, 23, 193 , \dodoi{10.1214/aoms/1177729437}

\bibitem[{{Antonini} {et~al.}(2015){Antonini}, {Barausse}, \& {Silk}}]{Antonini2015}
{Antonini}, F., {Barausse}, E., \& {Silk}, J. 2015, \apj, 812, 72, \dodoi{10.1088/0004-637X/812/1/72}

\bibitem[{{Arrigoni Battaia} {et~al.}(2012){Arrigoni Battaia}, {Gavazzi}, {Fumagalli}, {Boselli}, {Boissier}, {Cortese}, {Heinis}, {Ferrarese}, {C{\^o}t{\'e}}, {Mihos}, {Cuillandre}, {Duc}, {Durrell}, {Gwyn}, {Jord{\'a}n}, {Liu}, {Peng}, \& {Mei}}]{Arrigoni2012}
{Arrigoni Battaia}, F., {Gavazzi}, G., {Fumagalli}, M., {et~al.} 2012, \aap, 543, A112, \dodoi{10.1051/0004-6361/201218895}

\bibitem[{{Astropy Collaboration} {et~al.}(2013){Astropy Collaboration}, {Robitaille}, {Tollerud}, {Greenfield}, {Droettboom}, {Bray}, {Aldcroft}, {Davis}, {Ginsburg}, {Price-Whelan}, {Kerzendorf}, {Conley}, {Crighton}, {Barbary}, {Muna}, {Ferguson}, {Grollier}, {Parikh}, {Nair}, {Unther}, {Deil}, {Woillez}, {Conseil}, {Kramer}, {Turner}, {Singer}, {Fox}, {Weaver}, {Zabalza}, {Edwards}, {Azalee Bostroem}, {Burke}, {Casey}, {Crawford}, {Dencheva}, {Ely}, {Jenness}, {Labrie}, {Lim}, {Pierfederici}, {Pontzen}, {Ptak}, {Refsdal}, {Servillat}, \& {Streicher}}]{2013A&A...558A..33A}
{Astropy Collaboration}, {Robitaille}, T.~P., {Tollerud}, E.~J., {et~al.} 2013, \aap, 558, A33, \dodoi{10.1051/0004-6361/201322068}

\bibitem[{{Astropy Collaboration} {et~al.}(2018){Astropy Collaboration}, {Price-Whelan}, {Sip{\H{o}}cz}, {G{\"u}nther}, {Lim}, {Crawford}, {Conseil}, {Shupe}, {Craig}, {Dencheva}, {Ginsburg}, {VanderPlas}, {Bradley}, {P{\'e}rez-Su{\'a}rez}, {de Val-Borro}, {Aldcroft}, {Cruz}, {Robitaille}, {Tollerud}, {Ardelean}, {Babej}, {Bach}, {Bachetti}, {Bakanov}, {Bamford}, {Barentsen}, {Barmby}, {Baumbach}, {Berry}, {Biscani}, {Boquien}, {Bostroem}, {Bouma}, {Brammer}, {Bray}, {Breytenbach}, {Buddelmeijer}, {Burke}, {Calderone}, {Cano Rodr{\'\i}guez}, {Cara}, {Cardoso}, {Cheedella}, {Copin}, {Corrales}, {Crichton}, {D'Avella}, {Deil}, {Depagne}, {Dietrich}, {Donath}, {Droettboom}, {Earl}, {Erben}, {Fabbro}, {Ferreira}, {Finethy}, {Fox}, {Garrison}, {Gibbons}, {Goldstein}, {Gommers}, {Greco}, {Greenfield}, {Groener}, {Grollier}, {Hagen}, {Hirst}, {Homeier}, {Horton}, {Hosseinzadeh}, {Hu}, {Hunkeler}, {Ivezi{\'c}}, {Jain}, {Jenness}, {Kanarek}, {Kendrew}, {Kern}, {Kerzendorf}, {Khvalko}, {King}, {Kirkby}, {Kulkarni},
  {Kumar}, {Lee}, {Lenz}, {Littlefair}, {Ma}, {Macleod}, {Mastropietro}, {McCully}, {Montagnac}, {Morris}, {Mueller}, {Mumford}, {Muna}, {Murphy}, {Nelson}, {Nguyen}, {Ninan}, {N{\"o}the}, {Ogaz}, {Oh}, {Parejko}, {Parley}, {Pascual}, {Patil}, {Patil}, {Plunkett}, {Prochaska}, {Rastogi}, {Reddy Janga}, {Sabater}, {Sakurikar}, {Seifert}, {Sherbert}, {Sherwood-Taylor}, {Shih}, {Sick}, {Silbiger}, {Singanamalla}, {Singer}, {Sladen}, {Sooley}, {Sornarajah}, {Streicher}, {Teuben}, {Thomas}, {Tremblay}, {Turner}, {Terr{\'o}n}, {van Kerkwijk}, {de la Vega}, {Watkins}, {Weaver}, {Whitmore}, {Woillez}, {Zabalza}, \& {Astropy Contributors}}]{2018AJ....156..123A}
{Astropy Collaboration}, {Price-Whelan}, A.~M., {Sip{\H{o}}cz}, B.~M., {et~al.} 2018, \aj, 156, 123, \dodoi{10.3847/1538-3881/aabc4f}

\bibitem[{{Baldry} {et~al.}(2004){Baldry}, {Glazebrook}, {Brinkmann}, {Ivezi{\'c}}, {Lupton}, {Nichol}, \& {Szalay}}]{Baldry2004}
{Baldry}, I.~K., {Glazebrook}, K., {Brinkmann}, J., {et~al.} 2004, \apj, 600, 681, \dodoi{10.1086/380092}

\bibitem[{{Ball} {et~al.}(2006){Ball}, {Brunner}, {Myers}, \& {Tcheng}}]{Ball2006}
{Ball}, N.~M., {Brunner}, R.~J., {Myers}, A.~D., \& {Tcheng}, D. 2006, \apj, 650, 497, \dodoi{10.1086/507440}

\bibitem[{{Banerji} {et~al.}(2010){Banerji}, {Lahav}, {Lintott}, {Abdalla}, {Schawinski}, {Bamford}, {Andreescu}, {Murray}, {Raddick}, {Slosar}, {Szalay}, {Thomas}, \& {Vandenberg}}]{Banerji2010}
{Banerji}, M., {Lahav}, O., {Lintott}, C.~J., {et~al.} 2010, \mnras, 406, 342, \dodoi{10.1111/j.1365-2966.2010.16713.x}

\bibitem[{{Barchi} {et~al.}(2020){Barchi}, {de Carvalho}, {Rosa}, {Sautter}, {Soares-Santos}, {Marques}, {Clua}, {Gon{\c{c}}alves}, {de S{\'a}-Freitas}, \& {Moura}}]{Barchi2020}
{Barchi}, P.~H., {de Carvalho}, R.~R., {Rosa}, R.~R., {et~al.} 2020, Astronomy and Computing, 30, 100334, \dodoi{10.1016/j.ascom.2019.100334}

\bibitem[{{Beasley} {et~al.}(2016){Beasley}, {Romanowsky}, {Pota}, {Navarro}, {Martinez Delgado}, {Neyer}, \& {Deich}}]{Beasley2016}
{Beasley}, M.~A., {Romanowsky}, A.~J., {Pota}, V., {et~al.} 2016, \apjl, 819, L20, \dodoi{10.3847/2041-8205/819/2/L20}

\bibitem[{{Bekki}(2008)}]{Bekki2008}
{Bekki}, K. 2008, \mnras, 388, L10, \dodoi{10.1111/j.1745-3933.2008.00489.x}

\bibitem[{{Bekki} {et~al.}(2001){Bekki}, {Couch}, {Drinkwater}, \& {Gregg}}]{Bekki2001}
{Bekki}, K., {Couch}, W.~J., {Drinkwater}, M.~J., \& {Gregg}, M.~D. 2001, \apjl, 557, L39, \dodoi{10.1086/323075}

\bibitem[{{Bekki} {et~al.}(2003){Bekki}, {Couch}, {Drinkwater}, \& {Shioya}}]{Bekki2003}
{Bekki}, K., {Couch}, W.~J., {Drinkwater}, M.~J., \& {Shioya}, Y. 2003, \mnras, 344, 399, \dodoi{10.1046/j.1365-8711.2003.06916.x}

\bibitem[{{Bershady} {et~al.}(2000){Bershady}, {Jangren}, \& {Conselice}}]{Bershady2000}
{Bershady}, M.~A., {Jangren}, A., \& {Conselice}, C.~J. 2000, \aj, 119, 2645, \dodoi{10.1086/301386}

\bibitem[{{Bertin} \& {Arnouts}(1996)}]{1996A&AS..117..393B}
{Bertin}, E., \& {Arnouts}, S. 1996, \aaps, 117, 393, \dodoi{10.1051/aas:1996164}

\bibitem[{Bickley {et~al.}(2021)Bickley, Bottrell, Hani, Ellison, Teimoorinia, Yi, Wilkinson, Gwyn, \& Hudson}]{bickley2021convolutional}
Bickley, R.~W., Bottrell, C., Hani, M.~H., {et~al.} 2021, Monthly Notices of the Royal Astronomical Society, 504, 372

\bibitem[{{Binggeli} \& {Cameron}(1991)}]{Binggeli1991}
{Binggeli}, B., \& {Cameron}, L.~M. 1991, \aap, 252, 27

\bibitem[{{Binggeli} {et~al.}(1985){Binggeli}, {Sandage}, \& {Tammann}}]{Binggeli1985}
{Binggeli}, B., {Sandage}, A., \& {Tammann}, G.~A. 1985, \aj, 90, 1681, \dodoi{10.1086/113874}

\bibitem[{{Binggeli} {et~al.}(1987){Binggeli}, {Tammann}, \& {Sandage}}]{Binggeli1987}
{Binggeli}, B., {Tammann}, G.~A., \& {Sandage}, A. 1987, \aj, 94, 251, \dodoi{10.1086/114467}

\bibitem[{{Blakeslee} {et~al.}(2006){Blakeslee}, {Holden}, {Franx}, {Rosati}, {Bouwens}, {Demarco}, {Ford}, {Homeier}, {Illingworth}, {Jee}, {Mei}, {Menanteau}, {Meurer}, {Postman}, \& {Tran}}]{Blakeslee2006}
{Blakeslee}, J.~P., {Holden}, B.~P., {Franx}, M., {et~al.} 2006, \apj, 644, 30, \dodoi{10.1086/503539}

\bibitem[{{Blakeslee} {et~al.}(2009){Blakeslee}, {Jord{\'a}n}, {Mei}, {C{\^o}t{\'e}}, {Ferrarese}, {Infante}, {Peng}, {Tonry}, \& {West}}]{Blakeslee2009}
{Blakeslee}, J.~P., {Jord{\'a}n}, A., {Mei}, S., {et~al.} 2009, \apj, 694, 556, \dodoi{10.1088/0004-637X/694/1/556}

\bibitem[{{Blanton} {et~al.}(2005){Blanton}, {Lupton}, {Schlegel}, {Strauss}, {Brinkmann}, {Fukugita}, \& {Loveday}}]{Blanton2005}
{Blanton}, M.~R., {Lupton}, R.~H., {Schlegel}, D.~J., {et~al.} 2005, \apj, 631, 208, \dodoi{10.1086/431416}

\bibitem[{{Blumenthal} {et~al.}(1984){Blumenthal}, {Faber}, {Primack}, \& {Rees}}]{Blumenthal1984}
{Blumenthal}, G.~R., {Faber}, S.~M., {Primack}, J.~R., \& {Rees}, M.~J. 1984, \nat, 311, 517, \dodoi{10.1038/311517a0}

\bibitem[{{B{\"o}hringer} {et~al.}(1994){B{\"o}hringer}, {Briel}, {Schwarz}, {Voges}, {Hartner}, \& {Tr{\"u}mper}}]{Bohringer1994}
{B{\"o}hringer}, H., {Briel}, U.~G., {Schwarz}, R.~A., {et~al.} 1994, \nat, 368, 828, \dodoi{10.1038/368828a0}

\bibitem[{{B{\"o}ker} {et~al.}(2002){B{\"o}ker}, {Laine}, {van der Marel}, {Sarzi}, {Rix}, {Ho}, \& {Shields}}]{Boker2002}
{B{\"o}ker}, T., {Laine}, S., {van der Marel}, R.~P., {et~al.} 2002, \aj, 123, 1389, \dodoi{10.1086/339025}

\bibitem[{{Boselli} {et~al.}(2022){Boselli}, {Fossati}, \& {Sun}}]{Boselli2022}
{Boselli}, A., {Fossati}, M., \& {Sun}, M. 2022, \aapr, 30, 3, \dodoi{10.1007/s00159-022-00140-3}

\bibitem[{{Boselli} \& {Gavazzi}(2006)}]{Boselli2006}
{Boselli}, A., \& {Gavazzi}, G. 2006, \pasp, 118, 517, \dodoi{10.1086/500691}

\bibitem[{{Boselli} \& {Gavazzi}(2014)}]{Boselli2014}
---. 2014, \aapr, 22, 74, \dodoi{10.1007/s00159-014-0074-y}

\bibitem[{{Bottrell} {et~al.}(2019){Bottrell}, {Hani}, {Teimoorinia}, {Ellison}, {Moreno}, {Torrey}, {Hayward}, {Thorp}, {Simard}, \& {Hernquist}}]{Bottrell2019b}
{Bottrell}, C., {Hani}, M.~H., {Teimoorinia}, H., {et~al.} 2019, \mnras, 490, 5390, \dodoi{10.1093/mnras/stz2934}

\bibitem[{Bradley {et~al.}(2023)}]{photutils}
Bradley, L., {et~al.} 2023, photutils: Photometry Tools, \url{https://photutils.readthedocs.io/}, \dodoi{10.5281/zenodo.596036}

\bibitem[{{Bullock} \& {Boylan-Kolchin}(2017)}]{Bullock2017}
{Bullock}, J.~S., \& {Boylan-Kolchin}, M. 2017, \araa, 55, 343, \dodoi{10.1146/annurev-astro-091916-055313}

\bibitem[{{Buta} {et~al.}(2015){Buta}, {Sheth}, {Athanassoula}, {Bosma}, {Knapen}, {Laurikainen}, {Salo}, {Elmegreen}, {Ho}, {Zaritsky}, {Courtois}, {Hinz}, {Mu{\~n}oz-Mateos}, {Kim}, {Regan}, {Gadotti}, {Gil de Paz}, {Laine}, {Men{\'e}ndez-Delmestre}, {Comer{\'o}n}, {Erroz Ferrer}, {Seibert}, {Mizusawa}, {Holwerda}, \& {Madore}}]{Buta2015}
{Buta}, R.~J., {Sheth}, K., {Athanassoula}, E., {et~al.} 2015, \apjs, 217, 32, \dodoi{10.1088/0067-0049/217/2/32}

\bibitem[{{Butcher} \& {Oemler}(1984)}]{Butcher1984}
{Butcher}, H., \& {Oemler}, A., J. 1984, \apj, 285, 426, \dodoi{10.1086/162519}

\bibitem[{{Cair{\'o}s} {et~al.}(2001{\natexlab{a}}){Cair{\'o}s}, {Caon}, {V{\'\i}lchez}, {Gonz{\'a}lez-P{\'e}rez}, \& {Mu{\~n}oz-Tu{\~n}{\'o}n}}]{Cairos2001a}
{Cair{\'o}s}, L.~M., {Caon}, N., {V{\'\i}lchez}, J.~M., {Gonz{\'a}lez-P{\'e}rez}, J.~N., \& {Mu{\~n}oz-Tu{\~n}{\'o}n}, C. 2001{\natexlab{a}}, \apjs, 136, 393, \dodoi{10.1086/322532}

\bibitem[{{Cair{\'o}s} {et~al.}(2010){Cair{\'o}s}, {Caon}, {Zurita}, {Kehrig}, {Roth}, \& {Weilbacher}}]{Cairos2010}
{Cair{\'o}s}, L.~M., {Caon}, N., {Zurita}, C., {et~al.} 2010, \aap, 520, A90, \dodoi{10.1051/0004-6361/201014004}

\bibitem[{{Cair{\'o}s} {et~al.}(2001{\natexlab{b}}){Cair{\'o}s}, {V{\'\i}lchez}, {Gonz{\'a}lez P{\'e}rez}, {Iglesias-P{\'a}ramo}, \& {Caon}}]{Cairos2001b}
{Cair{\'o}s}, L.~M., {V{\'\i}lchez}, J.~M., {Gonz{\'a}lez P{\'e}rez}, J.~N., {Iglesias-P{\'a}ramo}, J., \& {Caon}, N. 2001{\natexlab{b}}, \apjs, 133, 321, \dodoi{10.1086/320350}

\bibitem[{{Cantiello} {et~al.}(2018){Cantiello}, {Blakeslee}, {Ferrarese}, {C{\^o}t{\'e}}, {Roediger}, {Raimondo}, {Peng}, {Gwyn}, {Durrell}, \& {Cuillandre}}]{Cantiello2018}
{Cantiello}, M., {Blakeslee}, J.~P., {Ferrarese}, L., {et~al.} 2018, \apj, 856, 126, \dodoi{10.3847/1538-4357/aab043}

\bibitem[{Cantiello {et~al.}(2024)Cantiello, Blakeslee, Ferrarese, Côté, Raimondo, Cuillandre, Durrell, Gwyn, Hazra, Peng, Roediger, Sánchez-Janssen, \& Kurzner}]{Cantiello2024}
Cantiello, M., Blakeslee, J.~P., Ferrarese, L., {et~al.} 2024, The Astrophysical Journal, 966, 145, \dodoi{10.3847/1538-4357/ad3453}

\bibitem[{{Cappellari}(2016)}]{Cappellari2016}
{Cappellari}, M. 2016, \araa, 54, 597, \dodoi{10.1146/annurev-astro-082214-122432}

\bibitem[{{Cappellari} {et~al.}(2006){Cappellari}, {Bacon}, {Bureau}, {Damen}, {Davies}, {de Zeeuw}, {Emsellem}, {Falc{\'o}n-Barroso}, {Krajnovi{\'c}}, {Kuntschner}, {McDermid}, {Peletier}, {Sarzi}, {van den Bosch}, \& {van de Ven}}]{Cappellari2006}
{Cappellari}, M., {Bacon}, R., {Bureau}, M., {et~al.} 2006, \mnras, 366, 1126, \dodoi{10.1111/j.1365-2966.2005.09981.x}

\bibitem[{{Carleton} {et~al.}(2019){Carleton}, {Errani}, {Cooper}, {Kaplinghat}, {Pe{\~n}arrubia}, \& {Guo}}]{Carleton2019}
{Carleton}, T., {Errani}, R., {Cooper}, M., {et~al.} 2019, \mnras, 485, 382, \dodoi{10.1093/mnras/stz383}

\bibitem[{{Casura} {et~al.}(2022){Casura}, {Liske}, {Robotham}, {Brough}, {Driver}, {Graham}, {H{\"a}u{\ss}ler}, {Holwerda}, {Hopkins}, {Kelvin}, {Moffett}, {Taranu}, \& {Taylor}}]{Casura2022}
{Casura}, S., {Liske}, J., {Robotham}, A. S.~G., {et~al.} 2022, \mnras, 516, 942, \dodoi{10.1093/mnras/stac2267}

\bibitem[{{Chambers} \& {Unions}(2023)}]{Unions}
{Chambers}, K., \& {Unions}. 2023, in American Astronomical Society Meeting Abstracts, Vol. 242, American Astronomical Society Meeting Abstracts, 118.01

\bibitem[{{Cheng} {et~al.}(2021){Cheng}, {Conselice}, {Arag{\'o}n-Salamanca}, {Aguena}, {Allam}, {Andrade-Oliveira}, {Annis}, {Bluck}, {Brooks}, {Burke}, {Carrasco Kind}, {Carretero}, {Choi}, {Costanzi}, {da Costa}, {Pereira}, {De Vicente}, {Diehl}, {Drlica-Wagner}, {Eckert}, {Everett}, {Evrard}, {Ferrero}, {Fosalba}, {Frieman}, {Garc{\'\i}a-Bellido}, {Gerdes}, {Giannantonio}, {Gruen}, {Gruendl}, {Gschwend}, {Gutierrez}, {Hinton}, {Hollowood}, {Honscheid}, {James}, {Krause}, {Kuehn}, {Kuropatkin}, {Lahav}, {Maia}, {March}, {Menanteau}, {Miquel}, {Morgan}, {Paz-Chinch{\'o}n}, {Pieres}, {Plazas Malag{\'o}n}, {Roodman}, {Sanchez}, {Scarpine}, {Serrano}, {Sevilla-Noarbe}, {Smith}, {Soares-Santos}, {Suchyta}, {Swanson}, {Tarle}, {Thomas}, \& {To}}]{Cheng2021a}
{Cheng}, T.-Y., {Conselice}, C.~J., {Arag{\'o}n-Salamanca}, A., {et~al.} 2021, \mnras, 507, 4425, \dodoi{10.1093/mnras/stab2142}

\bibitem[{{Chhatkuli} {et~al.}(2023){Chhatkuli}, {Paudel}, {Bachchan}, {Aryal}, \& {Yoo}}]{Chhatkuli2023}
{Chhatkuli}, D.~N., {Paudel}, S., {Bachchan}, R.~K., {Aryal}, B., \& {Yoo}, J. 2023, \mnras, 520, 4953, \dodoi{10.1093/mnras/stac3700}

\bibitem[{{Choi} {et~al.}(2002){Choi}, {Guhathakurta}, \& {Johnston}}]{Choi2002}
{Choi}, P.~I., {Guhathakurta}, P., \& {Johnston}, K.~V. 2002, \aj, 124, 310, \dodoi{10.1086/341041}

\bibitem[{{Conselice}(2003)}]{Conselice2003}
{Conselice}, C.~J. 2003, \apjs, 147, 1, \dodoi{10.1086/375001}

\bibitem[{{Conselice}(2006)}]{Conselice2006}
---. 2006, \mnras, 373, 1389, \dodoi{10.1111/j.1365-2966.2006.11114.x}

\bibitem[{Conselice(2014)}]{conselice2014evolution}
Conselice, C.~J. 2014, Annual Review of Astronomy and Astrophysics, 52, 291

\bibitem[{{Conselice} {et~al.}(2000){Conselice}, {Bershady}, \& {Jangren}}]{Conselice2000}
{Conselice}, C.~J., {Bershady}, M.~A., \& {Jangren}, A. 2000, \apj, 529, 886, \dodoi{10.1086/308300}

\bibitem[{{Cortese} {et~al.}(2021){Cortese}, {Catinella}, \& {Smith}}]{Cortese2021}
{Cortese}, L., {Catinella}, B., \& {Smith}, R. 2021, \pasa, 38, e035, \dodoi{10.1017/pasa.2021.18}

\bibitem[{Curtis-Lake {et~al.}(2016)Curtis-Lake, McLure, Dunlop, Rogers, Targett, Dekel, Ellis, Faber, Ferguson, Grogin, Kocevski, Koekemoer, Lai, Mármol-Queraltó, \& Robertson}]{Curtis-Lake2016}
Curtis-Lake, E., McLure, R.~J., Dunlop, J.~S., {et~al.} 2016, Monthly Notices of the Royal Astronomical Society, 457, 440, \dodoi{10.1093/mnras/stv3017}

\bibitem[{Côté {et~al.}(2006)Côté, Piatek, Ferrarese, Jordán, Merritt, Peng, Haşegan, Blakeslee, Mei, West, Milosavljević, \& Tonry}]{Côté2006}
Côté, P., Piatek, S., Ferrarese, L., {et~al.} 2006, The Astrophysical Journal Supplement Series, 165, 57, \dodoi{10.1086/504042}

\bibitem[{{Dalcanton} {et~al.}(1997){Dalcanton}, {Spergel}, {Gunn}, {Schmidt}, \& {Schneider}}]{Dalcanton1997}
{Dalcanton}, J.~J., {Spergel}, D.~N., {Gunn}, J.~E., {Schmidt}, M., \& {Schneider}, D.~P. 1997, \aj, 114, 635, \dodoi{10.1086/118499}

\bibitem[{{Darg} {et~al.}(2010){Darg}, {Kaviraj}, {Lintott}, {Schawinski}, {Sarzi}, {Bamford}, {Silk}, {Proctor}, {Andreescu}, {Murray}, {Nichol}, {Raddick}, {Slosar}, {Szalay}, {Thomas}, \& {Vandenberg}}]{Darg2010}
{Darg}, D.~W., {Kaviraj}, S., {Lintott}, C.~J., {et~al.} 2010, \mnras, 401, 1043, \dodoi{10.1111/j.1365-2966.2009.15686.x}

\bibitem[{{Davies} {et~al.}(2016){Davies}, {Davies}, \& {Keenan}}]{Davies2016}
{Davies}, J.~I., {Davies}, L.~J.~M., \& {Keenan}, O.~C. 2016, \mnras, 456, 1607, \dodoi{10.1093/mnras/stv2719}

\bibitem[{{de Vaucouleurs}(1948)}]{deVaucouleurs1948}
{de Vaucouleurs}, G. 1948, Annales d'Astrophysique, 11, 247

\bibitem[{{de Vaucouleurs}(1959)}]{deVaucouleurs1959}
---. 1959, Astrophysik iv: Sternsysteme/astrophysics iv: Stellar systems, 275

\bibitem[{{Dieleman} {et~al.}(2015){Dieleman}, {Willett}, \& {Dambre}}]{Dieleman2015}
{Dieleman}, S., {Willett}, K.~W., \& {Dambre}, J. 2015, \mnras, 450, 1441, \dodoi{10.1093/mnras/stv632}

\bibitem[{{Dimauro} {et~al.}(2018){Dimauro}, {Huertas-Company}, {Daddi}, {P{\'e}rez-Gonz{\'a}lez}, {Bernardi}, {Barro}, {Buitrago}, {Caro}, {Cattaneo}, {Dominguez-S{\'a}nchez}, {Faber}, {H{\"a}u{\ss}ler}, {Kocevski}, {Koekemoer}, {Koo}, {Lee}, {Mei}, {Margalef-Bentabol}, {Primack}, {Rodriguez-Puebla}, {Salvato}, {Shankar}, \& {Tuccillo}}]{Dimauro2018}
{Dimauro}, P., {Huertas-Company}, M., {Daddi}, E., {et~al.} 2018, \mnras, 478, 5410, \dodoi{10.1093/mnras/sty1379}

\bibitem[{{Djorgovski} \& {Davis}(1987)}]{Djorgovski1987}
{Djorgovski}, S., \& {Davis}, M. 1987, \apj, 313, 59, \dodoi{10.1086/164948}

\bibitem[{{Dom{\'\i}nguez S{\'a}nchez} {et~al.}(2018){Dom{\'\i}nguez S{\'a}nchez}, {Huertas-Company}, {Bernardi}, {Tuccillo}, \& {Fischer}}]{DominguezSanchez2018}
{Dom{\'\i}nguez S{\'a}nchez}, H., {Huertas-Company}, M., {Bernardi}, M., {Tuccillo}, D., \& {Fischer}, J.~L. 2018, \mnras, 476, 3661, \dodoi{10.1093/mnras/sty338}

\bibitem[{{Dom{\'\i}nguez S{\'a}nchez} {et~al.}(2019){Dom{\'\i}nguez S{\'a}nchez}, {Huertas-Company}, {Bernardi}, {Kaviraj}, {Fischer}, {Abbott}, {Abdalla}, {Annis}, {Avila}, {Brooks}, {Buckley-Geer}, {Carnero Rosell}, {Carrasco Kind}, {Carretero}, {Cunha}, {D'Andrea}, {da Costa}, {Davis}, {De Vicente}, {Doel}, {Evrard}, {Fosalba}, {Frieman}, {Garc{\'\i}a-Bellido}, {Gaztanaga}, {Gerdes}, {Gruen}, {Gruendl}, {Gschwend}, {Gutierrez}, {Hartley}, {Hollowood}, {Honscheid}, {Hoyle}, {James}, {Kuehn}, {Kuropatkin}, {Lahav}, {Maia}, {March}, {Melchior}, {Menanteau}, {Miquel}, {Nord}, {Plazas}, {Sanchez}, {Scarpine}, {Schindler}, {Schubnell}, {Smith}, {Smith}, {Soares-Santos}, {Sobreira}, {Suchyta}, {Swanson}, {Tarle}, {Thomas}, {Walker}, \& {Zuntz}}]{DominguezSanchez2019}
{Dom{\'\i}nguez S{\'a}nchez}, H., {Huertas-Company}, M., {Bernardi}, M., {et~al.} 2019, \mnras, 484, 93, \dodoi{10.1093/mnras/sty3497}

\bibitem[{Dressler(1980)}]{dressler1980galaxy}
Dressler, A. 1980, Astrophysical Journal, Part 1, vol. 236, Mar. 1, 1980, p. 351-365., 236, 351

\bibitem[{{Drinkwater} {et~al.}(2004){Drinkwater}, {Gregg}, {Couch}, {Ferguson}, {Hilker}, {Jones}, {Karick}, \& {Phillipps}}]{Drinkwater2004}
{Drinkwater}, M.~J., {Gregg}, M.~D., {Couch}, W.~J., {et~al.} 2004, \pasa, 21, 375, \dodoi{10.1071/AS04048}

\bibitem[{{Drinkwater} {et~al.}(2003){Drinkwater}, {Gregg}, {Hilker}, {Bekki}, {Couch}, {Ferguson}, {Jones}, \& {Phillipps}}]{Drinkwater2003}
{Drinkwater}, M.~J., {Gregg}, M.~D., {Hilker}, M., {et~al.} 2003, \nat, 423, 519, \dodoi{10.1038/nature01666}

\bibitem[{{Drinkwater} {et~al.}(2000){Drinkwater}, {Jones}, {Gregg}, \& {Phillipps}}]{Drinkwater2000}
{Drinkwater}, M.~J., {Jones}, J.~B., {Gregg}, M.~D., \& {Phillipps}, S. 2000, \pasa, 17, 227, \dodoi{10.1071/AS00034}

\bibitem[{{Driver} {et~al.}(2022){Driver}, {Bellstedt}, {Robotham}, {Baldry}, {Davies}, {Liske}, {Obreschkow}, {Taylor}, {Wright}, {Alpaslan}, {Bamford}, {Bauer}, {Bland-Hawthorn}, {Bilicki}, {Bravo}, {Brough}, {Casura}, {Cluver}, {Colless}, {Conselice}, {Croom}, {de Jong}, {D'Eugenio}, {De Propris}, {Dogruel}, {Drinkwater}, {Dvornik}, {Farrow}, {Frenk}, {Giblin}, {Graham}, {Grootes}, {Gunawardhana}, {Hashemizadeh}, {H{\"a}u{\ss}ler}, {Heymans}, {Hildebrandt}, {Holwerda}, {Hopkins}, {Jarrett}, {Heath Jones}, {Kelvin}, {Koushan}, {Kuijken}, {Lara-L{\'o}pez}, {Lange}, {L{\'o}pez-S{\'a}nchez}, {Loveday}, {Mahajan}, {Meyer}, {Moffett}, {Napolitano}, {Norberg}, {Owers}, {Radovich}, {Raouf}, {Peacock}, {Phillipps}, {Pimbblet}, {Popescu}, {Said}, {Sansom}, {Seibert}, {Sutherland}, {Thorne}, {Tuffs}, {Turner}, {van der Wel}, {van Kampen}, \& {Wilkins}}]{Driver2022}
{Driver}, S.~P., {Bellstedt}, S., {Robotham}, A. S.~G., {et~al.} 2022, \mnras, 513, 439, \dodoi{10.1093/mnras/stac472}

\bibitem[{{D'Souza} \& {Bell}(2018)}]{DSouza2018}
{D'Souza}, R., \& {Bell}, E.~F. 2018, Nature Astronomy, 2, 737, \dodoi{10.1038/s41550-018-0533-x}

\bibitem[{{Du} {et~al.}(2019){Du}, {Debattista}, {Ho}, {C{\^o}t{\'e}}, {Spengler}, {Erwin}, {Wadsley}, {Norris}, {Earp}, {Quinn}, {Fiteni}, \& {Caruana}}]{Du2019}
{Du}, M., {Debattista}, V.~P., {Ho}, L.~C., {et~al.} 2019, \apj, 875, 58, \dodoi{10.3847/1538-4357/ab0e0c}

\bibitem[{{Duc} {et~al.}(2015){Duc}, {Cuillandre}, {Karabal}, {Cappellari}, {Alatalo}, {Blitz}, {Bournaud}, {Bureau}, {Crocker}, {Davies}, {Davis}, {de Zeeuw}, {Emsellem}, {Khochfar}, {Krajnovi{\'c}}, {Kuntschner}, {McDermid}, {Michel-Dansac}, {Morganti}, {Naab}, {Oosterloo}, {Paudel}, {Sarzi}, {Scott}, {Serra}, {Weijmans}, \& {Young}}]{Duc2015}
{Duc}, P.-A., {Cuillandre}, J.-C., {Karabal}, E., {et~al.} 2015, \mnras, 446, 120, \dodoi{10.1093/mnras/stu2019}

\bibitem[{{Durrell} {et~al.}(2014){Durrell}, {C{\^o}t{\'e}}, {Peng}, {Blakeslee}, {Ferrarese}, {Mihos}, {Puzia}, {Lan{\c{c}}on}, {Liu}, {Zhang}, {Cuillandre}, {McConnachie}, {Jord{\'a}n}, {Accetta}, {Boissier}, {Boselli}, {Courteau}, {Duc}, {Emsellem}, {Gwyn}, {Mei}, \& {Taylor}}]{2014ApJ...794..103D}
{Durrell}, P.~R., {C{\^o}t{\'e}}, P., {Peng}, E.~W., {et~al.} 2014, \apj, 794, 103, \dodoi{10.1088/0004-637X/794/2/103}

\bibitem[{{Efstathiou} \& {Silk}(1983)}]{Efstathiou1983}
{Efstathiou}, G., \& {Silk}, J. 1983, \fcp, 9, 1

\bibitem[{{Ekta} \& {Chengalur}(2010)}]{Ekta2010}
{Ekta}, B., \& {Chengalur}, J.~N. 2010, \mnras, 406, 1238, \dodoi{10.1111/j.1365-2966.2010.16756.x}

\bibitem[{{Ellison} {et~al.}(2019){Ellison}, {Viswanathan}, {Patton}, {Bottrell}, {McConnachie}, {Gwyn}, \& {Cuillandre}}]{Ellison2019}
{Ellison}, S.~L., {Viswanathan}, A., {Patton}, D.~R., {et~al.} 2019, \mnras, 487, 2491, \dodoi{10.1093/mnras/stz1431}

\bibitem[{{Elmegreen} \& {Elmegreen}(1987)}]{Elmegreen1987}
{Elmegreen}, D.~M., \& {Elmegreen}, B.~G. 1987, \apj, 314, 3, \dodoi{10.1086/165034}

\bibitem[{{Erben} {et~al.}(2005){Erben}, {Schirmer}, {Dietrich}, {Cordes}, {Haberzettl}, {Hetterscheidt}, {Hildebrandt}, {Schmithuesen}, {Schneider}, {Simon}, {Deul}, {Hook}, {Kaiser}, {Radovich}, {Benoist}, {Nonino}, {Olsen}, {Prandoni}, {Wichmann}, {Zaggia}, {Bomans}, {Dettmar}, \& {Miralles}}]{Erben2005}
{Erben}, T., {Schirmer}, M., {Dietrich}, J.~P., {et~al.} 2005, Astronomische Nachrichten, 326, 432, \dodoi{10.1002/asna.200510396}

\bibitem[{{Euclid Collaboration} {et~al.}(2023){Euclid Collaboration}, {Merlin}, {Castellano}, {Bretonni{\`e}re}, {Huertas-Company}, {Kuchner}, {Tuccillo}, {Buitrago}, {Peterson}, {Conselice}, {Caro}, {Dimauro}, {Nemani}, {Fontana}, {K{\"u}mmel}, {H{\"a}u{\ss}ler}, {Hartley}, {Alvarez Ayllon}, {Bertin}, {Dubath}, {Ferrari}, {Ferreira}, {Gavazzi}, {Hern{\'a}ndez-Lang}, {Lucatelli}, {Robotham}, {Schefer}, {Tortora}, {Aghanim}, {Amara}, {Amendola}, {Auricchio}, {Baldi}, {Bender}, {Bodendorf}, {Branchini}, {Brescia}, {Camera}, {Capobianco}, {Carbone}, {Carretero}, {Castander}, {Cavuoti}, {Cimatti}, {Cledassou}, {Congedo}, {Conversi}, {Copin}, {Corcione}, {Courbin}, {Cropper}, {Da Silva}, {Degaudenzi}, {Dinis}, {Douspis}, {Dubath}, {Duncan}, {Dupac}, {Dusini}, {Farrens}, {Ferriol}, {Frailis}, {Franceschi}, {Franzetti}, {Galeotta}, {Garilli}, {Gillis}, {Giocoli}, {Grazian}, {Grupp}, {Haugan}, {Hoekstra}, {Holmes}, {Hormuth}, {Hornstrup}, {Hudelot}, {Jahnke}, {Kermiche}, {Kiessling}, {Kitching}, {Kohley}, {Kunz},
  {Kurki-Suonio}, {Ligori}, {Lilje}, {Lloro}, {Mansutti}, {Marggraf}, {Markovic}, {Marulli}, {Massey}, {McCracken}, {Medinaceli}, {Melchior}, {Meneghetti}, {Meylan}, {Moresco}, {Moscardini}, {Munari}, {Niemi}, {Padilla}, {Paltani}, {Pasian}, {Pedersen}, {Percival}, {Polenta}, {Poncet}, {Popa}, {Pozzetti}, {Raison}, {Rebolo}, {Renzi}, {Rhodes}, {Riccio}, {Romelli}, {Rossetti}, {Saglia}, {Sapone}, {Sartoris}, {Schneider}, {Secroun}, {Seidel}, {Sirignano}, {Sirri}, {Skottfelt}, {Starck}, {Tallada-Cresp{\'\i}}, {Taylor}, {Tereno}, {Toledo-Moreo}, {Tutusaus}, {Valenziano}, {Vassallo}, {Wang}, {Weller}, {Zacchei}, {Zamorani}, {Zoubian}, {Andreon}, {Bardelli}, {Boucaud}, {Colodro-Conde}, {Di Ferdinando}, {Graci{\'a}-Carpio}, {Lindholm}, {Mauri}, {Mei}, {Neissner}, {Scottez}, {Tramacere}, {Zucca}, {Baccigalupi}, {Balaguera-Antol{\'\i}nez}, {Ballardini}, {Bernardeau}, {Biviano}, {Borgani}, {Borlaff}, {Burigana}, {Cabanac}, {Cappi}, {Carvalho}, {Casas}, {Castignani}, {Cooray}, {Coupon}, {Courtois}, {Cucciati},
  {Davini}, {De Lucia}, {Desprez}, {Escartin}, {Escoffier}, {Farina}, {Ganga}, {Garcia-Bellido}, {George}, {Gozaliasl}, {Hildebrandt}, {Hook}, {Ilbert}, {Ili{\'c}}, {Joachimi}, {Kansal}, {Keihanen}, {Kirkpatrick}, {Loureiro}, {Macias-Perez}, {Magliocchetti}, {Mainetti}, {Maoli}, {Marcin}, {Martinelli}, {Martinet}, {Matthew}, {Maturi}, {Metcalf}, {Monaco}, {Morgante}, \& {Nadathur}}]{Euclid2023Morph}
{Euclid Collaboration}, {Merlin}, E., {Castellano}, M., {et~al.} 2023, \aap, 671, A101, \dodoi{10.1051/0004-6361/202245041}

\bibitem[{Faber \& Jackson(1976)}]{faber1976velocity}
Faber, S., \& Jackson, R.~E. 1976, The Astrophysical Journal, 204, 668

\bibitem[{{Faber} {et~al.}(2007){Faber}, {Willmer}, {Wolf}, {Koo}, {Weiner}, {Newman}, {Im}, {Coil}, {Conroy}, {Cooper}, {Davis}, {Finkbeiner}, {Gerke}, {Gebhardt}, {Groth}, {Guhathakurta}, {Harker}, {Kaiser}, {Kassin}, {Kleinheinrich}, {Konidaris}, {Kron}, {Lin}, {Luppino}, {Madgwick}, {Meisenheimer}, {Noeske}, {Phillips}, {Sarajedini}, {Schiavon}, {Simard}, {Szalay}, {Vogt}, \& {Yan}}]{Faber2007}
{Faber}, S.~M., {Willmer}, C.~N.~A., {Wolf}, C., {et~al.} 2007, \apj, 665, 265, \dodoi{10.1086/519294}

\bibitem[{{Fabjan} {et~al.}(2010){Fabjan}, {Borgani}, {Tornatore}, {Saro}, {Murante}, \& {Dolag}}]{Fabjan2010}
{Fabjan}, D., {Borgani}, S., {Tornatore}, L., {et~al.} 2010, \mnras, 401, 1670, \dodoi{10.1111/j.1365-2966.2009.15794.x}

\bibitem[{{Fahrion} {et~al.}(2020){Fahrion}, {M{\"u}ller}, {Rejkuba}, {Hilker}, {Lyubenova}, {van de Ven}, {Georgiev}, {Lelli}, {Pawlowski}, \& {Jerjen}}]{Fahrion2020}
{Fahrion}, K., {M{\"u}ller}, O., {Rejkuba}, M., {et~al.} 2020, \aap, 634, A53, \dodoi{10.1051/0004-6361/201937120}

\bibitem[{{Fellhauer} \& {Kroupa}(2002)}]{Fellhauer2002}
{Fellhauer}, M., \& {Kroupa}, P. 2002, \mnras, 330, 642, \dodoi{10.1046/j.1365-8711.2002.05087.x}

\bibitem[{{Fellhauer} \& {Kroupa}(2005)}]{Fellhauer2005}
---. 2005, \mnras, 359, 223, \dodoi{10.1111/j.1365-2966.2005.08891.x}

\bibitem[{{Ferguson} \& {Sandage}(1989)}]{Ferguson1989}
{Ferguson}, H.~C., \& {Sandage}, A. 1989, \apjl, 346, L53, \dodoi{10.1086/185577}

\bibitem[{{Ferrarese} {et~al.}(2006){Ferrarese}, {C{\^o}t{\'e}}, {Jord{\'a}n}, {Peng}, {Blakeslee}, {Piatek}, {Mei}, {Merritt}, {Milosavljevi{\'c}}, {Tonry}, \& {West}}]{Ferrarese2006}
{Ferrarese}, L., {C{\^o}t{\'e}}, P., {Jord{\'a}n}, A., {et~al.} 2006, \apjs, 164, 334, \dodoi{10.1086/501350}

\bibitem[{{Ferrarese} {et~al.}(2012){Ferrarese}, {C{\^o}t{\'e}}, {Cuillandre}, {Gwyn}, {Peng}, {MacArthur}, {Duc}, {Boselli}, {Mei}, {Erben}, {McConnachie}, {Durrell}, {Mihos}, {Jord{\'a}n}, {Lan{\c{c}}on}, {Puzia}, {Emsellem}, {Balogh}, {Blakeslee}, {van Waerbeke}, {Gavazzi}, {Vollmer}, {Kavelaars}, {Woods}, {Ball}, {Boissier}, {Courteau}, {Ferriere}, {Gavazzi}, {Hildebrandt}, {Hudelot}, {Huertas-Company}, {Liu}, {McLaughlin}, {Mellier}, {Milkeraitis}, {Schade}, {Balkowski}, {Bournaud}, {Carlberg}, {Chapman}, {Hoekstra}, {Peng}, {Sawicki}, {Simard}, {Taylor}, {Tully}, {van Driel}, {Wilson}, {Burdullis}, {Mahoney}, \& {Manset}}]{Ferrarese2012}
{Ferrarese}, L., {C{\^o}t{\'e}}, P., {Cuillandre}, J.-C., {et~al.} 2012, \apjs, 200, 4, \dodoi{10.1088/0067-0049/200/1/4}

\bibitem[{{Ferrarese} {et~al.}(2016){Ferrarese}, {C{\^o}t{\'e}}, {S{\'a}nchez-Janssen}, {Roediger}, {McConnachie}, {Durrell}, {MacArthur}, {Blakeslee}, {Duc}, {Boissier}, {Boselli}, {Courteau}, {Cuillandre}, {Emsellem}, {Gwyn}, {Guhathakurta}, {Jord{\'a}n}, {Lan{\c{c}}on}, {Liu}, {Mei}, {Mihos}, {Navarro}, {Peng}, {Puzia}, {Taylor}, {Toloba}, \& {Zhang}}]{Ferrarese2016}
{Ferrarese}, L., {C{\^o}t{\'e}}, P., {S{\'a}nchez-Janssen}, R., {et~al.} 2016, \apj, 824, 10, \dodoi{10.3847/0004-637X/824/1/10}

\bibitem[{Ferrarese {et~al.}(2020)Ferrarese, Côté, MacArthur, Durrell, Gwyn, Duc, Sánchez-Janssen, Santos, Blakeslee, Boselli, Boyer, Cantiello, Courteau, Cuillandre, Emsellem, Erben, Gavazzi, Guhathakurta, Huertas-Company, Jordán, Lançon, Liu, Mei, Mihos, Peng, Puzia, Roediger, Schade, Taylor, Toloba, \& Zhang}]{Ferrarese2020}
Ferrarese, L., Côté, P., MacArthur, L.~A., {et~al.} 2020, The Astrophysical Journal, 890, 128, \dodoi{10.3847/1538-4357/ab339f}

\bibitem[{{Ferreira} {et~al.}(2022){Ferreira}, {Adams}, {Conselice}, {Sazonova}, {Austin}, {Caruana}, {Ferrari}, {Verma}, {Trussler}, {Broadhurst}, {Diego}, {Frye}, {Pascale}, {Wilkins}, {Windhorst}, \& {Zitrin}}]{Ferreira2022}
{Ferreira}, L., {Adams}, N., {Conselice}, C.~J., {et~al.} 2022, \apjl, 938, L2, \dodoi{10.3847/2041-8213/ac947c}

\bibitem[{{Ferreira} {et~al.}(2023){Ferreira}, {Conselice}, {Sazonova}, {Ferrari}, {Caruana}, {Tohill}, {Lucatelli}, {Adams}, {Irodotou}, {Marshall}, {Roper}, {Lovell}, {Verma}, {Austin}, {Trussler}, \& {Wilkins}}]{Ferreira2023}
{Ferreira}, L., {Conselice}, C.~J., {Sazonova}, E., {et~al.} 2023, \apj, 955, 94, \dodoi{10.3847/1538-4357/acec76}

\bibitem[{{Gabor} \& {Dav{\'e}}(2015)}]{Gabor2015}
{Gabor}, J.~M., \& {Dav{\'e}}, R. 2015, \mnras, 447, 374, \dodoi{10.1093/mnras/stu2399}

\bibitem[{{Gaudet} {et~al.}(2010){Gaudet}, {Hill}, {Armstrong}, {Ball}, {Burke}, {Chapel}, {Chapin}, {Damian}, {Dowler}, {Gable}, {Goliath}, {Ghiurea}, {Fabbro}, {Gwyn}, {Jenkins}, {Kavelaars}, {Major}, {Ouellette}, {Paterson}, {Peddle}, {Penfold-Brown}, {Pritchet}, {Schade}, {Sobie}, {Woods}, {Yeung}, \& {Zhang}}]{gaudet10}
{Gaudet}, S., {Hill}, N., {Armstrong}, P., {et~al.} 2010, in Society of Photo-Optical Instrumentation Engineers (SPIE) Conference Series, Vol. 7740, Software and Cyberinfrastructure for Astronomy, ed. N.~M. {Radziwill} \& A.~{Bridger}, 77401I, \dodoi{10.1117/12.858026}

\bibitem[{{Georgiev} {et~al.}(2016){Georgiev}, {B{\"o}ker}, {Leigh}, {L{\"u}tzgendorf}, \& {Neumayer}}]{Georgiev2016}
{Georgiev}, I.~Y., {B{\"o}ker}, T., {Leigh}, N., {L{\"u}tzgendorf}, N., \& {Neumayer}, N. 2016, \mnras, 457, 2122, \dodoi{10.1093/mnras/stw093}

\bibitem[{{Gerola} {et~al.}(1980){Gerola}, {Seiden}, \& {Schulman}}]{Gerola1980}
{Gerola}, H., {Seiden}, P.~E., \& {Schulman}, L.~S. 1980, \apj, 242, 517, \dodoi{10.1086/158485}

\bibitem[{{Gil de Paz} {et~al.}(2003){Gil de Paz}, {Madore}, \& {Pevunova}}]{dePaz2003}
{Gil de Paz}, A., {Madore}, B.~F., \& {Pevunova}, O. 2003, \apjs, 147, 29, \dodoi{10.1086/374737}

\bibitem[{{Gilhuly} \& {Courteau}(2018)}]{Gilhuly2018}
{Gilhuly}, C., \& {Courteau}, S. 2018, \mnras, 477, 845, \dodoi{10.1093/mnras/sty756}

\bibitem[{{Goto} {et~al.}(2003){Goto}, {Yamauchi}, {Fujita}, {Okamura}, {Sekiguchi}, {Smail}, {Bernardi}, \& {Gomez}}]{Goto2003}
{Goto}, T., {Yamauchi}, C., {Fujita}, Y., {et~al.} 2003, \mnras, 346, 601, \dodoi{10.1046/j.1365-2966.2003.07114.x}

\bibitem[{{Graham} \& {Driver}(2005)}]{Graham2005}
{Graham}, A.~W., \& {Driver}, S.~P. 2005, \pasa, 22, 118, \dodoi{10.1071/AS05001}

\bibitem[{{Grossauer} {et~al.}(2015){Grossauer}, {Taylor}, {Ferrarese}, {MacArthur}, {C{\^o}t{\'e}}, {Roediger}, {Courteau}, {Cuillandre}, {Duc}, {Durrell}, {Gwyn}, {Jord{\'a}n}, {Mei}, \& {Peng}}]{Grossauer2015}
{Grossauer}, J., {Taylor}, J.~E., {Ferrarese}, L., {et~al.} 2015, \apj, 807, 88, \dodoi{10.1088/0004-637X/807/1/88}

\bibitem[{{Gu{\'e}rou} {et~al.}(2015){Gu{\'e}rou}, {Emsellem}, {McDermid}, {C{\^o}t{\'e}}, {Ferrarese}, {Blakeslee}, {Durrell}, {MacArthur}, {Peng}, {Cuillandre}, \& {Gwyn}}]{Guérou2015}
{Gu{\'e}rou}, A., {Emsellem}, E., {McDermid}, R.~M., {et~al.} 2015, \apj, 804, 70, \dodoi{10.1088/0004-637X/804/1/70}

\bibitem[{{Gunn} \& {Gott}(1972)}]{Gunn1972}
{Gunn}, J.~E., \& {Gott}, J.~Richard, I. 1972, \apj, 176, 1, \dodoi{10.1086/151605}

\bibitem[{{Gwyn} {et~al.}(2025){Gwyn}, {McConnachie}, {Cuillandre}, {Chambers}, {Magnier}, {Hudson}, {Oguri}, {Furusawa}, {Hildebrandt}, {Carlberg}, {Ellison}, {Furusawa}, {Gavazzi}, {Ibata}, {Mellier}, {Osato}, {Aussel}, {Baumont}, {Bayer}, {Boulade}, {C{\^o}t{\'e}}, {Chemaly}, {Daley}, {Duc}, {Ellien}, {Fabbro}, {Ferreira}, {Fitriana}, {Le Floc'h}, {Hammer}, {Francois}, {Fudamoto}, {Gao}, {Goh}, {Goto}, {Guerrini}, {Guinot}, {H{\'e}nault-Brunet}, {Harikane}, {Hayashi}, {Heesters}, {Ichikawa}, {Kilbinger}, {Kuzma}, {Li}, {Liaudat}, {Lin}, {M{\"u}ller}, {Martin}, {Matsuoka}, {Medina}, {Miyatake}, {Miyazaki}, {Mpetha}, {Nagao}, {Navarro}, {Niwano}, {Ogami}, {Okabe}, {Onoue}, {Paek}, {Parker}, {Patton}, {Hervas Peters}, {Prunet}, {S{\'a}nchez-Janssen}, {Schultheis}, {Sestito}, {Smith}, {Starck}, {Starkenburg}, {Stone}, {Storfer}, {Suzuki}, {Erben}, {T.}, {Taibi}, {Thomas}, {TianFang}, {Toba}, {Uchiyama}, {Valls-Gabaud}, {Venn}, {Van Waerbeke}, {Wainscoat}, {Wilkinson}, {Wittje}, {Yoshida}, \&
  {Zhong}}]{Gwyn2025}
{Gwyn}, S., {McConnachie}, A.~W., {Cuillandre}, J.-C., {et~al.} 2025, arXiv e-prints, arXiv:2503.13783.
\newblock \doarXiv{2503.13783}

\bibitem[{{Ha{\c{s}}egan} {et~al.}(2005){Ha{\c{s}}egan}, {Jord{\'a}n}, {C{\^o}t{\'e}}, {Djorgovski}, {McLaughlin}, {Blakeslee}, {Mei}, {West}, {Peng}, {Ferrarese}, {Milosavljevi{\'c}}, {Tonry}, \& {Merritt}}]{Hasegan2005}
{Ha{\c{s}}egan}, M., {Jord{\'a}n}, A., {C{\^o}t{\'e}}, P., {et~al.} 2005, \apj, 627, 203, \dodoi{10.1086/430342}

\bibitem[{{Haines} {et~al.}(2015){Haines}, {Pereira}, {Smith}, {Egami}, {Babul}, {Finoguenov}, {Ziparo}, {McGee}, {Rawle}, {Okabe}, \& {Moran}}]{Haines2015}
{Haines}, C.~P., {Pereira}, M.~J., {Smith}, G.~P., {et~al.} 2015, \apj, 806, 101, \dodoi{10.1088/0004-637X/806/1/101}

\bibitem[{Hambleton {et~al.}(2011)Hambleton, Gibson, Brook, Stinson, Conselice, Bailin, Couchman, \& Wadsley}]{Hambleton2011}
Hambleton, K.~M., Gibson, B.~K., Brook, C.~B., {et~al.} 2011, Monthly Notices of the Royal Astronomical Society, 418, 801, \dodoi{10.1111/j.1365-2966.2011.19532.x}

\bibitem[{{Harris} {et~al.}(2020){Harris}, {Millman}, {van der Walt}, {Gommers}, {Virtanen}, {Cournapeau}, {Wieser}, {Taylor}, {Berg}, {Smith}, {Kern}, {Picus}, {Hoyer}, {van Kerkwijk}, {Brett}, {Haldane}, {del R{\'\i}o}, {Wiebe}, {Peterson}, {G{\'e}rard-Marchant}, {Sheppard}, {Reddy}, {Weckesser}, {Abbasi}, {Gohlke}, \& {Oliphant}}]{numpy}
{Harris}, C.~R., {Millman}, K.~J., {van der Walt}, S.~J., {et~al.} 2020, \nat, 585, 357, \dodoi{10.1038/s41586-020-2649-2}

\bibitem[{{Hausen} \& {Robertson}(2020)}]{Hausen2020}
{Hausen}, R., \& {Robertson}, B.~E. 2020, \apjs, 248, 20, \dodoi{10.3847/1538-4365/ab8868}

\bibitem[{{Hilker} {et~al.}(1999){Hilker}, {Infante}, {Vieira}, {Kissler-Patig}, \& {Richtler}}]{Hilker1999}
{Hilker}, M., {Infante}, L., {Vieira}, G., {Kissler-Patig}, M., \& {Richtler}, T. 1999, \aaps, 134, 75, \dodoi{10.1051/aas:1999434}

\bibitem[{{Holmberg}(1958)}]{Holmberg1958}
{Holmberg}, E. 1958, Meddelanden fran Lunds Astronomiska Observatorium Serie II, 136, 1

\bibitem[{{Hsyu} {et~al.}(2018){Hsyu}, {Cooke}, {Prochaska}, \& {Bolte}}]{Hsyu2018}
{Hsyu}, T., {Cooke}, R.~J., {Prochaska}, J.~X., \& {Bolte}, M. 2018, \apj, 863, 134, \dodoi{10.3847/1538-4357/aad18a}

\bibitem[{Hubble(1927)}]{hubble1927classification}
Hubble, E. 1927, The Observatory, 50, 276

\bibitem[{Hubble(1936)}]{Hubble1936}
---. 1936, The Realm of the Nebulae (Yale University Press)

\bibitem[{{Hubble}(1926)}]{Hubble1926}
{Hubble}, E.~P. 1926, \apj, 64, 321, \dodoi{10.1086/143018}

\bibitem[{{Huertas-Company} {et~al.}(2011){Huertas-Company}, {Aguerri}, {Bernardi}, {Mei}, \& {S{\'a}nchez Almeida}}]{Huertas-Company2011}
{Huertas-Company}, M., {Aguerri}, J.~A.~L., {Bernardi}, M., {Mei}, S., \& {S{\'a}nchez Almeida}, J. 2011, \aap, 525, A157, \dodoi{10.1051/0004-6361/201015735}

\bibitem[{{Huertas-Company} {et~al.}(2008){Huertas-Company}, {Rouan}, {Tasca}, {Soucail}, \& {Le F{\`e}vre}}]{Huertas-Company2008}
{Huertas-Company}, M., {Rouan}, D., {Tasca}, L., {Soucail}, G., \& {Le F{\`e}vre}, O. 2008, \aap, 478, 971, \dodoi{10.1051/0004-6361:20078625}

\bibitem[{{Hunter} \& {Elmegreen}(2004)}]{Hunter2004}
{Hunter}, D.~A., \& {Elmegreen}, B.~G. 2004, \aj, 128, 2170, \dodoi{10.1086/424615}

\bibitem[{{Hunter}(2007)}]{Matplotlib}
{Hunter}, J.~D. 2007, Computing in Science and Engineering, 9, 90, \dodoi{10.1109/MCSE.2007.55}

\bibitem[{{Ibata} {et~al.}(2001){Ibata}, {Irwin}, {Lewis}, {Ferguson}, \& {Tanvir}}]{Ibata2001}
{Ibata}, R., {Irwin}, M., {Lewis}, G., {Ferguson}, A. M.~N., \& {Tanvir}, N. 2001, \nat, 412, 49, \dodoi{10.1038/35083506}

\bibitem[{{Impey} {et~al.}(1988){Impey}, {Bothun}, \& {Malin}}]{Impey1988}
{Impey}, C., {Bothun}, G., \& {Malin}, D. 1988, \apj, 330, 634, \dodoi{10.1086/166500}

\bibitem[{{Impey} {et~al.}(1996){Impey}, {Sprayberry}, {Irwin}, \& {Bothun}}]{Impey1996}
{Impey}, C.~D., {Sprayberry}, D., {Irwin}, M.~J., \& {Bothun}, G.~D. 1996, \apjs, 105, 209, \dodoi{10.1086/192313}

\bibitem[{{Ivezi{\'c}} {et~al.}(2019){Ivezi{\'c}}, {Kahn}, {Tyson}, {Abel}, {Acosta}, {Allsman}, {Alonso}, {AlSayyad}, {Anderson}, {Andrew}, {Angel}, {Angeli}, {Ansari}, {Antilogus}, {Araujo}, {Armstrong}, {Arndt}, {Astier}, {Aubourg}, {Auza}, {Axelrod}, {Bard}, {Barr}, {Barrau}, {Bartlett}, {Bauer}, {Bauman}, {Baumont}, {Bechtol}, {Bechtol}, {Becker}, {Becla}, {Beldica}, {Bellavia}, {Bianco}, {Biswas}, {Blanc}, {Blazek}, {Blandford}, {Bloom}, {Bogart}, {Bond}, {Booth}, {Borgland}, {Borne}, {Bosch}, {Boutigny}, {Brackett}, {Bradshaw}, {Brandt}, {Brown}, {Bullock}, {Burchat}, {Burke}, {Cagnoli}, {Calabrese}, {Callahan}, {Callen}, {Carlin}, {Carlson}, {Chandrasekharan}, {Charles-Emerson}, {Chesley}, {Cheu}, {Chiang}, {Chiang}, {Chirino}, {Chow}, {Ciardi}, {Claver}, {Cohen-Tanugi}, {Cockrum}, {Coles}, {Connolly}, {Cook}, {Cooray}, {Covey}, {Cribbs}, {Cui}, {Cutri}, {Daly}, {Daniel}, {Daruich}, {Daubard}, {Daues}, {Dawson}, {Delgado}, {Dellapenna}, {de Peyster}, {de Val-Borro}, {Digel}, {Doherty}, {Dubois},
  {Dubois-Felsmann}, {Durech}, {Economou}, {Eifler}, {Eracleous}, {Emmons}, {Fausti Neto}, {Ferguson}, {Figueroa}, {Fisher-Levine}, {Focke}, {Foss}, {Frank}, {Freemon}, {Gangler}, {Gawiser}, {Geary}, {Gee}, {Geha}, {Gessner}, {Gibson}, {Gilmore}, {Glanzman}, {Glick}, {Goldina}, {Goldstein}, {Goodenow}, {Graham}, {Gressler}, {Gris}, {Guy}, {Guyonnet}, {Haller}, {Harris}, {Hascall}, {Haupt}, {Hernandez}, {Herrmann}, {Hileman}, {Hoblitt}, {Hodgson}, {Hogan}, {Howard}, {Huang}, {Huffer}, {Ingraham}, {Innes}, {Jacoby}, {Jain}, {Jammes}, {Jee}, {Jenness}, {Jernigan}, {Jevremovi{\'c}}, {Johns}, {Johnson}, {Johnson}, {Jones}, {Juramy-Gilles}, {Juri{\'c}}, {Kalirai}, {Kallivayalil}, {Kalmbach}, {Kantor}, {Karst}, {Kasliwal}, {Kelly}, {Kessler}, {Kinnison}, {Kirkby}, {Knox}, {Kotov}, {Krabbendam}, {Krughoff}, {Kub{\'a}nek}, {Kuczewski}, {Kulkarni}, {Ku}, {Kurita}, {Lage}, {Lambert}, {Lange}, {Langton}, {Le Guillou}, {Levine}, {Liang}, {Lim}, {Lintott}, {Long}, {Lopez}, {Lotz}, {Lupton}, {Lust}, {MacArthur}, {Mahabal},
  {Mandelbaum}, {Markiewicz}, {Marsh}, {Marshall}, {Marshall}, {May}, {McKercher}, {McQueen}, {Meyers}, {Migliore}, {Miller}, \& {Mills}}]{Ivezic2019}
{Ivezi{\'c}}, {\v{Z}}., {Kahn}, S.~M., {Tyson}, J.~A., {et~al.} 2019, \apj, 873, 111, \dodoi{10.3847/1538-4357/ab042c}

\bibitem[{{Jaff{\'e}} {et~al.}(2018){Jaff{\'e}}, {Poggianti}, {Moretti}, {Gullieuszik}, {Smith}, {Vulcani}, {Fasano}, {Fritz}, {Tonnesen}, {Bettoni}, {Hau}, {Biviano}, {Bellhouse}, \& {McGee}}]{Jaffe2018}
{Jaff{\'e}}, Y.~L., {Poggianti}, B.~M., {Moretti}, A., {et~al.} 2018, \mnras, 476, 4753, \dodoi{10.1093/mnras/sty500}

\bibitem[{{Jedrzejewski}(1987)}]{Jedrzejewski1987}
{Jedrzejewski}, R.~I. 1987, \mnras, 226, 747, \dodoi{10.1093/mnras/226.4.747}

\bibitem[{{Jogee} {et~al.}(2009){Jogee}, {Miller}, {Penner}, {Skelton}, {Conselice}, {Somerville}, {Bell}, {Zheng}, {Rix}, {Robaina}, {Barazza}, {Barden}, {Borch}, {Beckwith}, {Caldwell}, {Peng}, {Heymans}, {McIntosh}, {H{\"a}u{\ss}ler}, {Jahnke}, {Meisenheimer}, {Sanchez}, {Wisotzki}, {Wolf}, \& {Papovich}}]{Jogee2009}
{Jogee}, S., {Miller}, S.~H., {Penner}, K., {et~al.} 2009, \apj, 697, 1971, \dodoi{10.1088/0004-637X/697/2/1971}

\bibitem[{Joye \& Mandel(2003)}]{DS9}
Joye, W.~A., \& Mandel, E. 2003, in Proceedings of SPIE, Vol. 4854, 741--745, \dodoi{10.1117/12.461523}

\bibitem[{{Junais} {et~al.}(2022){Junais}, {Boissier}, {Boselli}, {Ferrarese}, {C{\^o}t{\'e}}, {Gwyn}, {Roediger}, {Lim}, {Peng}, {Cuillandre}, {Longobardi}, {Fossati}, {Hensler}, {Koda}, {Bautista}, {Boquien}, {Ma{\l}ek}, {Amram}, \& {Roehlly}}]{Junais2022}
{Junais}, {Boissier}, S., {Boselli}, A., {et~al.} 2022, \aap, 667, A76, \dodoi{10.1051/0004-6361/202244237}

\bibitem[{{Kado-Fong} {et~al.}(2020){Kado-Fong}, {Greene}, {Greco}, {Beaton}, {Goulding}, {Johnson}, \& {Komiyama}}]{Kado-Fong2020}
{Kado-Fong}, E., {Greene}, J.~E., {Greco}, J.~P., {et~al.} 2020, \aj, 159, 103, \dodoi{10.3847/1538-3881/ab6ef3}

\bibitem[{{Kauffmann} {et~al.}(2004){Kauffmann}, {White}, {Heckman}, {M{\'e}nard}, {Brinchmann}, {Charlot}, {Tremonti}, \& {Brinkmann}}]{Kauffmann2004}
{Kauffmann}, G., {White}, S. D.~M., {Heckman}, T.~M., {et~al.} 2004, \mnras, 353, 713, \dodoi{10.1111/j.1365-2966.2004.08117.x}

\bibitem[{{Kauffmann} {et~al.}(2003){Kauffmann}, {Heckman}, {White}, {Charlot}, {Tremonti}, {Brinchmann}, {Bruzual}, {Peng}, {Seibert}, {Bernardi}, {Blanton}, {Brinkmann}, {Castander}, {Cs{\'a}bai}, {Fukugita}, {Ivezic}, {Munn}, {Nichol}, {Padmanabhan}, {Thakar}, {Weinberg}, \& {York}}]{Kauffmann2003}
{Kauffmann}, G., {Heckman}, T.~M., {White}, S. D.~M., {et~al.} 2003, \mnras, 341, 33, \dodoi{10.1046/j.1365-8711.2003.06291.x}

\bibitem[{{Kent}(1985)}]{Kent1985}
{Kent}, S.~M. 1985, \apjs, 59, 115, \dodoi{10.1086/191066}

\bibitem[{{Kim} {et~al.}(2024){Kim}, {Sheen}, {Jaff{\'e}}, {Kelkar}, {Ranjan}, {Piraino-Cerda}, {Crossett}, {Costa Louren{\c{c}}o}, {Martin}, {Nantais}, {Demarco}, {Treister}, \& {Yi}}]{Kim2024}
{Kim}, D., {Sheen}, Y.-K., {Jaff{\'e}}, Y.~L., {et~al.} 2024, \apj, 966, 124, \dodoi{10.3847/1538-4357/ad32ce}

\bibitem[{{Kim} {et~al.}(2014){Kim}, {Rey}, {Jerjen}, {Lisker}, {Sung}, {Lee}, {Chung}, {Pak}, {Yi}, \& {Lee}}]{Kim2014}
{Kim}, S., {Rey}, S.-C., {Jerjen}, H., {et~al.} 2014, \apjs, 215, 22, \dodoi{10.1088/0067-0049/215/2/22}

\bibitem[{{Kocevski} {et~al.}(2012){Kocevski}, {Faber}, {Mozena}, {Koekemoer}, {Nandra}, {Rangel}, {Laird}, {Brusa}, {Wuyts}, {Trump}, {Koo}, {Somerville}, {Bell}, {Lotz}, {Alexander}, {Bournaud}, {Conselice}, {Dahlen}, {Dekel}, {Donley}, {Dunlop}, {Finoguenov}, {Georgakakis}, {Giavalisco}, {Guo}, {Grogin}, {Hathi}, {Juneau}, {Kartaltepe}, {Lucas}, {McGrath}, {McIntosh}, {Mobasher}, {Robaina}, {Rosario}, {Straughn}, {van der Wel}, \& {Villforth}}]{Kocevski2012}
{Kocevski}, D.~D., {Faber}, S.~M., {Mozena}, M., {et~al.} 2012, \apj, 744, 148, \dodoi{10.1088/0004-637X/744/2/148}

\bibitem[{Kolmogorov(1933)}]{Kolmogorov1933}
Kolmogorov, A. 1933, Giornale dell’Istituto Italiano degli Attuari, 4, 83

\bibitem[{{Kormendy} {et~al.}(2009){Kormendy}, {Fisher}, {Cornell}, \& {Bender}}]{Kormendy2009}
{Kormendy}, J., {Fisher}, D.~B., {Cornell}, M.~E., \& {Bender}, R. 2009, \apjs, 182, 216, \dodoi{10.1088/0067-0049/182/1/216}

\bibitem[{{Kunth} {et~al.}(1988){Kunth}, {Maurogordato}, \& {Vigroux}}]{Kunth1988}
{Kunth}, D., {Maurogordato}, S., \& {Vigroux}, L. 1988, \aap, 204, 10

\bibitem[{{Lahav} {et~al.}(1995){Lahav}, {Naim}, {Buta}, {Corwin}, {de Vaucouleurs}, {Dressler}, {Huchra}, {van den Bergh}, {Raychaudhury}, {Sodre}, \& {Storrie-Lombardi}}]{Lahav1995}
{Lahav}, O., {Naim}, A., {Buta}, R.~J., {et~al.} 1995, Science, 267, 859, \dodoi{10.1126/science.267.5199.859}

\bibitem[{{Laureijs} {et~al.}(2011){Laureijs}, {Amiaux}, {Arduini}, {Augu{\`e}res}, {Brinchmann}, {Cole}, {Cropper}, {Dabin}, {Duvet}, {Ealet}, {Garilli}, {Gondoin}, {Guzzo}, {Hoar}, {Hoekstra}, {Holmes}, {Kitching}, {Maciaszek}, {Mellier}, {Pasian}, {Percival}, {Rhodes}, {Saavedra Criado}, {Sauvage}, {Scaramella}, {Valenziano}, {Warren}, {Bender}, {Castander}, {Cimatti}, {Le F{\`e}vre}, {Kurki-Suonio}, {Levi}, {Lilje}, {Meylan}, {Nichol}, {Pedersen}, {Popa}, {Rebolo Lopez}, {Rix}, {Rottgering}, {Zeilinger}, {Grupp}, {Hudelot}, {Massey}, {Meneghetti}, {Miller}, {Paltani}, {Paulin-Henriksson}, {Pires}, {Saxton}, {Schrabback}, {Seidel}, {Walsh}, {Aghanim}, {Amendola}, {Bartlett}, {Baccigalupi}, {Beaulieu}, {Benabed}, {Cuby}, {Elbaz}, {Fosalba}, {Gavazzi}, {Helmi}, {Hook}, {Irwin}, {Kneib}, {Kunz}, {Mannucci}, {Moscardini}, {Tao}, {Teyssier}, {Weller}, {Zamorani}, {Zapatero Osorio}, {Boulade}, {Foumond}, {Di Giorgio}, {Guttridge}, {James}, {Kemp}, {Martignac}, {Spencer}, {Walton}, {Bl{\"u}mchen}, {Bonoli},
  {Bortoletto}, {Cerna}, {Corcione}, {Fabron}, {Jahnke}, {Ligori}, {Madrid}, {Martin}, {Morgante}, {Pamplona}, {Prieto}, {Riva}, {Toledo}, {Trifoglio}, {Zerbi}, {Abdalla}, {Douspis}, {Grenet}, {Borgani}, {Bouwens}, {Courbin}, {Delouis}, {Dubath}, {Fontana}, {Frailis}, {Grazian}, {Koppenh{\"o}fer}, {Mansutti}, {Melchior}, {Mignoli}, {Mohr}, {Neissner}, {Noddle}, {Poncet}, {Scodeggio}, {Serrano}, {Shane}, {Starck}, {Surace}, {Taylor}, {Verdoes-Kleijn}, {Vuerli}, {Williams}, {Zacchei}, {Altieri}, {Escudero Sanz}, {Kohley}, {Oosterbroek}, {Astier}, {Bacon}, {Bardelli}, {Baugh}, {Bellagamba}, {Benoist}, {Bianchi}, {Biviano}, {Branchini}, {Carbone}, {Cardone}, {Clements}, {Colombi}, {Conselice}, {Cresci}, {Deacon}, {Dunlop}, {Fedeli}, {Fontanot}, {Franzetti}, {Giocoli}, {Garcia-Bellido}, {Gow}, {Heavens}, {Hewett}, {Heymans}, {Holland}, {Huang}, {Ilbert}, {Joachimi}, {Jennins}, {Kerins}, {Kiessling}, {Kirk}, {Kotak}, {Krause}, {Lahav}, {van Leeuwen}, {Lesgourgues}, {Lombardi}, {Magliocchetti}, {Maguire},
  {Majerotto}, {Maoli}, {Marulli}, {Maurogordato}, {McCracken}, {McLure}, {Melchiorri}, {Merson}, {Moresco}, {Nonino}, {Norberg}, {Peacock}, {Pello}, {Penny}, {Pettorino}, {Di Porto}, {Pozzetti}, {Quercellini}, {Radovich}, {Rassat}, {Roche}, {Ronayette}, {Rossetti}, {Sartoris}, {Schneider}, {Semboloni}, {Serjeant}, {Simpson}, {Skordis}, {Smadja}, {Smartt}, {Spano}, {Spiro}, {Sullivan}, {Tilquin}, {Trotta}, {Verde}, {Wang}, {Williger}, {Zhao}, {Zoubian}, \& {Zucca}}]{Euclid}
{Laureijs}, R., {Amiaux}, J., {Arduini}, S., {et~al.} 2011, arXiv e-prints, arXiv:1110.3193, \dodoi{10.48550/arXiv.1110.3193}

\bibitem[{{Leaman} \& {van de Ven}(2022)}]{Leaman2022}
{Leaman}, R., \& {van de Ven}, G. 2022, \mnras, 516, 4691, \dodoi{10.1093/mnras/stab1966}

\bibitem[{{Lee} {et~al.}(2021){Lee}, {Kim}, {Rey}, \& {Chung}}]{Lee2021}
{Lee}, Y., {Kim}, S., {Rey}, S.-C., \& {Chung}, J. 2021, \apj, 906, 68, \dodoi{10.3847/1538-4357/abcaa0}

\bibitem[{{Lelli} {et~al.}(2012){Lelli}, {Verheijen}, {Fraternali}, \& {Sancisi}}]{Lelli2012}
{Lelli}, F., {Verheijen}, M., {Fraternali}, F., \& {Sancisi}, R. 2012, \aap, 537, A72, \dodoi{10.1051/0004-6361/201117867}

\bibitem[{{Lequeux} \& {Viallefond}(1980)}]{Lequeux1980}
{Lequeux}, J., \& {Viallefond}, F. 1980, \aap, 91, 269

\bibitem[{{Lieder} {et~al.}(2012){Lieder}, {Lisker}, {Hilker}, {Misgeld}, \& {Durrell}}]{Lieder2012}
{Lieder}, S., {Lisker}, T., {Hilker}, M., {Misgeld}, I., \& {Durrell}, P. 2012, \aap, 538, A69, \dodoi{10.1051/0004-6361/201117163}

\bibitem[{{Lim} {et~al.}(2018){Lim}, {Peng}, {C{\^o}t{\'e}}, {Sales}, {den Brok}, {Blakeslee}, \& {Guhathakurta}}]{Lim2018}
{Lim}, S., {Peng}, E.~W., {C{\^o}t{\'e}}, P., {et~al.} 2018, \apj, 862, 82, \dodoi{10.3847/1538-4357/aacb81}

\bibitem[{{Lim} {et~al.}(2020){Lim}, {C{\^o}t{\'e}}, {Peng}, {Ferrarese}, {Roediger}, {Durrell}, {Mihos}, {Wang}, {Gwyn}, {Cuillandre}, {Liu}, {S{\'a}nchez-Janssen}, {Toloba}, {Sales}, {Guhathakurta}, {Lan{\c{c}}on}, \& {Puzia}}]{Lim2020}
{Lim}, S., {C{\^o}t{\'e}}, P., {Peng}, E.~W., {et~al.} 2020, \apj, 899, 69, \dodoi{10.3847/1538-4357/aba433}

\bibitem[{{Lim} {et~al.}(2024){Lim}, {Peng}, {C{\^o}t{\'e}}, {Ferrarese}, {Roediger}, {Liu}, {Spengler}, {Sola}, {Duc}, {Sales}, {Blakeslee}, {Cuillandre}, {Durrell}, {Emsellem}, {Gwyn}, {Lan{\c{c}}on}, {Marleau}, {Mihos}, {M{\"u}ller}, {Puzia}, \& {S{\'a}nchez-Janssen}}]{2024arXiv240309926L}
{Lim}, S., {Peng}, E.~W., {C{\^o}t{\'e}}, P., {et~al.} 2024, arXiv e-prints, arXiv:2403.09926, \dodoi{10.48550/arXiv.2403.09926}

\bibitem[{{Lim} {et~al.}(2025){Lim}, {Peng}, {C{\^o}t{\'e}}, {Ferrarese}, {Roediger}, {Liu}, {Spengler}, {Sola}, {Duc}, {Sales}, {Blakeslee}, {Cuillandre}, {Durrell}, {Emsellem}, {Gwyn}, {Lan{\c{c}}on}, {Marleau}, {Mihos}, {M{\"u}ller}, {Puzia}, \& {S{\'a}nchez-Janssen}}]{Lim2025}
---. 2025, \apjs, 276, 34, \dodoi{10.3847/1538-4365/ad97b7}

\bibitem[{{Lintott} {et~al.}(2011){Lintott}, {Schawinski}, {Bamford}, {Slosar}, {Land}, {Thomas}, {Edmondson}, {Masters}, {Nichol}, {Raddick}, {Szalay}, {Andreescu}, {Murray}, \& {Vandenberg}}]{Lintott2011}
{Lintott}, C., {Schawinski}, K., {Bamford}, S., {et~al.} 2011, \mnras, 410, 166, \dodoi{10.1111/j.1365-2966.2010.17432.x}

\bibitem[{Lintott {et~al.}(2008)Lintott, Schawinski, Slosar, Land, Bamford, Thomas, Raddick, Nichol, Szalay, Andreescu, {et~al.}}]{lintott2008galaxy}
Lintott, C.~J., Schawinski, K., Slosar, A., {et~al.} 2008, Monthly Notices of the Royal Astronomical Society, 389, 1179

\bibitem[{{Lisker} {et~al.}(2006){Lisker}, {Glatt}, {Westera}, \& {Grebel}}]{Lisker2006}
{Lisker}, T., {Glatt}, K., {Westera}, P., \& {Grebel}, E.~K. 2006, \aj, 132, 2432, \dodoi{10.1086/508414}

\bibitem[{Lisker {et~al.}(2007)Lisker, Grebel, Binggeli, \& Glatt}]{Lisker2007}
Lisker, T., Grebel, E.~K., Binggeli, B., \& Glatt, K. 2007, The Astrophysical Journal, 660, 1186, \dodoi{10.1086/513090}

\bibitem[{{Lisker} {et~al.}(2009){Lisker}, {Janz}, {Hensler}, {Kim}, {Rey}, {Weinmann}, {Mastropietro}, {Hielscher}, {Paudel}, \& {Kotulla}}]{Lisker2009}
{Lisker}, T., {Janz}, J., {Hensler}, G., {et~al.} 2009, \apjl, 706, L124, \dodoi{10.1088/0004-637X/706/1/L124}

\bibitem[{{Liu} {et~al.}(2015){Liu}, {Peng}, {C{\^o}t{\'e}}, {Ferrarese}, {Jord{\'a}n}, {Mihos}, {Zhang}, {Mu{\~n}oz}, {Puzia}, {Lan{\c{c}}on}, {Gwyn}, {Cuillandre}, {Blakeslee}, {Boselli}, {Durrell}, {Duc}, {Guhathakurta}, {MacArthur}, {Mei}, {S{\'a}nchez-Janssen}, \& {Xu}}]{Liu2015}
{Liu}, C., {Peng}, E.~W., {C{\^o}t{\'e}}, P., {et~al.} 2015, \apj, 812, 34, \dodoi{10.1088/0004-637X/812/1/34}

\bibitem[{{Liu} {et~al.}(2020){Liu}, {C{\^o}t{\'e}}, {Peng}, {Roediger}, {Zhang}, {Ferrarese}, {S{\'a}nchez-Janssen}, {Guhathakurta}, {Yang}, {Jing}, {Alamo-Mart{\'\i}nez}, {Blakeslee}, {Boselli}, {Cuilandre}, {Duc}, {Durrell}, {Gwyn}, {Jord{\'a}n}, {Ko}, {Lan{\c{c}}on}, {Lim}, {Longobardi}, {Mei}, {Mihos}, {Mu{\~n}oz}, {Powalka}, {Puzia}, {Spengler}, \& {Toloba}}]{Liu2020}
{Liu}, C., {C{\^o}t{\'e}}, P., {Peng}, E.~W., {et~al.} 2020, \apjs, 250, 17, \dodoi{10.3847/1538-4365/abad91}

\bibitem[{{Lotz} {et~al.}(2006){Lotz}, {Madau}, {Giavalisco}, {Primack}, \& {Ferguson}}]{Lotz2006}
{Lotz}, J.~M., {Madau}, P., {Giavalisco}, M., {Primack}, J., \& {Ferguson}, H.~C. 2006, \apj, 636, 592, \dodoi{10.1086/497950}

\bibitem[{{Lotz} {et~al.}(2004){Lotz}, {Primack}, \& {Madau}}]{Lotz2004}
{Lotz}, J.~M., {Primack}, J., \& {Madau}, P. 2004, \aj, 128, 163, \dodoi{10.1086/421849}

\bibitem[{{Lotz} {et~al.}(2008){Lotz}, {Davis}, {Faber}, {Guhathakurta}, {Gwyn}, {Huang}, {Koo}, {Le Floc'h}, {Lin}, {Newman}, {Noeske}, {Papovich}, {Willmer}, {Coil}, {Conselice}, {Cooper}, {Hopkins}, {Metevier}, {Primack}, {Rieke}, \& {Weiner}}]{Lotz2008}
{Lotz}, J.~M., {Davis}, M., {Faber}, S.~M., {et~al.} 2008, \apj, 672, 177, \dodoi{10.1086/523659}

\bibitem[{{Ma} {et~al.}(2014){Ma}, {Greene}, {McConnell}, {Janish}, {Blakeslee}, {Thomas}, \& {Murphy}}]{Ma2014}
{Ma}, C.-P., {Greene}, J.~E., {McConnell}, N., {et~al.} 2014, \apj, 795, 158, \dodoi{10.1088/0004-637X/795/2/158}

\bibitem[{{Mahani} {et~al.}(2021){Mahani}, {Zonoozi}, {Haghi}, {Je{\v{r}}{\'a}bkov{\'a}}, {Kroupa}, \& {Mieske}}]{Mahani2021}
{Mahani}, H., {Zonoozi}, A.~H., {Haghi}, H., {et~al.} 2021, \mnras, 502, 5185, \dodoi{10.1093/mnras/stab330}

\bibitem[{{Malin}(1977)}]{MalinUnsharp}
{Malin}, D.~F. 1977, AAS Photo Bulletin, 16, 10

\bibitem[{{Martin} {et~al.}(2020){Martin}, {Kaviraj}, {Hocking}, {Read}, \& {Geach}}]{Martin2020}
{Martin}, G., {Kaviraj}, S., {Hocking}, A., {Read}, S.~C., \& {Geach}, J.~E. 2020, \mnras, 491, 1408, \dodoi{10.1093/mnras/stz3006}

\bibitem[{{Mayer} {et~al.}(2001){Mayer}, {Governato}, {Colpi}, {Moore}, {Quinn}, {Wadsley}, {Stadel}, \& {Lake}}]{Mayer2001}
{Mayer}, L., {Governato}, F., {Colpi}, M., {et~al.} 2001, \apj, 559, 754, \dodoi{10.1086/322356}

\bibitem[{{Mayer} {et~al.}(2006){Mayer}, {Mastropietro}, {Wadsley}, {Stadel}, \& {Moore}}]{Mayer2006}
{Mayer}, L., {Mastropietro}, C., {Wadsley}, J., {Stadel}, J., \& {Moore}, B. 2006, \mnras, 369, 1021, \dodoi{10.1111/j.1365-2966.2006.10403.x}

\bibitem[{{Mayes} {et~al.}(2021){Mayes}, {Drinkwater}, {Pfeffer}, {Baumgardt}, {Liu}, {Ferrarese}, {C{\^o}t{\'e}}, \& {Peng}}]{Mayes2021}
{Mayes}, R.~J., {Drinkwater}, M.~J., {Pfeffer}, J., {et~al.} 2021, \mnras, 501, 1852, \dodoi{10.1093/mnras/staa3731}

\bibitem[{{McConnell} \& {Ma}(2013)}]{McConnell2013}
{McConnell}, N.~J., \& {Ma}, C.-P. 2013, \apj, 764, 184, \dodoi{10.1088/0004-637X/764/2/184}

\bibitem[{{Meert} {et~al.}(2015){Meert}, {Vikram}, \& {Bernardi}}]{Meert2015}
{Meert}, A., {Vikram}, V., \& {Bernardi}, M. 2015, \mnras, 446, 3943, \dodoi{10.1093/mnras/stu2333}

\bibitem[{{Mei} {et~al.}(2007){Mei}, {Blakeslee}, {C{\^o}t{\'e}}, {Tonry}, {West}, {Ferrarese}, {Jord{\'a}n}, {Peng}, {Anthony}, \& {Merritt}}]{Mei2007}
{Mei}, S., {Blakeslee}, J.~P., {C{\^o}t{\'e}}, P., {et~al.} 2007, \apj, 655, 144, \dodoi{10.1086/509598}

\bibitem[{{Meza} {et~al.}(2003){Meza}, {Navarro}, {Steinmetz}, \& {Eke}}]{Meza2003}
{Meza}, A., {Navarro}, J.~F., {Steinmetz}, M., \& {Eke}, V.~R. 2003, \apj, 590, 619, \dodoi{10.1086/375151}

\bibitem[{{Mieske} {et~al.}(2008){Mieske}, {Hilker}, {Jord{\'a}n}, {Infante}, {Kissler-Patig}, {Rejkuba}, {Richtler}, {C{\^o}t{\'e}}, {Baumgardt}, {West}, {Ferrarese}, \& {Peng}}]{Mieske2008}
{Mieske}, S., {Hilker}, M., {Jord{\'a}n}, A., {et~al.} 2008, \aap, 487, 921, \dodoi{10.1051/0004-6361:200810077}

\bibitem[{{Mihos} {et~al.}(2005){Mihos}, {Harding}, {Feldmeier}, \& {Morrison}}]{Mihos2005}
{Mihos}, J.~C., {Harding}, P., {Feldmeier}, J., \& {Morrison}, H. 2005, \apjl, 631, L41, \dodoi{10.1086/497030}

\bibitem[{{Mihos} {et~al.}(2017){Mihos}, {Harding}, {Feldmeier}, {Rudick}, {Janowiecki}, {Morrison}, {Slater}, \& {Watkins}}]{Mihos2017}
{Mihos}, J.~C., {Harding}, P., {Feldmeier}, J.~J., {et~al.} 2017, \apj, 834, 16, \dodoi{10.3847/1538-4357/834/1/16}

\bibitem[{{Mihos} {et~al.}(2015){Mihos}, {Durrell}, {Ferrarese}, {Feldmeier}, {C{\^o}t{\'e}}, {Peng}, {Harding}, {Liu}, {Gwyn}, \& {Cuillandre}}]{Mihos2015}
{Mihos}, J.~C., {Durrell}, P.~R., {Ferrarese}, L., {et~al.} 2015, \apjl, 809, L21, \dodoi{10.1088/2041-8205/809/2/L21}

\bibitem[{{Moore} {et~al.}(1996){Moore}, {Katz}, {Lake}, {Dressler}, \& {Oemler}}]{Moore1996}
{Moore}, B., {Katz}, N., {Lake}, G., {Dressler}, A., \& {Oemler}, A. 1996, \nat, 379, 613, \dodoi{10.1038/379613a0}

\bibitem[{{Moore} {et~al.}(1998){Moore}, {Lake}, \& {Katz}}]{Moore1998}
{Moore}, B., {Lake}, G., \& {Katz}, N. 1998, \apj, 495, 139, \dodoi{10.1086/305264}

\bibitem[{{Morgan}(1959)}]{Morgan1959}
{Morgan}, W.~W. 1959, \pasp, 71, 394, \dodoi{10.1086/127415}

\bibitem[{{Naim} {et~al.}(1995){Naim}, {Lahav}, {Sodre}, \& {Storrie-Lombardi}}]{Naim1995}
{Naim}, A., {Lahav}, O., {Sodre}, Jr., L., \& {Storrie-Lombardi}, M.~C. 1995, \mnras, 275, 567, \dodoi{10.1093/mnras/275.3.567}

\bibitem[{{Napolitano} {et~al.}(2022){Napolitano}, {Gatto}, {Spiniello}, {Cantiello}, {Hilker}, {Arnaboldi}, {Tortora}, {Chaturvedi}, {D'Abrusco}, {Li}, {Paolillo}, {Peletier}, {Saifollahi}, {Spavone}, {Venhola}, {Capaccioli}, \& {Longo}}]{Napolitano2022}
{Napolitano}, N.~R., {Gatto}, M., {Spiniello}, C., {et~al.} 2022, \aap, 657, A94, \dodoi{10.1051/0004-6361/202141872}

\bibitem[{{Nelson} {et~al.}(2019){Nelson}, {Springel}, {Pillepich}, {Rodriguez-Gomez}, {Torrey}, {Genel}, {Vogelsberger}, {Pakmor}, {Marinacci}, {Weinberger}, {Kelley}, {Lovell}, {Diemer}, \& {Hernquist}}]{Nelson2019}
{Nelson}, D., {Springel}, V., {Pillepich}, A., {et~al.} 2019, Computational Astrophysics and Cosmology, 6, 2, \dodoi{10.1186/s40668-019-0028-x}

\bibitem[{{Neumayer} {et~al.}(2020){Neumayer}, {Seth}, \& {B{\"o}ker}}]{Neumayer2020}
{Neumayer}, N., {Seth}, A., \& {B{\"o}ker}, T. 2020, \aapr, 28, 4, \dodoi{10.1007/s00159-020-00125-0}

\bibitem[{{Nogueira-Cavalcante} {et~al.}(2018){Nogueira-Cavalcante}, {Gon{\c{c}}alves}, {Men{\'e}ndez-Delmestre}, \& {Sheth}}]{Nogueira-Cavalcante2018}
{Nogueira-Cavalcante}, J.~P., {Gon{\c{c}}alves}, T.~S., {Men{\'e}ndez-Delmestre}, K., \& {Sheth}, K. 2018, \mnras, 473, 1346, \dodoi{10.1093/mnras/stx2399}

\bibitem[{{Oh} {et~al.}(2016){Oh}, {Yi}, {Cortese}, {van de Sande}, {Mahajan}, {Jeong}, {Sheen}, {Allen}, {Bekki}, {Bland-Hawthorn}, {Bloom}, {Brough}, {Bryant}, {Colless}, {Croom}, {Fogarty}, {Goodwin}, {Green}, {Konstantopoulos}, {Lawrence}, {L{\'o}pez-S{\'a}nchez}, {Lorente}, {Medling}, {Owers}, {Richards}, {Scott}, {Sharp}, \& {Sweet}}]{Oh2016}
{Oh}, S., {Yi}, S.~K., {Cortese}, L., {et~al.} 2016, \apj, 832, 69, \dodoi{10.3847/0004-637X/832/1/69}

\bibitem[{{Oh} {et~al.}(2018){Oh}, {Kim}, {Lee}, {Sheen}, {Kim}, {Ree}, {Ho}, {Kyeong}, {Sung}, {Park}, \& {Yi}}]{Oh2018}
{Oh}, S., {Kim}, K., {Lee}, J.~H., {et~al.} 2018, \apjs, 237, 14, \dodoi{10.3847/1538-4365/aacd47}

\bibitem[{{Oh} {et~al.}(2019){Oh}, {Kim}, {Lee}, {Kim}, {Sheen}, {Rhee}, {Ree}, {Jeong}, {Ho}, {Kyeong}, {Sung}, {Park}, \& {Yi}}]{Oh2019}
---. 2019, \mnras, 488, 4169, \dodoi{10.1093/mnras/stz1920}

\bibitem[{{Paccagnella} {et~al.}(2016){Paccagnella}, {Vulcani}, {Poggianti}, {Moretti}, {Fritz}, {Gullieuszik}, {Couch}, {Bettoni}, {Cava}, {D'Onofrio}, \& {Fasano}}]{Paccagnella2016}
{Paccagnella}, A., {Vulcani}, B., {Poggianti}, B.~M., {et~al.} 2016, \apjl, 816, L25, \dodoi{10.3847/2041-8205/816/2/L25}

\bibitem[{{Pandey} {et~al.}(2024){Pandey}, {Kaviraj}, {Saha}, \& {Sharma}}]{Pandey2024}
{Pandey}, D., {Kaviraj}, S., {Saha}, K., \& {Sharma}, S. 2024, arXiv e-prints, arXiv:2403.12160, \dodoi{10.48550/arXiv.2403.12160}

\bibitem[{{Papaderos} {et~al.}(1996){Papaderos}, {Loose}, {Thuan}, \& {Fricke}}]{Papaderos1996}
{Papaderos}, P., {Loose}, H.~H., {Thuan}, T.~X., \& {Fricke}, K.~J. 1996, \aaps, 120, 207

\bibitem[{{Paudel} {et~al.}(2013){Paudel}, {Duc}, {C{\^o}t{\'e}}, {Cuillandre}, {Ferrarese}, {Ferriere}, {Gwyn}, {Mihos}, {Vollmer}, {Balogh}, {Carlberg}, {Boissier}, {Boselli}, {Durrell}, {Emsellem}, {MacArthur}, {Mei}, {Michel-Dansac}, \& {van Driel}}]{Paudel2013}
{Paudel}, S., {Duc}, P.-A., {C{\^o}t{\'e}}, P., {et~al.} 2013, \apj, 767, 133, \dodoi{10.1088/0004-637X/767/2/133}

\bibitem[{{Paudel} {et~al.}(2017){Paudel}, {Smith}, {Duc}, {C{\^o}t{\'e}}, {Cuillandre}, {Ferrarese}, {Blakeslee}, {Boselli}, {Cantiello}, {Gwyn}, {Guhathakurta}, {Mei}, {Mihos}, {Peng}, {Powalka}, {S{\'a}nchez-Janssen}, {Toloba}, \& {Zhang}}]{Paudel2017}
{Paudel}, S., {Smith}, R., {Duc}, P.-A., {et~al.} 2017, \apj, 834, 66, \dodoi{10.3847/1538-4357/834/1/66}

\bibitem[{{Pawlik} {et~al.}(2016){Pawlik}, {Wild}, {Walcher}, {Johansson}, {Villforth}, {Rowlands}, {Mendez-Abreu}, \& {Hewlett}}]{Pawlik2016}
{Pawlik}, M.~M., {Wild}, V., {Walcher}, C.~J., {et~al.} 2016, \mnras, 456, 3032, \dodoi{10.1093/mnras/stv2878}

\bibitem[{{Pearson} {et~al.}(2019){Pearson}, {Wang}, {Alpaslan}, {Baldry}, {Bilicki}, {Brown}, {Grootes}, {Holwerda}, {Kitching}, {Kruk}, \& {van der Tak}}]{Pearson2019}
{Pearson}, W.~J., {Wang}, L., {Alpaslan}, M., {et~al.} 2019, \aap, 631, A51, \dodoi{10.1051/0004-6361/201936337}

\bibitem[{{Peng} {et~al.}(2002){Peng}, {Ho}, {Impey}, \& {Rix}}]{Peng2002}
{Peng}, C.~Y., {Ho}, L.~C., {Impey}, C.~D., \& {Rix}, H.-W. 2002, \aj, 124, 266, \dodoi{10.1086/340952}

\bibitem[{{Peng} \& {Lim}(2016)}]{Peng2016}
{Peng}, E.~W., \& {Lim}, S. 2016, \apjl, 822, L31, \dodoi{10.3847/2041-8205/822/2/L31}

\bibitem[{{Peng} {et~al.}(2010){Peng}, {Lilly}, {Kova{\v{c}}}, {Bolzonella}, {Pozzetti}, {Renzini}, {Zamorani}, {Ilbert}, {Knobel}, {Iovino}, {Maier}, {Cucciati}, {Tasca}, {Carollo}, {Silverman}, {Kampczyk}, {de Ravel}, {Sanders}, {Scoville}, {Contini}, {Mainieri}, {Scodeggio}, {Kneib}, {Le F{\`e}vre}, {Bardelli}, {Bongiorno}, {Caputi}, {Coppa}, {de la Torre}, {Franzetti}, {Garilli}, {Lamareille}, {Le Borgne}, {Le Brun}, {Mignoli}, {Perez Montero}, {Pello}, {Ricciardelli}, {Tanaka}, {Tresse}, {Vergani}, {Welikala}, {Zucca}, {Oesch}, {Abbas}, {Barnes}, {Bordoloi}, {Bottini}, {Cappi}, {Cassata}, {Cimatti}, {Fumana}, {Hasinger}, {Koekemoer}, {Leauthaud}, {Maccagni}, {Marinoni}, {McCracken}, {Memeo}, {Meneux}, {Nair}, {Porciani}, {Presotto}, \& {Scaramella}}]{Peng2010}
{Peng}, Y.-j., {Lilly}, S.~J., {Kova{\v{c}}}, K., {et~al.} 2010, \apj, 721, 193, \dodoi{10.1088/0004-637X/721/1/193}

\bibitem[{{Phillipps} {et~al.}(2001){Phillipps}, {Drinkwater}, {Gregg}, \& {Jones}}]{Phillipps2001}
{Phillipps}, S., {Drinkwater}, M.~J., {Gregg}, M.~D., \& {Jones}, J.~B. 2001, \apj, 560, 201, \dodoi{10.1086/322517}

\bibitem[{{Phillipps} {et~al.}(1998){Phillipps}, {Parker}, {Schwartzenberg}, \& {Jones}}]{Phillips1998}
{Phillipps}, S., {Parker}, Q.~A., {Schwartzenberg}, J.~M., \& {Jones}, J.~B. 1998, \apjl, 493, L59, \dodoi{10.1086/311144}

\bibitem[{{Piraino-Cerda} {et~al.}(2024){Piraino-Cerda}, {Jaff{\'e}}, {Louren{\c{c}}o}, {Crossett}, {Salinas}, {Kim}, {Sheen}, {Kelkar}, {Pallero}, \& {Bravo-Alfaro}}]{Piraino-Cerda2024}
{Piraino-Cerda}, F., {Jaff{\'e}}, Y.~L., {Louren{\c{c}}o}, A.~C., {et~al.} 2024, \mnras, 528, 919, \dodoi{10.1093/mnras/stad3957}

\bibitem[{{Postman} \& {Geller}(1984)}]{Postman1984}
{Postman}, M., \& {Geller}, M.~J. 1984, \apj, 281, 95, \dodoi{10.1086/162078}

\bibitem[{{Press} \& {Schechter}(1974)}]{Press1974}
{Press}, W.~H., \& {Schechter}, P. 1974, \apj, 187, 425, \dodoi{10.1086/152650}

\bibitem[{{Reaves}(1983)}]{Reaves1983}
{Reaves}, G. 1983, \apjs, 53, 375, \dodoi{10.1086/190895}

\bibitem[{Reback {et~al.}(2020)Reback, McKinney, {et~al.}}]{Reback2020}
Reback, J., McKinney, W., {et~al.} 2020, pandas-dev/pandas: Pandas, \url{https://pandas.pydata.org/}, \dodoi{10.5281/zenodo.3509134}

\bibitem[{{Roberts} \& {Parker}(2020)}]{Roberts2020}
{Roberts}, I.~D., \& {Parker}, L.~C. 2020, \mnras, 495, 554, \dodoi{10.1093/mnras/staa1213}

\bibitem[{{Roberts}(1963)}]{Roberts1963}
{Roberts}, M.~S. 1963, \araa, 1, 149, \dodoi{10.1146/annurev.aa.01.090163.001053}

\bibitem[{{Roberts} \& {Haynes}(1994)}]{Roberts1994}
{Roberts}, M.~S., \& {Haynes}, M.~P. 1994, \araa, 32, 115, \dodoi{10.1146/annurev.aa.32.090194.000555}

\bibitem[{{Rodriguez-Gomez} {et~al.}(2019){Rodriguez-Gomez}, {Snyder}, {Lotz}, {Nelson}, {Pillepich}, {Springel}, {Genel}, {Weinberger}, {Tacchella}, {Pakmor}, {Torrey}, {Marinacci}, {Vogelsberger}, {Hernquist}, \& {Thilker}}]{RodriguezGomez2019}
{Rodriguez-Gomez}, V., {Snyder}, G.~F., {Lotz}, J.~M., {et~al.} 2019, \mnras, 483, 4140, \dodoi{10.1093/mnras/sty3345}

\bibitem[{Roediger {et~al.}(2017)Roediger, Ferrarese, Côté, MacArthur, Sánchez-Janssen, Blakeslee, Peng, Liu, Munoz, Cuillandre, Gwyn, Mei, Boissier, Boselli, Cantiello, Courteau, Duc, Lançon, Mihos, Puzia, Taylor, Durrell, Toloba, Guhathakurta, \& Zhang}]{Roediger2017}
Roediger, J.~C., Ferrarese, L., Côté, P., {et~al.} 2017, The Astrophysical Journal, 836, 120, \dodoi{10.3847/1538-4357/836/1/120}

\bibitem[{{Rong} {et~al.}(2017){Rong}, {Guo}, {Gao}, {Liao}, {Xie}, {Puzia}, {Sun}, \& {Pan}}]{Rong2017}
{Rong}, Y., {Guo}, Q., {Gao}, L., {et~al.} 2017, \mnras, 470, 4231, \dodoi{10.1093/mnras/stx1440}

\bibitem[{{Rong} {et~al.}(2020){Rong}, {Dong}, {Puzia}, {Galaz}, {S{\'a}nchez-Janssen}, {Cao}, {van der Burg}, {Sif{\'o}n}, {Mancera Pi{\~n}a}, {Marcelo}, {D'Ago}, {Zhang}, {Johnston}, \& {Eigenthaler}}]{Rong2020}
{Rong}, Y., {Dong}, X.-Y., {Puzia}, T.~H., {et~al.} 2020, \apj, 899, 78, \dodoi{10.3847/1538-4357/aba74a}

\bibitem[{{Saifollahi} {et~al.}(2021){Saifollahi}, {Janz}, {Peletier}, {Cantiello}, {Hilker}, {Mieske}, {Valentijn}, {Venhola}, \& {Verdoes Kleijn}}]{Saifollahi2021}
{Saifollahi}, T., {Janz}, J., {Peletier}, R.~F., {et~al.} 2021, \mnras, 504, 3580, \dodoi{10.1093/mnras/stab1118}

\bibitem[{{S{\'a}nchez-Janssen} {et~al.}(2008){S{\'a}nchez-Janssen}, {Aguerri}, \& {Mu{\~n}oz-Tu{\~n}{\'o}n}}]{Sanchez-Jannsen2008}
{S{\'a}nchez-Janssen}, R., {Aguerri}, J. A.~L., \& {Mu{\~n}oz-Tu{\~n}{\'o}n}, C. 2008, \apjl, 679, L77, \dodoi{10.1086/589617}

\bibitem[{{S{\'a}nchez-Janssen} {et~al.}(2019{\natexlab{a}}){S{\'a}nchez-Janssen}, {C{\^o}t{\'e}}, {Ferrarese}, {Peng}, {Roediger}, {Blakeslee}, {Emsellem}, {Puzia}, {Spengler}, {Taylor}, {{\'A}lamo-Mart{\'\i}nez}, {Boselli}, {Cantiello}, {Cuillandre}, {Duc}, {Durrell}, {Gwyn}, {MacArthur}, {Lan{\c{c}}on}, {Lim}, {Liu}, {Mei}, {Miller}, {Mu{\~n}oz}, {Mihos}, {Paudel}, {Powalka}, \& {Toloba}}]{Sanchez-Janssen2019}
{S{\'a}nchez-Janssen}, R., {C{\^o}t{\'e}}, P., {Ferrarese}, L., {et~al.} 2019{\natexlab{a}}, \apj, 878, 18, \dodoi{10.3847/1538-4357/aaf4fd}

\bibitem[{{S{\'a}nchez-Janssen} {et~al.}(2019{\natexlab{b}}){S{\'a}nchez-Janssen}, {Puzia}, {Ferrarese}, {C{\^o}t{\'e}}, {Eigenthaler}, {Miller}, {Ordenes-Brice{\~n}o}, {Peng}, {Ribbeck}, {Roediger}, {Spengler}, \& {Taylor}}]{Sanchez-JanssenMNRAS2019}
{S{\'a}nchez-Janssen}, R., {Puzia}, T.~H., {Ferrarese}, L., {et~al.} 2019{\natexlab{b}}, \mnras, 486, L1, \dodoi{10.1093/mnrasl/slz008}

\bibitem[{Sandage(1961)}]{Sandage1961}
Sandage, A. 1961, The Hubble Atlas of Galaxies (Carnegie Institution of Washington)

\bibitem[{Sandage(1981)}]{Sandage1981}
---. 1981, The Hubble Atlas (supplement/related work) (Carnegie Institution of Washington)

\bibitem[{{Sandage} \& {Binggeli}(1984)}]{Sandage1984}
{Sandage}, A., \& {Binggeli}, B. 1984, \aj, 89, 919, \dodoi{10.1086/113588}

\bibitem[{{Sandage} {et~al.}(1985){Sandage}, {Binggeli}, \& {Tammann}}]{Sandage1985a}
{Sandage}, A., {Binggeli}, B., \& {Tammann}, G.~A. 1985, \aj, 90, 1759, \dodoi{10.1086/113875}

\bibitem[{Sandage \& Tammann(1981)}]{RSA1981}
Sandage, A., \& Tammann, G.~A. 1981, A Revised Shapley-Ames Catalog of Bright Galaxies (Carnegie Institution of Washington)

\bibitem[{{Sargent} \& {Searle}(1970)}]{Sargent1970}
{Sargent}, W. L.~W., \& {Searle}, L. 1970, \apjl, 162, L155, \dodoi{10.1086/180644}

\bibitem[{{Sazonova} {et~al.}(2024){Sazonova}, {Morgan}, {Balogh}, {Alatalo}, {Benavides}, {Bluck}, {Brough}, {Busa}, {Demarco}, {Donevski}, {Figueira}, {Martin}, {Rodriguez-Gomez}, {Rom{\'a}n}, \& {Rowlands}}]{Sazonva2024}
{Sazonova}, E., {Morgan}, C., {Balogh}, M., {et~al.} 2024, arXiv e-prints, arXiv:2404.05792, \dodoi{10.48550/arXiv.2404.05792}

\bibitem[{{Schade} {et~al.}(1995){Schade}, {Lilly}, {Crampton}, {Hammer}, {Le Fevre}, \& {Tresse}}]{Schade1995}
{Schade}, D., {Lilly}, S.~J., {Crampton}, D., {et~al.} 1995, \apjl, 451, L1, \dodoi{10.1086/309677}

\bibitem[{{Schawinski} {et~al.}(2014){Schawinski}, {Urry}, {Simmons}, {Fortson}, {Kaviraj}, {Keel}, {Lintott}, {Masters}, {Nichol}, {Sarzi}, {Skibba}, {Treister}, {Willett}, {Wong}, \& {Yi}}]{Schawinski2014}
{Schawinski}, K., {Urry}, C.~M., {Simmons}, B.~D., {et~al.} 2014, \mnras, 440, 889, \dodoi{10.1093/mnras/stu327}

\bibitem[{{Schindler} {et~al.}(1999){Schindler}, {Binggeli}, \& {B{\"o}hringer}}]{Schindler1999}
{Schindler}, S., {Binggeli}, B., \& {B{\"o}hringer}, H. 1999, \aap, 343, 420, \dodoi{10.48550/arXiv.astro-ph/9811464}

\bibitem[{Scholz \& Stephens(1987)}]{Scholz1987}
Scholz, F.~W., \& Stephens, M.~A. 1987, Journal of the American Statistical Association, 82, 918.
\newblock \url{http://www.jstor.org/stable/2288805}

\bibitem[{{Searle} {et~al.}(1973){Searle}, {Sargent}, \& {Bagnuolo}}]{Searle1973}
{Searle}, L., {Sargent}, W.~L.~W., \& {Bagnuolo}, W.~G. 1973, \apj, 179, 427, \dodoi{10.1086/151882}

\bibitem[{{Secker}(1995)}]{Secker1995}
{Secker}, J. 1995, \pasp, 107, 496, \dodoi{10.1086/133580}

\bibitem[{S{\'e}rsic(1963)}]{sersic1963influence}
S{\'e}rsic, J. 1963, Boletin de la Asociacion Argentina de Astronomia La Plata Argentina, 6, 41

\bibitem[{{Sheen} {et~al.}(2012){Sheen}, {Yi}, {Ree}, \& {Lee}}]{Sheen2012}
{Sheen}, Y.-K., {Yi}, S.~K., {Ree}, C.~H., \& {Lee}, J. 2012, \apjs, 202, 8, \dodoi{10.1088/0067-0049/202/1/8}

\bibitem[{{Shen} {et~al.}(2014){Shen}, {Madau}, {Conroy}, {Governato}, \& {Mayer}}]{Shen2014}
{Shen}, S., {Madau}, P., {Conroy}, C., {Governato}, F., \& {Mayer}, L. 2014, \apj, 792, 99, \dodoi{10.1088/0004-637X/792/2/99}

\bibitem[{{Silk} \& {Wyse}(1993)}]{Silk1993}
{Silk}, J., \& {Wyse}, R. F.~G. 1993, \physrep, 231, 293, \dodoi{10.1016/0370-1573(93)90174-C}

\bibitem[{{Simard} {et~al.}(2011){Simard}, {Mendel}, {Patton}, {Ellison}, \& {McConnachie}}]{Simard2011}
{Simard}, L., {Mendel}, J.~T., {Patton}, D.~R., {Ellison}, S.~L., \& {McConnachie}, A.~W. 2011, \apjs, 196, 11, \dodoi{10.1088/0067-0049/196/1/11}

\bibitem[{Smirnov(1948)}]{Smirnov1948}
Smirnov, N. 1948, The Annals of Mathematical Statistics, 19, 279 , \dodoi{10.1214/aoms/1177730256}

\bibitem[{{Smith} {et~al.}(2013){Smith}, {Duc}, {Candlish}, {Fellhauer}, {Sheen}, \& {Gibson}}]{Smith2013}
{Smith}, R., {Duc}, P.~A., {Candlish}, G.~N., {et~al.} 2013, \mnras, 436, 839, \dodoi{10.1093/mnras/stt1619}

\bibitem[{{Smith} {et~al.}(2012){Smith}, {Lucey}, {Price}, {Hudson}, \& {Phillipps}}]{Smith2012}
{Smith}, R.~J., {Lucey}, J.~R., {Price}, J., {Hudson}, M.~J., \& {Phillipps}, S. 2012, \mnras, 419, 3167, \dodoi{10.1111/j.1365-2966.2011.19956.x}

\bibitem[{Spengler {et~al.}(2017)Spengler, Côté, Roediger, Ferrarese, Sánchez-Janssen, Toloba, Liu, Guhathakurta, Cuillandre, Gwyn, Zirm, Muñoz, Puzia, Lançon, Peng, Mei, \& Powalka}]{Spengler2017}
Spengler, C., Côté, P., Roediger, J., {et~al.} 2017, The Astrophysical Journal, 849, 55, \dodoi{10.3847/1538-4357/aa8a78}

\bibitem[{{Spergel} {et~al.}(2013){Spergel}, {Gehrels}, {Breckinridge}, {Donahue}, {Dressler}, {Gaudi}, {Greene}, {Guyon}, {Hirata}, {Kalirai}, {Kasdin}, {Moos}, {Perlmutter}, {Postman}, {Rauscher}, {Rhodes}, {Wang}, {Weinberg}, {Centrella}, {Traub}, {Baltay}, {Colbert}, {Bennett}, {Kiessling}, {Macintosh}, {Merten}, {Mortonson}, {Penny}, {Rozo}, {Savransky}, {Stapelfeldt}, {Zu}, {Baker}, {Cheng}, {Content}, {Dooley}, {Foote}, {Goullioud}, {Grady}, {Jackson}, {Kruk}, {Levine}, {Melton}, {Peddie}, {Ruffa}, \& {Shaklan}}]{WFIRST}
{Spergel}, D., {Gehrels}, N., {Breckinridge}, J., {et~al.} 2013, arXiv e-prints, arXiv:1305.5422, \dodoi{10.48550/arXiv.1305.5422}

\bibitem[{{Strateva} {et~al.}(2001){Strateva}, {Ivezi{\'c}}, {Knapp}, {Narayanan}, {Strauss}, {Gunn}, {Lupton}, {Schlegel}, {Bahcall}, {Brinkmann}, {Brunner}, {Budav{\'a}ri}, {Csabai}, {Castander}, {Doi}, {Fukugita}, {Gy{\H{o}}ry}, {Hamabe}, {Hennessy}, {Ichikawa}, {Kunszt}, {Lamb}, {McKay}, {Okamura}, {Racusin}, {Sekiguchi}, {Schneider}, {Shimasaku}, \& {York}}]{Strateva2001}
{Strateva}, I., {Ivezi{\'c}}, {\v{Z}}., {Knapp}, G.~R., {et~al.} 2001, \aj, 122, 1861, \dodoi{10.1086/323301}

\bibitem[{Student(1908)}]{Student1908}
Student. 1908, Biometrika, 1

\bibitem[{{Sun} {et~al.}(2025){Sun}, {Zhang}, {Smith}, {Brinks}, {C{\^o}t{\'e}}, {Oh}, {Lin}, {Boselli}, {Ferrarese}, {Li}, {Sun}, {Chen}, {Zhang}, {Kim}, {Kim}, {Li}, {Tao}, {Taylor}, {Duc}, {S{\'a}nchez-Janss{\'e}n}, {Zhao}, {Paudel}, {Peng}, {Wang}, {Gwyn}, {Fossati}, \& {Cuillandre}}]{Sun2025}
{Sun}, W., {Zhang}, H.-X., {Smith}, R., {et~al.} 2025, \aap, 700, A113, \dodoi{10.1051/0004-6361/202555455}

\bibitem[{{Taranu} {et~al.}(2014){Taranu}, {Hudson}, {Balogh}, {Smith}, {Power}, {Oman}, \& {Krane}}]{Taranu2014}
{Taranu}, D.~S., {Hudson}, M.~J., {Balogh}, M.~L., {et~al.} 2014, \mnras, 440, 1934, \dodoi{10.1093/mnras/stu389}

\bibitem[{{Thomas} {et~al.}(2005){Thomas}, {Maraston}, {Bender}, \& {Mendes de Oliveira}}]{Thomas2005}
{Thomas}, D., {Maraston}, C., {Bender}, R., \& {Mendes de Oliveira}, C. 2005, \apj, 621, 673, \dodoi{10.1086/426932}

\bibitem[{{Thomas} {et~al.}(2010){Thomas}, {Maraston}, {Schawinski}, {Sarzi}, \& {Silk}}]{Thomas2010}
{Thomas}, D., {Maraston}, C., {Schawinski}, K., {Sarzi}, M., \& {Silk}, J. 2010, \mnras, 404, 1775, \dodoi{10.1111/j.1365-2966.2010.16427.x}

\bibitem[{{Thuan}(1991)}]{Thuan1991}
{Thuan}, T.~X. 1991, in Massive Stars in Starbursts, ed. C.~{Leitherer}, N.~{Walborn}, T.~{Heckman}, \& C.~{Norman}, 349

\bibitem[{{Thuan} \& {Martin}(1981)}]{Thuan1981}
{Thuan}, T.~X., \& {Martin}, G.~E. 1981, \apj, 247, 823, \dodoi{10.1086/159094}

\bibitem[{{Tinsley}(1980)}]{Tinsley1980}
{Tinsley}, B.~M. 1980, \fcp, 5, 287, \dodoi{10.48550/arXiv.2203.02041}

\bibitem[{{Tohill} {et~al.}(2021){Tohill}, {Ferreira}, {Conselice}, {Bamford}, \& {Ferrari}}]{Tohill2021}
{Tohill}, C., {Ferreira}, L., {Conselice}, C.~J., {Bamford}, S.~P., \& {Ferrari}, F. 2021, \apj, 916, 4, \dodoi{10.3847/1538-4357/ac033c}

\bibitem[{{Toloba} {et~al.}(2018){Toloba}, {Lim}, {Peng}, {Sales}, {Guhathakurta}, {Mihos}, {C{\^o}t{\'e}}, {Boselli}, {Cuillandre}, {Ferrarese}, {Gwyn}, {Lan{\c{c}}on}, {Mu{\~n}oz}, \& {Puzia}}]{Toloba2018}
{Toloba}, E., {Lim}, S., {Peng}, E., {et~al.} 2018, \apjl, 856, L31, \dodoi{10.3847/2041-8213/aab603}

\bibitem[{{Toloba} {et~al.}(2023){Toloba}, {Sales}, {Lim}, {Peng}, {Guhathakurta}, {Roediger}, {Wang}, {Mihos}, {C{\^o}t{\'e}}, {Durrell}, \& {Ferrarese}}]{Toloba2023}
{Toloba}, E., {Sales}, L.~V., {Lim}, S., {et~al.} 2023, \apj, 951, 77, \dodoi{10.3847/1538-4357/acd336}

\bibitem[{{Trentham} \& {Hodgkin}(2002)}]{Trentham2002}
{Trentham}, N., \& {Hodgkin}, S. 2002, \mnras, 333, 423, \dodoi{10.1046/j.1365-8711.2002.05440.x}

\bibitem[{{Trujillo-Gomez} {et~al.}(2022){Trujillo-Gomez}, {Kruijssen}, \& {Reina-Campos}}]{Trujillo2021}
{Trujillo-Gomez}, S., {Kruijssen}, J.~M.~D., \& {Reina-Campos}, M. 2022, \mnras, 510, 3356, \dodoi{10.1093/mnras/stab3401}

\bibitem[{{Tully} \& {Fisher}(1977)}]{Tully1977}
{Tully}, R.~B., \& {Fisher}, J.~R. 1977, \aap, 54, 661

\bibitem[{{van den Bergh}(1960{\natexlab{a}})}]{vandenBergh1960a}
{van den Bergh}, S. 1960{\natexlab{a}}, \apj, 131, 558, \dodoi{10.1086/146869}

\bibitem[{{van den Bergh}(1960{\natexlab{b}})}]{vandenBergh1960b}
---. 1960{\natexlab{b}}, \apj, 131, 215, \dodoi{10.1086/146821}

\bibitem[{{van den Bergh}(1976)}]{vandenBergh1976}
---. 1976, \apj, 206, 883, \dodoi{10.1086/154452}

\bibitem[{{van der Burg} {et~al.}(2016){van der Burg}, {Muzzin}, \& {Hoekstra}}]{vanderBurg2016}
{van der Burg}, R. F.~J., {Muzzin}, A., \& {Hoekstra}, H. 2016, \aap, 590, A20, \dodoi{10.1051/0004-6361/201628222}

\bibitem[{{van der Wel} {et~al.}(2010){van der Wel}, {Bell}, {Holden}, {Skibba}, \& {Rix}}]{vanderWel2010}
{van der Wel}, A., {Bell}, E.~F., {Holden}, B.~P., {Skibba}, R.~A., \& {Rix}, H.-W. 2010, \apj, 714, 1779, \dodoi{10.1088/0004-637X/714/2/1779}

\bibitem[{{van Dokkum} {et~al.}(2019){van Dokkum}, {Danieli}, {Abraham}, {Conroy}, \& {Romanowsky}}]{vanDokkum2019}
{van Dokkum}, P., {Danieli}, S., {Abraham}, R., {Conroy}, C., \& {Romanowsky}, A.~J. 2019, \apjl, 874, L5, \dodoi{10.3847/2041-8213/ab0d92}

\bibitem[{{van Dokkum} {et~al.}(2017){van Dokkum}, {Abraham}, {Romanowsky}, {Brodie}, {Conroy}, {Danieli}, {Lokhorst}, {Merritt}, {Mowla}, \& {Zhang}}]{vanDokkum2017}
{van Dokkum}, P., {Abraham}, R., {Romanowsky}, A.~J., {et~al.} 2017, \apjl, 844, L11, \dodoi{10.3847/2041-8213/aa7ca2}

\bibitem[{{van Dokkum} {et~al.}(2018){van Dokkum}, {Danieli}, {Cohen}, {Merritt}, {Romanowsky}, {Abraham}, {Brodie}, {Conroy}, {Lokhorst}, {Mowla}, {O'Sullivan}, \& {Zhang}}]{vanDokkum2018}
{van Dokkum}, P., {Danieli}, S., {Cohen}, Y., {et~al.} 2018, \nat, 555, 629, \dodoi{10.1038/nature25767}

\bibitem[{{van Dokkum} {et~al.}(2015){van Dokkum}, {Abraham}, {Merritt}, {Zhang}, {Geha}, \& {Conroy}}]{vanDokkum2015}
{van Dokkum}, P.~G., {Abraham}, R., {Merritt}, A., {et~al.} 2015, \apjl, 798, L45, \dodoi{10.1088/2041-8205/798/2/L45}

\bibitem[{{Vega-Acevedo} \& {Hidalgo-G{\'a}mez}(2022)}]{Vega-Acevedo2022}
{Vega-Acevedo}, I., \& {Hidalgo-G{\'a}mez}, A.~M. 2022, \rmxaa, 58, 61, \dodoi{10.22201/ia.01851101p.2022.58.01.05}

\bibitem[{{Venhola} {et~al.}(2019){Venhola}, {Peletier}, {Laurikainen}, {Salo}, {Iodice}, {Mieske}, {Hilker}, {Wittmann}, {Paolillo}, {Cantiello}, {Janz}, {Spavone}, {D'Abrusco}, {van de Ven}, {Napolitano}, {Verdoes Kleijn}, {Capaccioli}, {Grado}, {Valentijn}, {Falc{\'o}n-Barroso}, \& {Limatola}}]{Venhola2019}
{Venhola}, A., {Peletier}, R., {Laurikainen}, E., {et~al.} 2019, \aap, 625, A143, \dodoi{10.1051/0004-6361/201935231}

\bibitem[{{Vulcani} {et~al.}(2021){Vulcani}, {Poggianti}, {Moretti}, {Franchetto}, {Bacchini}, {McGee}, {Jaff{\'e}}, {Mingozzi}, {Werle}, {Tomi{\v{c}}i{\'c}}, {Fritz}, {Bettoni}, {Wolter}, \& {Gullieuszik}}]{Vulcani2021}
{Vulcani}, B., {Poggianti}, B.~M., {Moretti}, A., {et~al.} 2021, \apj, 914, 27, \dodoi{10.3847/1538-4357/abf655}

\bibitem[{Walmsley {et~al.}(2020)Walmsley, Smith, Lintott, Gal, Bamford, Dickinson, Fortson, Kruk, Masters, Scarlata, {et~al.}}]{walmsley2020galaxy}
Walmsley, M., Smith, L., Lintott, C., {et~al.} 2020, Monthly Notices of the Royal Astronomical Society, 491, 1554

\bibitem[{{Walmsley} {et~al.}(2022){Walmsley}, {Scaife}, {Lintott}, {Lochner}, {Etsebeth}, {G{\'e}ron}, {Dickinson}, {Fortson}, {Kruk}, {Masters}, {Mantha}, \& {Simmons}}]{Walmsley2022}
{Walmsley}, M., {Scaife}, A. M.~M., {Lintott}, C., {et~al.} 2022, \mnras, 513, 1581, \dodoi{10.1093/mnras/stac525}

\bibitem[{Walmsley {et~al.}(2022)Walmsley, Lintott, Tobias, Kruk, Krawczyk, Willett, Bamford, Keel, Kelvin, Fortson, Masters, Mehta, Simmons, Smethurst, Baeten, \& Macmillan}]{Walmsley2022decals}
Walmsley, M., Lintott, C., Tobias, G., {et~al.} 2022, Monthly Notices of the Royal Astronomical Society, 509, 3966.
\newblock \url{https://arxiv.org/abs/2102.08414}

\bibitem[{Walmsley {et~al.}(2023)Walmsley, Allen, Aussel, Bowles, Gregorowicz, Slijepcevic, Lintott, m.~Scaife, Jabłońska, Karchev, Lanzieri, Mohan, O’Ryan, Saiguhan, Suárez, Guerra-Varas, \& Velu}]{Walmsley2023}
Walmsley, M., Allen, C., Aussel, B., {et~al.} 2023, Journal of Open Source Software, 8, 5312, \dodoi{10.21105/joss.05312}

\bibitem[{{Wang} {et~al.}(2023){Wang}, {Peng}, {Liu}, {Mihos}, {C{\^o}t{\'e}}, {Ferrarese}, {Taylor}, {Blakeslee}, {Cuillandre}, {Duc}, {Guhathakurta}, {Gwyn}, {Ko}, {Lan{\c{c}}on}, {Lim}, {MacArthur}, {Puzia}, {Roediger}, {Sales}, {S{\'a}nchez-Janssen}, {Spengler}, {Toloba}, {Zhang}, \& {Zhu}}]{Wang2023}
{Wang}, K., {Peng}, E.~W., {Liu}, C., {et~al.} 2023, \nat, 623, 296, \dodoi{10.1038/s41586-023-06650-z}

\bibitem[{{Waskom}(2021)}]{seaborn}
{Waskom}, M. 2021, The Journal of Open Source Software, 6, 3021, \dodoi{10.21105/joss.03021}

\bibitem[{{Wen} \& {Zheng}(2016)}]{Wen2016}
{Wen}, Z.~Z., \& {Zheng}, X.~Z. 2016, \apj, 832, 90, \dodoi{10.3847/0004-637X/832/1/90}

\bibitem[{Wen {et~al.}(2014)Wen, Zheng, \& An}]{Wen2014}
Wen, Z.~Z., Zheng, X.~Z., \& An, F.~X. 2014, The Astrophysical Journal, 787, 130, \dodoi{10.1088/0004-637X/787/2/130}

\bibitem[{{W}es {M}c{K}inney(2010)}]{McKinney2010}
{W}es {M}c{K}inney. 2010, in {P}roceedings of the 9th {P}ython in {S}cience {C}onference, ed. {S}t\'efan van~der {W}alt \& {J}arrod {M}illman, 56 -- 61, \dodoi{10.25080/Majora-92bf1922-00a}

\bibitem[{{Wetzel} {et~al.}(2013){Wetzel}, {Tinker}, {Conroy}, \& {van den Bosch}}]{Wetzel2013}
{Wetzel}, A.~R., {Tinker}, J.~L., {Conroy}, C., \& {van den Bosch}, F.~C. 2013, \mnras, 432, 336, \dodoi{10.1093/mnras/stt469}

\bibitem[{{White} \& {Rees}(1978)}]{White1978}
{White}, S.~D.~M., \& {Rees}, M.~J. 1978, \mnras, 183, 341, \dodoi{10.1093/mnras/183.3.341}

\bibitem[{{Whitney} {et~al.}(2021){Whitney}, {Ferreira}, {Conselice}, \& {Duncan}}]{Whitney2021}
{Whitney}, A., {Ferreira}, L., {Conselice}, C.~J., \& {Duncan}, K. 2021, \apj, 919, 139, \dodoi{10.3847/1538-4357/ac1422}

\bibitem[{{Wilkinson} {et~al.}(2024){Wilkinson}, {Ellison}, {Bottrell}, {Bickley}, {Byrne-Mamahit}, {Ferreira}, \& {Patton}}]{Wilkinson2024}
{Wilkinson}, S., {Ellison}, S.~L., {Bottrell}, C., {et~al.} 2024, \mnras, 528, 5558, \dodoi{10.1093/mnras/stae287}

\bibitem[{{Wilkinson} {et~al.}(2022){Wilkinson}, {Ellison}, {Bottrell}, {Bickley}, {Gwyn}, {Cuillandre}, \& {Wild}}]{Wilkinson2022}
---. 2022, \mnras, 516, 4354, \dodoi{10.1093/mnras/stac1962}

\bibitem[{{Willett} {et~al.}(2013){Willett}, {Lintott}, {Bamford}, {Masters}, {Simmons}, {Casteels}, {Edmondson}, {Fortson}, {Kaviraj}, {Keel}, {Melvin}, {Nichol}, {Raddick}, {Schawinski}, {Simpson}, {Skibba}, {Smith}, \& {Thomas}}]{Willet2013}
{Willett}, K.~W., {Lintott}, C.~J., {Bamford}, S.~P., {et~al.} 2013, \mnras, 435, 2835, \dodoi{10.1093/mnras/stt1458}

\bibitem[{{Yin} {et~al.}(2011){Yin}, {Matteucci}, \& {Vladilo}}]{Yin2011}
{Yin}, J., {Matteucci}, F., \& {Vladilo}, G. 2011, \aap, 531, A136, \dodoi{10.1051/0004-6361/201015022}

\bibitem[{{York} {et~al.}(2000){York}, {Adelman}, {Anderson}, {Anderson}, {Annis}, {Bahcall}, \& {SDSS Collaboration}}]{York2000}
{York}, D.~G., {Adelman}, J., {Anderson}, John~E., J., {et~al.} 2000, \aj, 120, 1579, \dodoi{10.1086/301513}

\bibitem[{{Zhang} {et~al.}(2015){Zhang}, {Peng}, {C{\^o}t{\'e}}, {Liu}, {Ferrarese}, {Cuillandre}, {Caldwell}, {Gwyn}, {Jord{\'a}n}, {Lan{\c{c}}on}, {Li}, {Mu{\~n}oz}, {Puzia}, {Bekki}, {Blakeslee}, {Boselli}, {Drinkwater}, {Duc}, {Durrell}, {Emsellem}, {Firth}, \& {S{\'a}nchez-Janssen}}]{Zhang2015}
{Zhang}, H.-X., {Peng}, E.~W., {C{\^o}t{\'e}}, P., {et~al.} 2015, \apj, 802, 30, \dodoi{10.1088/0004-637X/802/1/30}

\bibitem[{{Zhang} {et~al.}(2018){Zhang}, {Puzia}, {Peng}, {Liu}, {C{\^o}t{\'e}}, {Ferrarese}, {Duc}, {Eigenthaler}, {Lim}, {Lan{\c{c}}on}, {Mu{\~n}oz}, {Roediger}, {S{\'a}nchez-Janssen}, {Taylor}, \& {Yu}}]{Zhang2018}
{Zhang}, H.-X., {Puzia}, T.~H., {Peng}, E.~W., {et~al.} 2018, \apj, 858, 37, \dodoi{10.3847/1538-4357/aab88a}

\bibitem[{{Zhang} {et~al.}(2020{\natexlab{a}}){Zhang}, {Paudel}, {Smith}, {Duc}, {Puzia}, {Peng}, {C{\^o}te}, {Ferrarese}, {Boselli}, {Wang}, \& {Oh}}]{Zhang2020a}
{Zhang}, H.-X., {Paudel}, S., {Smith}, R., {et~al.} 2020{\natexlab{a}}, \apjl, 891, L23, \dodoi{10.3847/2041-8213/ab7825}

\bibitem[{{Zhang} {et~al.}(2020{\natexlab{b}}){Zhang}, {Smith}, {Oh}, {Paudel}, {Duc}, {Boselli}, {C{\^o}t{\'e}}, {Ferrarese}, {Gao}, {Hunter}, {Puzia}, {Peng}, {Rong}, {Shin}, \& {Zhao}}]{Zhang2020b}
{Zhang}, H.-X., {Smith}, R., {Oh}, S.-H., {et~al.} 2020{\natexlab{b}}, \apj, 900, 152, \dodoi{10.3847/1538-4357/abab96}

\end{thebibliography}

%
%

\begin{deluxetable*}{cccc}

\tabletypesize{\scriptsize}
\tablecaption{Comparison of C1 NGVS and C2 Codes for Structural and Star Formation Classification\label{tab:ngvs_combined_comparison}}
\tablehead{
\colhead{C1 Code} & \colhead{C2 Code} & \colhead{N (Random 100)} & \colhead{N (Faint Third)}
}
\startdata
\multicolumn{4}{c}{{\bf Structural Code Comparison}} \\
E  & E   & 84   & 1194 \\
ES & E   & 1    & 5    \\
I  & I   & 3    & -    \\
EI & E   & 1    & 21   \\
E  & ES  & 2    & -    \\
ES & ES  & 4    & 9    \\
ES & EI  & 1    & 1    \\
EI & ES  & 2    & 2    \\
S  & S   & -    & 3    \\
E  & EI  & -    & 9    \\
EI & EI  & -    & 22   \\
\tableline
\multicolumn{4}{c}{{\bf Star Formation Code Comparison}} \\
Q  & Q   & 87   & 1163 \\
I  & Q   & 5    & 51   \\
A  & Q   & 1    & 1    \\
Q  & I   & -    & 3    \\
I  & I   & 3    & 37   \\
I  & A   & 1    & 3    \\
A  & A   & 3    & 11   \\
A  & I   & -    & 2    \\
\tableline
\enddata
\end{deluxetable*}

\begin{deluxetable*}{cccccc}
\tabletypesize{\scriptsize}
\tablecaption{Distribution of Star Formation and Structural Codes\label{tab:full_ngvs_morphologies}}
\tablehead{\colhead{Star Formation Code} & \colhead{N} & \colhead{P (\%)} &
            \colhead{Structural Code} & \colhead{N} & \colhead{P (\%)}}
\startdata
\multicolumn{6}{c}{{\bf Full Sample (N=3689)}}\\
Q & 3037 & 82.33\% $\pm$ 1.49 & E  & 3099 & 84.01\% $\pm$ 1.51 \\
I &  510 & 13.82\% $\pm$ 0.61 & ES &  276 &  7.48\% $\pm$ 0.45 \\
A &  138 &  3.74\% $\pm$ 0.32 & S  &   86 &  2.33\% $\pm$ 0.25 \\
& & & SI &  10 & 0.27\% $\pm$ 0.09 \\
& & & EI & 192 & 5.20\% $\pm$ 0.38 \\
& & & I  &  25 & 0.68\% $\pm$ 0.14 \\
& & & & & \\
\multicolumn{6}{c}{{\bf High-Mass Sample (N=236)}}\\
Q & 104 & 44.07\% $\pm$ 4.32 & E  &  52 & 22.03\% $\pm$ 3.06 \\
I &  75 & 31.78\% $\pm$ 3.67 & ES & 115 & 48.73\% $\pm$ 4.54 \\
A &  55 & 23.31\% $\pm$ 3.14 & S  &  62 & 26.27\% $\pm$ 3.34 \\
& & & SI &   1 & 0.42\% $\pm$ 0.42 \\
& & & EI &   5 & 2.12\% $\pm$ 0.95 \\
& & & & & \\
\multicolumn{6}{c}{{\bf Low-Mass Sample (N=3431)}}\\
Q & 2921 & 85.14\% $\pm$ 1.58 & E  & 3034 & 88.43\% $\pm$ 1.61 \\
I &  429 & 12.50\% $\pm$ 0.60 & ES &  155 &  4.52\% $\pm$ 0.36 \\
A &   79 &  2.30\% $\pm$ 0.26 & S  &   22 &  0.64\% $\pm$ 0.14 \\
& & & SI &    9 & 0.26\% $\pm$ 0.09 \\
& & & EI &  186 & 5.42\% $\pm$ 0.40 \\
& & & I  &   25 & 0.73\% $\pm$ 0.15 \\
\enddata
\end{deluxetable*}




\begin{deluxetable*}{cccc}
\tabletypesize{\scriptsize}
\tablecaption{Profile Results for Structural and Star Formation Codes \label{tab:profile_results}}
\tablehead{ \colhead{Type} & \colhead{$r_{50}$ (arcmin)} & \colhead{$r_{10}$ (arcmin)} & \colhead{Peak Density (N deg$^{-2}$)} }
\startdata
\multicolumn{4}{c}{\textbf{Full Sample}} \\
 & & & \\
All Structural Codes & $138.4 \pm 6.3$ & $360.1 \pm 4.4$ & 94.3 \\
All Star Formation Codes & $138.4 \pm 6.3$ & $360.1 \pm 4.4$ & 94.0 \\
 & & & \\
\multicolumn{4}{c}{\textbf{Structural Code}} \\
 & & & \\
E & $135.9 \pm 4.3$ & $346.1 \pm 5.6$ & 85.7 \\
ES & $143.6 \pm 55.6$ & $405.8 \pm 9.8$ & 5.1 \\
S & $95.7 \pm 121.8$ & $408.9 \pm 15.2$ & 0.9 \\
SI & $110.2 \pm 44.2$ & $322.5 \pm 139.3$ & 0.1 \\
EI & $288.9 \pm 92.3$ & $413.4 \pm 6.2$ & 2.5 \\
I & $198.3 \pm 107.6$ & $397.4 \pm 42.5$ & 0.3 \\
 & & & \\
\multicolumn{4}{c}{\textbf{Star Formation Code}} \\
 & & & \\
Q & $135.9 \pm 4.5$ & $344.4 \pm 5.8$ & 84.0 \\
I & $161.0 \pm 55.6$ & $405.0 \pm 7.0$ & 9.2 \\
A & $293.4 \pm 130.3$ & $413.9 \pm 5.7$ & 1.4 \\
\tableline
\enddata
\tablecomments{$r_{10}$ and $r_{50}$ denote the projected radii where the observed
density falls to one-tenth and one-half its peak value, respectively.}
\end{deluxetable*}
\FloatBarrier



\clearpage
\startlongtable
\setlength{\tabcolsep}{2.0pt}
\begin{deluxetable*}{r p{2.7cm}@{\hspace{1.8em}} c c c c c p{1.9cm} >{\raggedright\arraybackslash\hspace{0pt}\sloppy}p{8.6cm}}
\tabletypesize{\scriptsize}
\tablewidth{\textwidth}
\tablecaption{Candidate Compact Galaxies\label{tab:_types_ce}}
\tablehead{
 \colhead{\#} &
 \colhead{ID [NGVSJ]} &
 \colhead{NGVS} &
 \colhead{VCC} &
 \colhead{$\alpha$(2000)} &
 \colhead{$\delta$(2000)} &
 \colhead{$g^\prime$} &
 \colhead{Type} &
 \colhead{Notes} \\
 \colhead{} & \colhead{} & \colhead{} & \colhead{} &
 \colhead{(deg)} & \colhead{(deg)} & \colhead{(mag)} &
 \colhead{} & \colhead{}
}
\startdata
1 & 12:10:35.68+11:45:38.9 & 26 & 24 & 182.65 & 11.76 & 15.37 & ES(di\textbar{}du)-I(c\textbar{}d) & ES galaxy with dust and star formation. Possible compact central region offset from photocenter, probably due to dust absorption. Candidate UCG. Classified as ``certain" member. Cross ID: VCC24 \\
2 & 12:11:02.73+12:06:14.4 & 34 & 32 & 182.76 & 12.10 & 13.94 & E(n:)-Q & Smooth ETG with possible compact nucleus and hint of other nuclear substructure. Candidate UCG. Classified as ``certain" member. Cross ID: VCC32 \\
3 & 12:11:45.94+13:17:07.9 & 50 & - & 182.94 & 13.29 & 18.18 & E(n)-Q & Nucleated, compact, round  ETG. Candidate UCG. Classified as ``certain" member \\
4 & 12:13:13.68+13:31:22.1 & 89 & - & 183.31 & 13.52 & 16.07 & E(du:\textbar{}b:)-I(c:) & Compact, intermediate luminosity ETG with boxy isophotes. SF in central region with possible bar-like structure. Candidate UCG. Classified as ``certain" member \\
5 & 12:14:43.05+10:31:59.3 & 131 & - & 183.68 & 10.53 & 18.07 & ES-I(c:) & Compact, possible faint ES galaxy seen face-on, with possible blue core and clumpy central structure surrounded by lower surface brightness, regular envelope. Candidate UCG. Classified as ``certain" member \\
6 & 12:14:53.50+09:40:11.8 & 140 & - & 183.72 & 9.67 & 17.69 & E-I(c) & Compact, elongated  ETG with possible irregular structure in the core. Candidate UCG. Classified as ``certain" member \\
7 & 12:15:27.29+10:30:44.5 & 164 & - & 183.86 & 10.51 & 17.60 & EI-I(c\textbar{}f) & Compact EI galaxy with SF in inner core, in patches and filaments. Candidate UCG. Classified as ``certain" member \\
8 & 12:16:34.04+10:12:22.8 & 231 & - & 184.14 & 10.21 & 17.48 & EI-I(c\textbar{}g:) & Compact EI galaxy with complex, irregular morphology and possible central star formation. Candidate UCG. Classified as ``certain" member \\
9 & 12:16:47.81+08:02:57.0 & 244 & 207 & 184.20 & 8.05 & 17.02 & EI(di:)-I(c\textbar{}p\textbar{}d) & Compact EI galaxy with active star formation. Peak of emission is offset from photocenter. The galaxy, which has velocity of 2588 km/s, might be associated with VCC180 (11.2' to the SW, v=2239 km/s) and VCC190 (16.2' to the SW, v=2354 km/s). Candidate UCG. Classified as ``certain" member. Cross ID: VCC207 \\
10 & 12:17:28.07+12:55:56.0 & 279 & - & 184.37 & 12.93 & 17.07 & ES(di\textbar{}du:)-I(d\textbar{}c\textbar{}p) & Compact ES galaxy seen edge-on with star formation and dust in disk. Candidate UCG. Classified as ``certain" member \\
11 & 12:18:17.99+07:07:41.0 & 332 & - & 184.57 & 7.13 & 16.60 & E(du:)-I(c) & Compact  ETG with star formation and possible dust in core. Candidate UCG. Classified as ``certain" member \\
12 & 12:19:23.09+05:47:41.3 & 400 & 344 & 184.85 & 5.79 & 15.11 & E(n:)-Q & Compact  ETG with possible nucleus. The galaxy, which has a measured velocity of 2038 km/s might be part of a small group with VCC345 as the brightest member (v = 2202 km/s), VCC340 (v~1506 kms/s), VCC336 (v~3016 kms/s) and VCC358 (v~2514 kms/s), all of which are between 2 and 7 effective radii of VCC345 (108 to 308\arcsec). No sign of tidal interaction, but there is some isophotal twisting. Galaxy is elongated in the inner regions, becoming round at large radius. Candidate UCG. Classified as ``certain" member. Cross ID: VCC344 \\
13 & 12:20:21.89+12:11:15.4 & 485 & 410 & 185.09 & 12.19 & 17.21 & EI-I(f\textbar{}p) & Compact EI galaxy with active star formation. Primary star formation is in patches and filaments offset from galaxy photocenter. Candidate UCG. Classified as ``certain" member. Cross ID: VCC410 \\
14 & 12:20:30.91+06:38:23.9 & 503 & - & 185.13 & 6.64 & 18.09 & E-I(c:) & Compact ETG with high-surface brightness blue core, central substructure, and extended, lower surface brightness elongated envelope. Candidate UCG. Classified as ``certain" member \\
15 & 12:23:18.39+05:36:25.5 & 827 & - & 185.83 & 5.61 & 16.58 & ES(di)-I(d\textbar{}c) & Compact, high-surface brightness ES galaxy seen edge-on, with SF in disk and patches. Candidate UCG. Classified as ``certain" member \\
16 & 12:24:35.32+06:01:58.8 & 1006 & - & 186.15 & 6.03 & 16.71 & E-I(c) & Compact  ETG with SF in core. Candidate UCG. Classified as ``certain" member \\
17 & 12:24:54.85+16:48:38.9 & 1041 & - & 186.23 & 16.81 & 17.37 & E(n:)-I(c) & Compact, round  ETG with star formation in blue core and possible nucleus. Candidate UCG. Classified as ``certain" member \\
18 & 12:26:21.55+06:40:10.4 & 1289 & 890 & 186.59 & 6.67 & 16.57 & EI(du)-I(c) & Compact EI  galaxy with active star formation in its core. Candidate UCG. Classified as ``certain" member. Cross ID: VCC890 \\
19 & 12:27:26.96+05:11:16.9 & 1515 & - & 186.86 & 5.19 & 16.55 & E-I(c\textbar{}f) & Compact, elongated  ETG with star formation in core and filaments. Candidate UCG. Classified as ``certain" member \\
20 & 12:29:21.29+08:09:23.8 & 1840 & 1178 & 187.34 & 8.16 & 12.95 & E(n)-Q & Nucleated ETG with strong isophote twisting suggestive of tidal influence from VCC1226. The galaxy, which has velocity v=1243 km/s, is located 677\arcsec to the NW of M49 (VCC1226, v=989 km/s). Candidate UCG. Classified as ``certain" member. Cross ID: VCC1178 \\
21 & 12:29:34.99+08:03:28.8 & 1877 & 1199 & 187.40 & 8.06 & 15.94 & E(n)-Q & Compact  ETG with small envelope and bright nucleus. The galaxy, which has velocity v=1395 km/s, is located 271\arcsec to the NW of M49 (VCC1226, v=989 km/s). Candidate UCG. Classified as ``certain" member. Cross ID: VCC1199 \\
22 & 12:30:31.97+12:29:24.6 & 2065 & 1297 & 187.63 & 12.49 & 13.76 & E(n)-Q & Compact  ETG with low-surface brightness envelope and bright nucleus. The galaxy, which has velocity v=1554 km/s, is located 438\arcsec to the NW of M87 (VCC1316, v=1284 km/s) and is considered a classical example of a Ultra Compact Galaxy (UCG) due to interaction with M87 itself. Classified as ``certain" member. Cross ID: VCC1297, NGC4486B \\
23 & 12:30:42.00+05:52:43.8 & 2090 & - & 187.68 & 5.88 & 17.23 & EI(du\textbar{}di:)-I(c\textbar{}f\textbar{}d:) & Compact EI galaxy with SF in disk or filament embedded in slightly elliptical isophotes. Candidate UCG.  Classified as ``certain" member \\
24 & 12:30:48.58+12:02:42.7 & 2112 & 1313 & 187.70 & 12.05 & 16.94 & EI-I(p) & Very compact EI galaxy with very active star formation displaced from the photocenter of the older, more smoothly distributed component. The galaxy, which has velocity v=1260 km/s, is located 1245\arcsec to the SW of M87 (VCC1316, v=1284 km/s). Candidate UCG. Classified as ``certain" member. Cross ID: VCC1313 \\
25 & 12:30:57.77+12:16:15.5 & 2141 & 1327 & 187.74 & 12.27 & 12.87 & E(di:\textbar{}du:)-Q & ETG projected ~3\arcsec from a very bright star. ACSVCS images reveal compact edge-on disk with dust. The galaxy, which has velocity v=758 km/s, is located in the Virgo core, 449\arcsec to the SE of M87 (VCC1316, v=1284 km/s). Candidate UCG. Classified as ``certain" member. Cross ID: VCC1327 \\
26 & 12:31:52.90+12:15:59.1 & 2275 & - & 187.97 & 12.27 & 17.80 & E(n)-Q & Round, compact, very LSB ETG with compact envelope and bright nucleus. Candidate UCG. Classified as ``certain" member \\
27 & 12:32:33.52+09:10:25.2 & 2392 & 1437 & 188.14 & 9.17 & 14.79 & EI(n:\textbar{}sh\textbar{}du)-I(c) & Possibly nucleated compact EI galaxy with dust and active star formation in its core. Candidate UCG. Classified as ``certain" member. Cross ID: VCC1437 \\
28 & 12:35:37.25+12:22:55.4 & 2784 & 1627 & 188.91 & 12.38 & 14.94 & E(n)-Q & Nucleated ETG. May be associated with VCC1619, which is located 9.8' to the SW and has similar velocity (266 km/s for VCC1627 and 420 km/s for VCC1619), and VCC1632 (v=339 km/s), located 10.4' to the N. Candidate UCG. Classified as ``certain" member. Cross ID: VCC1627 \\
29 & 12:38:15.49+06:59:39.1 & 3040 & 1750 & 189.56 & 6.99 & 16.66 & EI(st:)-I(c\textbar{}p\textbar{}g) & Compact EI with low level star formation throughout, and a more intense patch of star formation offset from core. Isophotes appear distorted at large radius. May be associated with VCC1726, which is located 10.0' to the NW and has similar velocity (-112 km/s for VCC1750 and 45 km/s for VCC1726). Candidate UCG. Classified as ``certain" member. Cross ID: VCC1750 \\
30 & 12:41:01.08+09:43:06.3 & 3238 & - & 190.25 & 9.72 & 17.78 & EI-I(c) & Compact EI galaxy with complex morphology, central star formation, and low-surface brightness regular, elongated, envelope. Candidate UCG. Classified as ``certain" member \\
\enddata
\end{deluxetable*}

\clearpage




\begin{longrotatetable}
\tabletypesize{\scriptsize}
\setlength{\tabcolsep}{2pt}

\begin{deluxetable*}{%
  r               
  c               
  p{4.4cm}        
  c               
  c               
  c               
  c               
  c               
  p{1.8cm}        
  c               
  c               
  c               
  p{0.32\textwidth} 
}
\tablecaption{Candidate Ultra-Diffuse Galaxies \label{tab:_types_udg}}
\tablehead{
\colhead{\#} & \colhead{Sample} & \colhead{ID [NGVSJ]} & \colhead{NGVS} & \colhead{VCC} &
\colhead{$\alpha$ (2000)} & \colhead{$\delta$ (2000)} & \colhead{$g^\prime$} &
\colhead{Type} & \colhead{$\log R_e$} & \colhead{$\mu_{e,g}$} &
\colhead{$\langle \mu_{e,g} \rangle$} & \colhead{Notes}
}
\startdata
\multicolumn{13}{c}{\textbf{Certain NGVS Members}} \\
\tableline
1 & Primary & 12:16:32.59+13:09:51.3 & 227 & {197} & 184.14 & 13.16 & 16.77 & E(n\vbar st)-Q & 1.79 & 28.85 & 27.72 & Nucleated, very elongated, very LSB ETG, possibly due to tidal stripping event. Candidate UDG.  Classified as ``certain" member. Cross ID: VCC197 \\
2 & Primary & 12:19:36.96+15:27:16.8 & 421 & {360} & 184.90 & 15.45 & 16.61 & EI-I(g:) & 1.53 & 26.80 & 26.25 & Very low-surface brightness EI galaxy with irregular morphology. Candidate UDG. Classified as ``certain" member. Cross ID: VCC360 \\
3 & Primary & 12:25:37.61+10:14:58.6 & 1160 & {811} & 186.41 & 10.25 & 16.96 & E-I(g) & 1.53 & 27.10 & 26.60 & Extended, very LSB ETG. A faint and compact LSB galaxy, NGVS1148, is located 90\arcsec to the SW, although the velocity of either galaxy is unknown. Candidate UDG. Classified as ``certain" member. Cross ID: VCC811 \\
4 & Primary & 12:27:31.55+09:35:44.3 & 1528 & 1017 & 186.88 & 9.60 & 14.55 & EI(sh\vbar st:\vbar du:\vbar a)-I(g) & 1.73 & 25.83 & 25.19 & Very LSB EI galaxy that is star forming at a relatively low level. Shell-like structures (and generally disturbed appearance) at large radius. May be interacting with VCC983 (264\arcsec to the NW) and VCC993 (163 to the SW), with connecting diffuse light, although neither galaxy has a measured velocity. Candidate UDG. Classified as ``certain" member. Cross ID: VCC1017 \\
5 & Primary & 12:27:55.22+12:22:09.5 & 1593 & {1052} & 186.98 & 12.37 & 16.06 & E-I(g\vbar f) & 1.68 & 27.00 & 26.43 & Extended, extremely LSB EI galaxy with disturbed morphology. Possible low level star formation.  Located 80\arcsec to the S of NGVS1552 (VCC1036) and NGVS1588 (VCC1047), which likely form a physical pair. A compact, very LSB ETG, NGVS1586 is located 57\arcsec to the NW. Candidate UDG. Classified as ``certain" member. Cross ID: VCC1052 \\
6 & Primary & 12:30:08.68+09:42:56.2 & 1993 & - & 187.54 & 9.72 & 17.09 & E-Q & 1.55 & 27.27 & 26.84 & Extended, extremely LSB ETG. Apparent member of an LSB pair, with the second member, NGVS2001, located 57\arcsec to the E. Neither galaxy has a measured velocity. A third, possibly connected LSB ETG, NGVS2031, is located 175\arcsec to the E. Candidate UDG. Classified as ``certain" member \\
7 & Primary & 12:30:24.43+13:58:54.5 & 2045 & {1287} & 187.60 & 13.98 & 15.71 & E-Q & 1.66 & 26.58 & 26.01 & Extended very LSB ETG with no obvious nucleus. Candidate UDG. Classified as ``certain" member. Cross ID: VCC1287 \\
8 & Secondary & 12:33:29.59+15:14:02.8 & 2531 & - & 188.37 & 15.23 & 17.51 & E-Q & 1.43 & 27.24 & 26.67 & Extended, extremely LSB ETG. Located 356\arcsec to the W of the bright LTG NGVS2568 (VCC1524), with which it shares almost identical velocity (260 vs 296 km/s). Candidate UDG. Classified as ``certain" member \\
9 & Secondary & 12:40:21.21+12:43:02.8 & 3190 & {1835} & 190.09 & 12.72 & 18.23 & E-Q & 1.39 & 27.62 & 27.16 & Extended, extremely LSB ETG. Candidate UDG. Classified as ``certain" member. Cross ID: VCC1835 \\
10 & Primary & 12:41:39.34+09:12:30.5 & 3272 & {1884} & 190.41 & 9.21 & 16.20 & E-Q & 1.59 & 26.69 & 26.16 & Extended, very LSB ETG. Candidate UDG. Classified as ``certain" member. Cross ID: VCC1884 \\
\tableline
\multicolumn{13}{c}{\textbf{Probable NGVS Members}} \\
\tableline
11 & Secondary & 12:20:11.38+11:53:56.5 & 466 & - & 185.05 & 11.90 & 19.78 & E(n:\vbar st:)-Q & 1.24 & 28.65 & 27.96 & Extended, extremely LSB ETG with possible tidal stream. Candidate UDG. Classified as ``probable" member \\
12 & Primary & 12:24:01.84+13:51:56.4 & 935 & - & 186.01 & 13.87 & 18.39 & E(n)-Q & 1.52 & 28.32 & 27.95 & Extended, extremely  LSB ETG with possible bright nucleus. Several faint galaxies (NGVS878, NGVS952, NGVS971 and NGVS978) are located within 153\arcsec and 327\arcsec, while a brighter galaxy (NGVS904 == VCC674) is located 147\arcsec to the N. Candidate UCD transition object. Candidate UDG. Classified as ``probable" member \\
13 & Primary & 12:24:42.06+13:31:00.6 & 1017 & - & 186.18 & 13.52 & 17.60 & E(n:)-Q & 1.51 & 27.85 & 27.15 & Extended, extremely LSB ETG with possible bright nucleus. Candidate UCD transition object. Candidate UDG. Classified as ``probable" member \\
14 & Primary & 12:26:38.25+13:04:44.2 & 1353 & {927} & 186.66 & 13.08 & 18.19 & E-Q & 1.41 & 28.15 & 27.25 & Extended, extremely LSB ETG with bright nucleus. Candidate UCD transition object. Candidate UDG. Classified as ``probable" member. Cross ID: VCC927 \\
15 & Primary & 12:26:48.36+13:21:17.7 & 1397 & - & 186.70 & 13.35 & 18.97 & E-Q & 1.45 & 28.88 & 28.19 & Extended, extremely LSB ETG. Candidate UDG. Classified as ``probable" member \\
16 & Secondary & 12:27:15.75+13:26:56.1 & 1478 & - & 186.82 & 13.45 & 18.91 & E-Q & 1.32 & 28.28 & 27.51 & Extended, extremely LSB ETG. Another very LSB ETG, NGVS1476, is located 135\arcsec to the S. Candidate UDG. NGVS1478 and NGVS1479 are identified here as two separate galaxies, with NGVS1479  located 4\arcsec  to the N.  However, this could also be a single, asymmetric galaxy. . Classified as ``probable" member \\
17 & Primary & 12:28:10.07+12:43:29.4 & 1632 & - & 187.04 & 12.72 & 18.42 & E-Q & 1.50 & 28.30 & 27.89 & Extended, extremely LSB ETG. Candidate UDG. Classified as ``probable" member \\
18 & Primary & 12:28:37.88+12:51:42.0 & 1721 & - & 187.16 & 12.86 & 18.75 & E-Q & 1.46 & 28.46 & 28.02 & Extended, extremely LSB ETG. Three bright galaxies are found nearby, although physical association cannot be assessed given that NGVS1721 does not have a measured velocity: NGVS1693 (VCC1104, 198\arcsec to the SW, v=1709 km/s), NGVS1731 (VCC1122, 203\arcsec to the N, v= 465 km/s), and NGVS1742 (VCC1129, 214\arcsec to the SE, v=12 km/s). Candidate UDG. Classified as ``probable" member \\
19 & Primary & 12:30:37.30+10:20:53.0 & 2079 & - & 187.66 & 10.35 & 17.45 & E-Q & 1.67 & 28.25 & 27.77 & Extended, extremely LSB ETG. Candidate UDG. Classified as ``probable" member \\
20 & Primary & 12:31:48.01+12:21:33.1 & 2268 & - & 187.95 & 12.36 & 19.75 & E-Q & 1.33 & 28.83 & 28.38 & Extended, extremely LSB ETG. Candidate UDG. Classified as ``probable" member \\
21 & Primary & 12:32:22.52+12:19:32.1 & 2365 & - & 188.09 & 12.33 & 19.20 & E-Q & 1.57 & 29.75 & 29.05 & Extended, extremely LSB ETG. Candidate UDG. Classified as ``probable" member \\
22 & Primary & 12:33:02.20+13:42:14.0 & 2458 & - & 188.26 & 13.70 & 20.03 & E-Q & 1.35 & 29.13 & 28.74 & Extended, extremely LSB ETG. Candidate UDG. Classified as ``probable" member \\
23 & Primary & 12:35:12.86+07:03:22.4 & 2731 & - & 188.80 & 7.06 & 18.09 & E-Q & 1.64 & 28.82 & 28.26 & Extended, extremely LSB ETG. Candidate UDG. Classified as ``probable" member \\
24 & Primary & 12:38:09.93+10:47:17.6 & 3032 & - & 189.54 & 10.79 & 20.66 & E-Q & 1.19 & 29.25 & 28.61 & Extremely LSB ETG. Candidate UDG. Classified as ``probable" member \\
25 & Secondary & 12:39:21.99+12:05:34.6 & 3116 & - & 189.84 & 12.09 & 20.76 & E(n:)-Q & 1.10 & 29.10 & 28.24 & Extremely LSB ETG with possible bright nucleus. Candidate UCD transition object. Candidate UDG.  Classified as ``probable" member \\
26 & Secondary & 12:39:48.00+07:18:47.2 & 3146 & - & 189.95 & 7.31 & 18.12 & E(n:)-Q & 1.41 & 27.68 & 27.13 & Extended, extremely LSB ETG. Candidate UDG. Classified as ``probable" member \\
27 & Primary & 12:40:56.41+14:15:16.3 & 3225 & - & 190.24 & 14.25 & 19.46 & E-Q & 1.31 & 28.66 & 27.98 & Extended, extremely LSB ETG. Candidate UDG. Seemingly part of a triplet of galaxies, although none has a measured velocity: NGVS3225 is at the  farther western position, the more compact and brighter NGVS3231 (VCC1862) is 58\arcsec to the SE, and NGVS3233 is 55\arcsec to the NE. Classified as ``probable" member \\
28 & Secondary & 12:40:58.86+14:15:57.6 & 3233 & - & 190.25 & 14.27 & 19.50 & E-Q & 1.29 & 28.26 & 27.94 & Extended, extremely LSB ETG. Seemingly part of a triplet of galaxies, although none has a measured velocity: NGVS3225 is at the  farther western position, the more compact and brighter NGVS3231 (VCC1862) is 58\arcsec to the SE, and NGVS3233 is 55\arcsec to the NE.  Candidate UDG. Classified as ``probable" member \\
29 & Secondary & 12:43:07.09+16:28:42.0 & 3365 & - & 190.78 & 16.48 & 19.38 & E-Q & 1.32 & 28.24 & 27.97 & Extended, extremely LSB ETG. Candidate UDG. Classified as ``probable" member \\
\tableline
\multicolumn{13}{c}{\textbf{Possible NGVS Members}} \\
\tableline
30 & Primary & 12:16:20.63+06:56:52.9 & 213 & - & 184.09 & 6.95 & 18.47 &{E-Q} & 1.57 & 28.80 & 28.31 & Extended, extremely LSB ETG. Candidate UDG. Classified as ``possible" member \\
31 & Secondary & 12:24:11.52+12:11:41.0 & 960 & - & 186.05 & 12.19 & 21.69 & E-Q & 1.04 & 29.45 & 28.89 & Extended, extremely LSB ETG. Located 151\arcsec to the W of the bright LTG NGVS934 (VCC692). Another extended, extremely LSB ETG, NGVS955, is located 108\arcsec to the SW. Candidate UDG. Classified as ``possible" member \\
32 & Primary & 12:26:55.95+12:59:40.0 & 1418 & - & 186.73 & 12.99 & 19.21 & E-Q & 1.46 & 29.08 & 28.49 & Extended, extremely LSB ETG. Candidate UDG. Classified as ``possible" member \\
33 & Primary & 12:27:15.01+12:50:55.9 & 1473 & - & 186.81 & 12.85 & 18.98 & E-Q & 1.49 & 28.67 & 28.40 & Highly elongated, extended, extremely LSB ETG with possible offset nucleus. Candidate UDG. Classified as ``possible" member \\
34 & Primary & 12:29:20.27+12:01:16.4 & 1836 & - & 187.33 & 12.02 & 18.58 & E-Q & 1.48 & 28.28 & 27.98 & Extended, extremely LSB ETG. Candidate UDG. Classified as ``possible" member \\
35 & Secondary & 12:30:09.41+09:16:45.3 & 1995 & - & 187.54 & 9.28 & 21.56 & E-Q & 1.01 & 29.41 & 28.59 & Extremely LSB ETG. Candidate UDG. Classified as ``possible" member \\
36 & Primary & 12:32:18.20+10:17:29.0 & 2355 & - & 188.08 & 10.29 & 15.89 & E-Q & 1.91 & 27.81 & 27.41 & Extended, extremely LSB ETG. A bright ES galaxy, NGVS2348 (VCC1422, v=1300 km/s) is located 155\arcsec to the NE. Candidate UDG. Classified as ``possible" member \\
37 & Secondary & 12:32:18.91+13:44:35.6 & 2359 & - & 188.08 & 13.74 & 21.90 & E-Q & 0.96 & 29.84 & 28.70 & Extremely LSB ETG. Candidate UDG. Classified as ``possible" member \\
38 & Secondary & 12:35:47.22+12:47:40.8 & 2802 & - & 188.95 & 12.79 & 21.88 & E-Q & 1.02 & 29.55 & 28.94 & Extended, extremely LSB ETG. Candidate UDG. Classified as ``possible" member \\
39 & Primary & 12:37:38.91+14:18:45.6 & 2981 & - & 189.41 & 14.31 & 22.34 & E-Q & 1.11 & 31.20 & 29.88 & Extremely LSB ETG. Candidate UDG. Classified as ``possible" member \\
40 & Primary & 12:37:51.81+16:11:45.7 & 3004 & - & 189.47 & 16.20 & 18.41 & E-Q & 1.70 & 29.47 & 28.89 & Extended, extremely LSB ETG. Two compact LSB ETG, NGVS2992 and NGV3002, are located 97\arcsec and 102\arcsec, respectively, to the SW. Candidate UDG. Classified as ``possible" member \\
41 & Secondary & 12:41:03.65+09:40:54.5 & 3240 & - & 190.27 & 9.68 & 21.17 & E-Q & 1.11 & 29.27 & 28.72 & Extended, extremely LSB ETG. Candidate UDG. Classified as ``possible" member \\
42 & Secondary & 12:43:49.88+14:26:54.0 & 3404 & - & 190.96 & 14.45 & 19.12 & EI(sh:st)-I(f) & 1.33 & 27.94 & 27.75 & Extremely low-surface brightness EI galaxy with irregular structure and possible filaments or streams. Candidate UDG. Classified as ``possible" member \\
43 & Primary & 12:45:05.47+13:54:28.4 & 3477 & - & 191.27 & 13.91 & 18.15 & E-Q & 1.67 & 28.72 & 28.50 & Extended, extremely LSB ETG. Candidate UDG. Classified as ``possible" member \\
44 & Primary & 12:45:17.82+15:31:56.1 & 3492 & - & 191.32 & 15.53 & 18.31 & E-Q & 1.58 & 28.72 & 28.19 & Extended, extremely LSB ETG. Candidate UDG. Classified as ``possible" member \\
45 & Primary & 12:45:44.41+12:35:14.3 & 3516 & - & 191.44 & 12.59 & 18.78 & E-Q & 1.61 & 29.24 & 28.82 & Extended, extremely LSB ETG. Candidate UDG. Classified as ``possible" member \\
46 & Secondary & 12:47:04.44+09:48:45.4 & 3551 & - & 191.77 & 9.81 & 21.59 & E-Q & 1.00 & 29.55 & 28.57 & Extremely LSB ETG. Candidate UDG. Classified as ``possible" member \\
47 & Secondary & 12:49:19.72+11:52:54.2 & 3626 & - & 192.33 & 11.88 & 20.77 & E-Q & 1.24 & 28.98 & 28.96 & Extended, extremely LSB ETG. Candidate UDG. Classified as ``possible" member \\
48 & Secondary & 12:49:24.40+13:23:22.3 & 3628 & - & 192.35 & 13.39 & 21.31 & E-Q & 1.13 & 29.18 & 28.96 & Extremely LSB ETG. Candidate UDG. Classified as ``possible" member \\
49 & Primary & 12:51:08.90+11:28:36.4 & 3659 & - & 192.79 & 11.48 & 18.81 & E-Q & 1.48 & 28.88 & 28.20 & Extended, extremely LSB ETG. Located 118\arcsec to the W of the very LSB ETG NGVS3654. Candidate UDG. Classified as ``possible" member \\
50 & Primary & 12:51:12.85+13:05:45.1 & 3661 & - & 192.80 & 13.10 & 19.61 & E-Q & 1.33 & 28.96 & 28.27 & Extended, extremely LSB ETG. Candidate UDG. Classified as ``possible" member \\
\enddata
\end{deluxetable*}
\end{longrotatetable}

\clearpage



\startlongtable
\begin{deluxetable*}{%
  r
  p{3.5cm} @{\hspace{0.8em}}
  c c c c
  >{\centering\arraybackslash}p{1.7cm}
  >{\raggedright\arraybackslash}p{2.2cm}
  >{\raggedright\arraybackslash\hspace{0pt}\sloppy}p{6.0cm}
}
\tabletypesize{\scriptsize}
\tablecaption{Candidate UCD Transition Objects\label{tab:_types_ucdtrans}}
\tablehead{
  \colhead{\#} &
  \colhead{ID [NGVSJ]} &
  \colhead{NGVS} &
  \colhead{VCC} &
  \colhead{$\alpha$(2000)} &
  \colhead{$\delta$(2000)} &
  \colhead{$g^\prime$} &
  \colhead{Type} &
  \colhead{Notes} \\
 \colhead{} & \colhead{} & \colhead{} & \colhead{} &
 \colhead{(deg)} & \colhead{(deg)} & \colhead{(mag)} &
 \colhead{} & \colhead{}
}
\startdata
1 & 12:15:04.86+13:24:33.5 & 149 & - & 183.77 & 13.41 & 20.38 & E(n)-Q & Very compact, extremely LSB ETG and bright nucleus. Candidate UCD transition object. Classified as ``possible" member \\
2 & 12:15:19.21+07:58:56.2 & 159 & - & 183.83 & 7.98 & 20.34 & E(n)-Q & Compact, extremely LSB ETG with small envelope and bright, offset nucleus. Candidate UCD transition object. Classified as ``possible" member \\
3 & 12:17:55.31+07:33:24.3 & 308 & - & 184.48 & 7.56 & 20.63 & E(n:)-Q & Compact, very elongated, very LSB ETG with possible bright, offset nucleus. Candidate UCD transition object. Classified as ``probable" member \\
4 & 12:18:04.41+15:10:43.7 & 317 & - & 184.52 & 15.18 & 18.48 & E(n:)-Q & Extended, very LSB ETG with possible offset (double) nucleus. Candidate UCD transition object. Classified as ``certain" member \\
5 & 12:20:22.73+16:43:17.0 & 486 & - & 185.09 & 16.72 & 20.41 & E(n:)-Q & Extremely LSB ETG with possible bright nucleus. Candidate UCD transition object. Classified as ``possible" member \\
6 & 12:20:26.70+11:31:24.7 & 495 & - & 185.11 & 11.52 & 17.80 & E(n)-Q & Very LSB ETG with bright nucleus. Candidate UCD transition object. Classified as ``certain" member \\
7 & 12:20:38.25+07:18:43.1 & 514 & - & 185.16 & 7.31 & 20.34 & E(n)-Q & Compact, extremely LSB ETG with bright nucleus. Candidate UCD transition object. Classified as ``probable" member \\
8 & 12:21:51.50+07:01:26.6 & 630 & - & 185.46 & 7.02 & 20.23 & E(n)-Q & Compact, extremely LSB ETG with small envelope and bright nucleus. Candidate UCD transition object. Classified as ``probable" member \\
9 & 12:22:22.50+11:27:45.2 & 685 & - & 185.59 & 11.46 & 20.96 & E(n)-Q & Nucleated, very compact, extremely LSB ETG. Candidate UCD transition object. Classified as ``possible" member \\
10 & 12:22:53.78+12:23:20.2 & 763 & 595 & 185.72 & 12.39 & 19.12 & E(n)-Q & Compact, very LSB ETG with small envelope and bright nucleus. Candidate UCD transition object. Classified as ``probable" member. Cross ID: VCC595 \\
11 & 12:22:56.69+14:21:46.7 & 778 & - & 185.74 & 14.36 & 19.49 & E(n)-Q & Extremely LSB ETG with bright nucleus. Candidate UCD transition object. Classified as ``probable" member \\
12 & 12:23:01.30+13:59:19.1 & 787 & - & 185.76 & 13.99 & 21.98 & E(n:)-Q & Very compact, extremely LSB ETG with possible bright nucleus. Candidate UCD transition object. Classified as ``possible" member \\
13 & 12:23:04.54+12:20:43.8 & 795 & - & 185.77 & 12.35 & 20.57 & E(n)-Q & Nucleated, compact, extremely LSB ETG with faint, small envelope. Candidate UCD transition object. Classified as ``probable" member \\
14 & 12:23:04.59+12:00:53.4 & 796 & 615 & 185.77 & 12.01 & 17.25 & E(n:)-Q & Very LSB ETG with possible bright nucleus. Candidate UCD transition object. Classified as ``certain" member. Cross ID: VCC615 \\
15 & 12:23:11.59+13:25:06.4 & 812 & 624 & 185.80 & 13.42 & 18.65 & E(n)-Q & Compact LSB ETG with round isophotes and bright nucleus. Candidate UCD transition object. Classified as ``certain" member. Cross ID: VCC624 \\
16 & 12:23:22.80+13:00:07.8 & 835 & - & 185.84 & 13.00 & 18.95 & E(n)-Q & Compact, very LSB ETG with small low-surface brightness envelope and bright nucleus. Candidate UCD transition object. Classified as ``certain" member \\
17 & 12:23:39.21+07:47:43.9 & 876 & - & 185.91 & 7.80 & 19.98 & E(n:)-Q & Compact, extremely LSB ETG with possible bright nucleus. Candidate UCD transition object. Classified as ``probable" member \\
18 & 12:23:52.90+12:46:23.1 & 906 & 678 & 185.97 & 12.77 & 17.58 & E(n)-Q & Very  LSB ETG with round isophotes and bright nucleus. Candidate UCD transition object. Classified as ``certain" member. Cross ID: VCC678 \\
19 & 12:24:01.84+13:51:56.4 & 935 & 674 & 186.01 & 13.87 & 18.39 & E(n:)-Q & Extended, extremely  LSB ETG with possible bright nucleus. Several faint galaxies (NGVS878, NGVS952, NGVS971 and NGVS978) are located within 153\arcsec and 327\arcsec, while a brighter galaxy (NGVS904 == VCC674) is located 147\arcsec to the N. Candidate UCD transition object. Candidate UDG. Classified as ``probable" member \\
20 & 12:24:42.06+13:31:00.6 & 1017 & - & 186.18 & 13.52 & 17.60 & E(n:)-Q & Extended, extremely LSB ETG with possible bright nucleus. Candidate UCD transition object. Candidate UDG. Classified as ``probable" member \\
21 & 12:24:48.19+07:32:11.3 & 1032 & - & 186.20 & 7.54 & 20.47 & E(n)-Q & Nucleated, very compact, elongated, extremely LSB ETG with small envelope. Candidate UCD transition object. Classified as ``possible" member \\
22 & 12:24:58.09+09:37:17.9 & 1048 & - & 186.24 & 9.62 & 20.42 & E(n:)-Q & Extremely LSB ETG. Possible double nucleus. Candidate UCD transition object. Classified as ``possible" member \\
23 & 12:25:09.21+12:09:18.5 & 1075 & - & 186.29 & 12.16 & 21.65 & E(n:\vbar{}st:)-Q & Compact, extremely LSB ETG with possible nucleus and faint streams. Candidate UCD transition object. Classified as ``possible" member \\
24 & 12:25:23.69+12:40:30.3 & 1119 & 800 & 186.35 & 12.68 & 19.42 & E(n)-Q & Compact, very LSB ETG with small envelope and bright nucleus. Located 43\arcsec to the E of NGVS1128 (VCC800), although the velocity of either galaxy is unknown. Additionally, both galaxies are located \textasciitilde{}300\arcsec to the W of VCC836 and might be part of a small group containing, besides VCC836, NGVS1133, VCC871, and VCC838. Candidate UCD transition object. Classified as ``certain" member. Cross ID: VCC796 \\
25 & 12:25:35.64+12:10:01.8 & 1153 & - & 186.40 & 12.17 & 20.05 & E(n:)-Q & Extremely LSB ETG with very low-surface brightness envelope and possible bright. offset nucleus. Candidate UCD transition object. Classified as ``probable" member \\
26 & 12:25:36.62+12:50:58.9 & 1157 & 814 & 186.40 & 12.85 & 18.24 & E(n)-Q & Very LSB ETG with bright nucleus. Located 499\arcsec to the SE of M84. Candidate UCD transition object. Classified as ``certain" member. Cross ID: VCC814 \\
27 & 12:25:55.70+11:48:03.8 & 1212 & 853 & 186.48 & 11.80 & 18.92 & E(n)-Q & Compact, very LSB ETG with faint envelope and bright nucleus. Candidate UCD transition object. Classified as ``certain" member. Cross ID: VCC853 \\
28 & 12:25:55.79+10:43:13.2 & 1214 & 852 & 186.48 & 10.72 & 18.50 & E(n:)-Q & Extended, extremely LSB ETG with possible offset nucleus. Candidate UCD transition object.  Classified as ``certain" member. Cross ID: VCC852 \\
29 & 12:26:05.19+12:03:01.3 & 1239 & - & 186.52 & 12.05 & 19.75 & E(n)-Q & Compact, very LSB ETG with small, faint envelope and bright nucleus. Candidate UCD transition object. Classified as ``probable" member \\
30 & 12:26:25.15+11:18:36.9 & 1298 & 915 & 186.60 & 11.31 & 21.01 & E(n:)-Q & Very compact, extremely LSB ETG with possible bright nucleus. Located 158\arcsec to the NW of the brighter ETG NGVS1331 (VCC915), although neither galaxy has a measured velocity. Candidate UCD transition object. Classified as ``possible" member \\
31 & 12:26:36.32+12:48:10.0 & 1342 & 923 & 186.65 & 12.80 & 17.89 & E(n)-Q & Nucleated, very LSB ETG with round isophotes. The galaxy is located 629\arcsec to the SE of M86 (v=-233 km/s) and has similar velocity (168 km/s), suggesting it might be part of the M86 group. Candidate UCD transition object. Classified as ``certain" member. Cross ID: VCC923 \\
32 & 12:26:38.25+13:04:44.2 & 1353 & 927 & 186.66 & 13.08 & 18.19 & E(n)-Q & Extended, extremely LSB ETG with bright nucleus. Candidate UCD transition object. Candidate UDG. Classified as ``probable" member. Cross ID: VCC927 \\
33 & 12:26:42.50+09:53:54.8 & 1368 & - & 186.68 & 9.90 & 20.22 & E(n:)-Q & Compact, extremely LSB ETG with compact envelope and possible bright nucleus. Candidate UCD transition object. Classified as ``possible" member \\
34 & 12:26:43.31+12:17:44.0 & 1370 & - & 186.68 & 12.30 & 18.39 & E(n)-Q & Very LSB ETG with bright nucleus. Candidate UCD transition object. Classified as ``probable" member \\
35 & 12:26:51.99+12:39:08.2 & 1407 & - & 186.72 & 12.65 & 20.54 & E(n)-Q & Compact, extremely LSB ETG with bright, possibly offset nucleus. Candidate UCD transition object. Classified as ``probable" member \\
36 & 12:27:03.75+14:48:54.0 & 1439 & - & 186.77 & 14.81 & 20.73 & E(n:)-Q & Extremely LSB ETG with possible bright nucleus. Candidate UCD transition object. Classified as ``probable" member \\
37 & 12:27:16.78+12:32:07.8 & 1482 & - & 186.82 & 12.54 & 21.27 & E(n)-Q & Very compact, very LSB ETG with small low-surface brightness envelope and bright nucleus. Candidate UCD transition object. Classified as ``possible" member \\
38 & 12:27:17.61+06:46:05.7 & 1484 & 988 & 186.82 & 6.77 & 18.77 & E(n)-Q & Extremely LSB ETG with small envelope and bright nucleus. Candidate UCD transition object. Classified as ``certain" member. Cross ID: VCC988 \\
39 & 12:27:19.26+13:51:01.9 & 1488 & - & 186.83 & 13.85 & 21.07 & E(n)-Q & Very compact, extremely LSB ETG with very small envelope and bright nucleus. Candidate UCD transition object. Classified as ``probable" member \\
40 & 12:27:44.67+07:56:43.0 & 1568 & - & 186.94 & 7.95 & 19.03 & E(n)-Q & Extremely LSB ETG with bright nucleus. Candidate UCD transition object. Classified as ``possible" member \\
41 & 12:28:03.74+12:46:41.2 & 1612 & - & 187.02 & 12.78 & 21.04 & E(n)-Q & Very compact, extremely LSB ETG with small envelope and bright nucleus. Candidate UCD transition object. Classified as ``possible" member \\
42 & 12:28:06.26+08:03:43.7 & 1620 & 1067 & 187.03 & 8.06 & 18.14 & E(n:)-Q & Extended, extremely LSB ETG with possible extended blue nucleus and boxy structure. Candidate UCD transition object.  Classified as ``certain" member. Cross ID: VCC1067 \\
43 & 12:28:06.77+12:58:43.2 & 1622 & 1070 & 187.03 & 12.98 & 18.99 & E(n)-Q & Very LSB ETG with bright nucleus. Candidate UCD transition object. Classified as ``certain" member. Cross ID: VCC1070 \\
44 & 12:28:14.47+08:59:51.3 & 1649 & - & 187.06 & 9.00 & 18.12 & E(n)-Q & Round, very LSB ETG with small envelope and bright nucleus. Candidate UCD transition object. Classified as ``certain" member \\
45 & 12:28:15.85+12:52:12.0 & 1653 & - & 187.07 & 12.87 & 19.09 & E(n)-Q & Extremely LSB ETG with bright nucleus. Candidate UCD transition object. Classified as ``probable" member \\
46 & 12:28:18.72+10:59:33.2 & 1658 & - & 187.08 & 10.99 & 19.97 & E(n:)-Q & Extremely LSB ETG with possible extended and bright nucleus. Candidate UCD transition object. Classified as ``possible" member \\
47 & 12:28:25.96+09:53:13.8 & 1686 & - & 187.11 & 9.89 & 18.65 & E(n:)-Q & Very LSB ETG with possible offset, extremely bright blue nucleus. Candidate UCD transition object. Classified as ``certain" member \\
48 & 12:28:34.01+07:34:16.7 & 1712 & 1113 & 187.14 & 7.57 & 18.77 & E(n)-Q & Very LSB ETG with bright nucleus. Candidate UCD transition object. Classified as ``certain" member. Cross ID: VCC1113 \\
49 & 12:28:37.53+07:54:05.1 & 1719 & - & 187.16 & 7.90 & 19.04 & E(n)-Q & Nucleated, elongated, extremely LSB ETG with very low-surface brightness and compact envelope. Candidate UCD transition object. Classified as ``probable" member \\
50 & 12:28:46.92+12:38:31.5 & 1747 & - & 187.20 & 12.64 & 19.94 & E(n)-Q & Compact, extremely LSB ETG with bright nucleus. Candidate UCD transition object. Classified as ``certain" member \\
51 & 12:28:59.15+12:02:30.4 & 1784 & - & 187.25 & 12.04 & 20.41 & E(n)-Q & Nucleated, compact, extremely LSB ETG with bright nucleus. Candidate UCD transition object. Classified as ``possible" member \\
52 & 12:28:59.50+11:55:23.4 & 1786 & - & 187.25 & 11.92 & 19.47 & E(n:)-Q & Round, extremely LSB ETG with possible bright nucleus. Candidate UCD transition object. Classified as ``probable" member \\
53 & 12:29:05.13+12:09:13.6 & 1801 & 1162 & 187.27 & 12.15 & 19.38 & E(n)-Q & Very LSB ETG with bright nucleus. Candidate UCD transition object.  Classified as ``certain" member. Cross ID: VCC1162 \\
54 & 12:29:10.21+07:56:27.7 & 1813 & - & 187.29 & 7.94 & 20.57 & E(n)-Q & Extremely LSB ETG with small envelope and bright nucleus. Projected in the Virgo B subcluster. Candidate UCD transition object. Classified as ``probable" member \\
55 & 12:29:17.39+15:44:56.7 & 1831 & - & 187.32 & 15.75 & 19.84 & E(n:)-Q & Compact, extremely LSB ETG with possible nucleus and round isophotes. Candidate UCD transition object. Classified as ``probable" member \\
56 & 12:29:23.56+14:53:48.9 & 1849 & - & 187.35 & 14.90 & 20.83 & E(n)-Q & Very compact, very elongated, very LSB ETG with elongated, low-surface brightness envelope and bright blue nucleus. Candidate UCD transition object. Classified as ``probable" member \\
57 & 12:29:38.10+13:05:18.2 & 1889 & - & 187.41 & 13.09 & 20.88 & E(n)-Q & Compact, extremely LSB ETG with bright nucleus. Candidate UCD transition object. Classified as ``possible" member \\
58 & 12:29:38.67+10:58:51.0 & 1894 & - & 187.41 & 10.98 & 21.10 & E(n)-Q & Very compact, extremely LSB ETG with small envelope and bright nucleus. Candidate UCD transition object. Classified as ``possible" member \\
59 & 12:29:41.39+11:34:19.3 & 1911 & - & 187.42 & 11.57 & 19.72 & E(n:)-Q & Extremely LSB ETG with possible bright nucleus. Candidate UCD transition object. Classified as ``probable" member \\
60 & 12:29:52.34+06:03:32.1 & 1943 & 1234 & 187.47 & 6.06 & 19.41 & E(n:)-Q & Very LSB ETG with possible bright nucleus. Candidate UCD transition object. Classified as ``certain" member. Cross ID: VCC1234 \\
61 & 12:29:56.34+13:13:12.3 & 1956 & 1244 & 187.48 & 13.22 & 17.70 & E(n)-Q & Compact, very LSB ETG and bright nucleus. Candidate UCD transition object. Classified as ``certain" member. Cross ID: VCC1244 \\
62 & 12:29:57.10+11:24:03.1 & 1957 & 1217 & 187.49 & 11.40 & 19.47 & E(n:)-Q & Extremely LSB ETG with possible bright nucleus. Although its velocity is unknown, the galaxy is located 199\arcsec to the W of the bright LTG NGVS1916 (VCC1217). Candidate UCD transition object. Classified as ``certain" member \\
63 & 12:30:12.55+09:42:56.3 & 2001 & - & 187.55 & 9.72 & 18.70 & E(n:)-Q & Extremely LSB ETG with possible nucleus. Apparent member of an LSB pair, with the second member, NGVS1993, located 57\arcsec to the W. Neither galaxy has a measured velocity. A third, possibly connected LSB ETG, NGVS2031, is located 124\arcsec to the E. Candidate UCD transition object. Classified as ``certain" member \\
64 & 12:30:18.15+14:40:52.7 & 2024 & 1274 & 187.58 & 14.68 & 19.40 & E(n)-Q & Nucleated, very elongated, very LSB ETG. A fainter LSB ETG, NGVS2038, is located 63\arcsec to the SE, while the brighter LSB ETG NGVS2016 (VCC1274) is located 83\arcsec to the N. None of the galaxies has a measured velocity. Candidate UCD transition object. Classified as ``certain" member. Cross ID: VCC1276 \\
65 & 12:30:31.24+06:23:19.2 & 2063 & - & 187.63 & 6.39 & 19.68 & E(n)-Q & Extremely LSB ETG with bright nucleus. Candidate UCD transition object. Classified as ``certain" member \\
66 & 12:30:33.15+08:07:47.6 & 2069 & 1226 & 187.64 & 8.13 & 19.94 & E(n:)-Q & Compact, extremely LSB ETG with possible bright offset nucleus.  The galaxy, which has velocity v=830.11 km/s, is located 831\arcsec to the NE of M49 (VCC1226, v=989 km/s). Candidate UCD transition object. Classified as ``probable" member \\
67 & 12:30:36.82+06:34:49.0 & 2075 & - & 187.65 & 6.58 & 19.36 & E(n:)-Q & Extremely LSB ETG with possible bright nucleus. Although its velocity is unknown, the galaxy is located in the Virgo core,  close (324\arcsec to the NW) of M87. Candidate UCD transition object. Classified as ``probable" member \\
68 & 12:30:45.94+09:00:44.9 & 2101 & 1303 & 187.69 & 9.01 & 18.29 & E(n)-Q & Compact, very LSB ETG with bright nucleus. A bright ES galaxy (NGVS2087, VCC1303) is located very close, 79\arcsec to the W. Candidate UCD transition object. Classified as ``certain" member. Cross ID: VCC1306 \\
69 & 12:30:53.59+06:36:32.7 & 2128 & 1320 & 187.72 & 6.61 & 18.62 & E(n)-Q & Compact, very LSB ETG with small envelope and bright nucleus. Candidate UCD transition object. Classified as ``certain" member. Cross ID: VCC1320 \\
70 & 12:31:08.79+14:36:04.0 & 2162 & - & 187.79 & 14.60 & 20.34 & E(n:)-Q & Extremely LSB ETG with possible bright nucleus. Candidate UCD transition object. Classified as ``probable" member \\
71 & 12:31:19.44+07:14:29.4 & 2191 & - & 187.83 & 7.24 & 20.36 & E(n)-Q & Compact, extremely LSB ETG with bright nucleus. Candidate UCD transition object. Classified as ``certain" member \\
72 & 12:31:25.90+08:20:02.4 & 2204 & 1360 & 187.86 & 8.33 & 18.40 & E(n)-Q & Very LSB ETG with bright nucleus and nearly round isophotes. Candidate UCD transition object. Classified as ``certain" member. Cross ID: VCC1360 \\
73 & 12:31:37.17+05:23:02.5 & 2244 & - & 187.90 & 5.38 & 20.07 & E(n:)-Q & Elongated, extremely LSB ETG with possible bright nucleus. Candidate UCD transition object. Classified as ``probable" member \\
74 & 12:31:38.75+11:49:44.7 & 2249 & - & 187.91 & 11.83 & 21.14 & E(n)-Q & Very compact, extremely LSB ETG with small envelope and bright nucleus. Candidate UCD transition object. Classified as ``possible" member \\
75 & 12:31:39.76+13:38:53.7 & 2253 & - & 187.92 & 13.65 & 19.30 & E(n)-Q & Compact, very LSB ETG with small, very diffuse envelope and bright nucleus. Candidate UCD transition object. Classified as ``probable" member \\
76 & 12:31:44.79+11:13:59.1 & 2259 & - & 187.94 & 11.23 & 19.37 & E(n)-Q & Compact, very LSB ETG with bright nucleus. Candidate UCD transition object. Classified as ``certain" member \\
77 & 12:31:47.77+06:55:56.9 & 2266 & - & 187.95 & 6.93 & 19.11 & E(n)-Q & Very compact, extremely LSB ETG with bright nucleus. Candidate UCD transition object. Classified as ``possible" member \\
78 & 12:31:56.03+08:42:02.2 & 2291 & - & 187.98 & 8.70 & 21.73 & E(n:)-Q & Compact, extremely LSB ETG with small envelope and possible bright nucleus. Candidate UCD transition object. Classified as ``possible" member \\
79 & 12:32:12.37+13:35:21.6 & 2341 & - & 188.05 & 13.59 & 19.09 & E(n)-Q & Extremely LSB ETG with bright nucleus. Candidate UCD transition object. Classified as ``probable" member \\
80 & 12:32:12.92+09:18:56.7 & 2343 & 1421 & 188.05 & 9.32 & 17.16 & E(n:)-Q & Extremely LSB ETG with possible bright and blue nucleus. Candidate UCD transition object. Classified as ``certain" member. Cross ID: VCC1421 \\
81 & 12:32:54.13+07:05:53.5 & 2442 & - & 188.23 & 7.10 & 20.57 & E(n:)-Q & Compact, extremely  LSB ETG with possible bright nucleus. A second LSB ETG, NGVS2436, is located 54\arcsec to the NW, although neither galaxy has a measured velocity. Candidate UCD transition object. Classified as ``certain" member \\
82 & 12:33:06.41+13:18:11.1 & 2472 & 1477 & 188.28 & 13.30 & 18.22 & E(n)-Q & Round, very LSB ETG with bright nucleus. Candidate UCD transition object. Classified as ``certain" member. Cross ID: VCC1477 \\
83 & 12:33:15.05+11:13:49.6 & 2499 & - & 188.31 & 11.23 & 18.39 & E(n)-Q & Extremely LSB ETG with bright blue nucleus. Located 86\arcsec to the NW of the LSB ETG NGVS2491, although the velocity of either galaxy is unknown. Candidate UCD transition object. Classified as ``certain" member \\
84 & 12:33:21.31+10:22:25.2 & 2514 & - & 188.34 & 10.37 & 19.40 & E(n)-Q & Compact, extremely LSB ETG with bright nucleus. Candidate UCD transition object. Classified as ``probable" member \\
85 & 12:34:07.83+11:45:48.1 & 2606 & - & 188.53 & 11.76 & 20.60 & E(n)-Q & Very compact, extremely LSB ETG with bright nucleus. Candidate UCD transition object. Classified as ``probable" member \\
86 & 12:34:23.11+12:20:21.3 & 2641 & - & 188.60 & 12.34 & 20.59 & E(n:)-Q & Extremely LSB ETG with possible very bright nucleus. Candidate UCD transition object. Classified as ``possible" member \\
87 & 12:34:25.91+12:45:38.2 & 2647 & - & 188.61 & 12.76 & 20.89 & E(n:)-Q & Extremely LSB ETG with possible bright nucleus. Candidate UCD transition object. Classified as ``possible" member \\
88 & 12:34:44.66+13:43:09.5 & 2681 & - & 188.69 & 13.72 & 20.31 & E(n)-Q & Compact, extremely LSB ETG with extremely diffuse, elongated envelope and bright nucleus. Candidate UCD transition object. Classified as ``possible" member \\
89 & 12:34:52.53+08:40:42.1 & 2695 & - & 188.72 & 8.68 & 21.42 & E(n)-Q & Very compact, extremely LSB ETG with small low-surface brightness envelope and bright nucleus. Candidate UCD transition object. Classified as ``probable" member \\
90 & 12:35:37.87+08:05:55.1 & 2785 & - & 188.91 & 8.10 & 21.13 & E(n)-Q & Compact, extremely elongated, extremely LSB ETG with bright nucleus. Candidate UCD transition object. Classified as ``probable" member \\
91 & 12:36:32.59+12:31:05.8 & 2878 & 1672 & 189.14 & 12.52 & 18.11 & E(n)-Q & Compact, extremely LSB ETG with very low-surface brightness isophotes envelope and bright nucleus. Candidate UCD transition object. Classified as ``probable" member. Cross ID: VCC1672 \\
92 & 12:36:59.28+07:52:22.5 & 2918 & - & 189.25 & 7.87 & 18.25 & E(n:)-Q & Very LSB ETG with diffuse, compact envelope and possible nucleus. Candidate UCD transition object. Classified as ``certain" member \\
93 & 12:37:25.09+12:35:11.0 & 2960 & - & 189.35 & 12.59 & 20.47 & E(n:)-Q & Very compact, extremely LSB ETG with very small envelope and possible bright nucleus. Candidate UCD transition object. Classified as ``possible" member \\
94 & 12:37:36.26+09:12:24.8 & 2978 & 1723 & 189.40 & 9.21 & 18.75 & E(n)-Q & Compact, very LSB ETG with bright nucleus. Candidate UCD transition object. Classified as ``certain" member. Cross ID: VCC1723 \\
95 & 12:37:49.73+07:49:23.1 & 2999 & - & 189.46 & 7.82 & 19.50 & E(n:)-Q & Extremely LSB ETG with possible offset bright nucleus. Candidate UCD transition object. Classified as ``probable" member \\
96 & 12:38:12.15+10:42:06.6 & 3038 & 1749 & 189.55 & 10.70 & 18.37 & E(n)-Q & Compact, very LSB ETG with small envelope and bright nucleus. Candidate UCD transition object. Classified as ``certain" member. Cross ID: VCC1749 \\
97 & 12:39:12.23+08:09:20.6 & 3099 & - & 189.80 & 8.16 & 20.32 & E(n:)-Q & Compact, very elongated, extremely LSB ETG with possible bright nucleus. Candidate UCD transition object. Classified as ``certain" member \\
98 & 12:39:20.34+11:26:20.0 & 3112 & 1798 & 189.83 & 11.44 & 20.16 & E(n)-Q & Extremely LSB ETG with faint envelope and bright nucleus. Candidate UCD transition object. Located 177\arcsec to the SW of the extended, extremely LSB ETG BGVS3128 (VCC1798). Classified as ``probable" member \\
99 & 12:39:21.99+12:05:34.6 & 3116 & - & 189.84 & 12.09 & 20.76 & E(n:)-Q & Extremely LSB ETG with possible bright nucleus. Candidate UCD transition object. Candidate UDG.  Classified as ``probable" member \\
100 & 12:39:24.84+10:29:56.2 & 3117 & - & 189.85 & 10.50 & 19.49 & E(n:)-Q & Extremely LSB ETG with possible bright nucleus. One member of an apparent pair of candidate UCD transition objects, the second one being NGVS3119, located 20\arcsec to the NE, although neither galaxy has a measured velocity. Classified as ``possible" member \\
101 & 12:39:25.72+10:30:12.3 & 3119 & - & 189.86 & 10.50 & 21.00 & E(n:)-Q & Very compact, extremely LSB ETG with small low-surface brightness envelope and possible bright nucleus. Candidate UCD transition object. Member of an apparent pair of candidate UCD transition objects, the second one being NGVS3117, located 20\arcsec to the SW, although neither galaxy has a measured velocity.  Classified as ``possible" member \\
102 & 12:39:48.45+09:49:46.6 & 3149 & - & 189.95 & 9.83 & 19.14 & E(n:)-Q & Compact, unusual, very LSB ETG with small envelope and possible bright, possibly resolved, nucleus. Candidate UCD transition object. Classified as ``certain" member \\
103 & 12:40:01.63+11:52:44.7 & 3164 & 1815 & 190.01 & 11.88 & 19.20 & E(n)-Q & Compact, very LSB ETG with bright nucleus (and possible offset dual nucleus). NGVS3156 (VCC1815) is located 119\arcsec to the NW. Candidate UCD transition object. Classified as ``certain" member \\
104 & 12:40:01.95+09:57:25.1 & 3167 & 1826 & 190.01 & 9.96 & 19.06 & E(n)-Q & Very LSB ETG with round isophotes and bright nucleus. Located 30\arcsec to the S of the fainter LSB ETG NGVS3165, whose velocity is unknown. May be associated with VCC1826, which is located 4.3' to the SE and has similar velocity (1974 km/s for VCC1826 and 1845 km/s for VCC1817). Candidate UCD transition object. Classified as ``certain" member. Cross ID: VCC1817 \\
105 & 12:41:32.64+15:55:54.2 & 3267 & - & 190.39 & 15.93 & 20.62 & E(n:)-Q & Extremely LSB ETG with small envelope and possible bright nucleus. Candidate UCD transition object. Classified as ``probable" member \\
106 & 12:42:22.80+11:41:06.2 & 3317 & 1903 & 190.60 & 11.69 & 19.74 & E(n)-Q & Nucleated, compact, very LSB ETG. Candidate UCD transition object. Located 331\arcsec to the NE of NGVS3296 (VCC1903, M59). Classified as ``certain" member \\
107 & 12:43:20.88+14:02:02.9 & 3379 & - & 190.84 & 14.03 & 18.77 & E(n:)-Q & Extended, extremely  LSB ETG with possible bright blue nucleus. Candidate UCD transition object. Classified as ``probable" member \\
108 & 12:43:57.77+11:52:50.5 & 3413 & 1986 & 190.99 & 11.88 & 17.94 & E(n)-Q & Compact, very LSB ETG with bright nucleus. Possible tidal streams. Candidate UCD transition object. Classified as ``certain" member. Cross ID: VCC1986 \\
109 & 12:44:23.92+08:47:29.1 & 3434 & - & 191.10 & 8.79 & 18.88 & E(n:)-Q & Very LSB ETG with possible offset blue nucleus. Candidate UCD transition object. Classified as ``certain" member \\
110 & 12:44:45.00+11:48:03.5 & 3455 & - & 191.19 & 11.80 & 18.94 & E(n)-Q & Extremely LSB ETG with round isophotes and bright nucleus. Candidate UCD transition object. Classified as ``certain" member \\
111 & 12:45:54.08+10:39:29.5 & 3522 & - & 191.48 & 10.66 & 19.44 & E(n)-Q & Compact, very LSB ETG with small envelope and bright nucleus. Candidate UCD transition object. Classified as ``probable" member \\
112 & 12:46:31.37+13:19:33.0 & 3538 & - & 191.63 & 13.33 & 19.07 & E(n)-Q & Extremely LSB ETG with bright nucleus. Candidate UCD transition object. Classified as ``probable" member \\
113 & 12:47:47.50+12:41:42.5 & 3582 & - & 191.95 & 12.70 & 19.60 & E(n)-Q & Extremely LSB ETG with bright nucleus. Candidate UCD transition object. Classified as ``probable" member \\
114 & 12:47:49.44+15:43:28.9 & 3584 & - & 191.96 & 15.72 & 20.53 & E(n:)-Q & Compact, extremely LSB ETG with small envelope and possible nucleus. Candidate UCD transition object. Classified as ``probable" member \\
115 & 12:49:00.69+09:00:42.2 & 3619 & - & 192.25 & 9.01 & 19.47 & E(n:)-Q & Compact LSB ETG with small, round envelope and possible extended nucleus. Candidate UCD transition object. Classified as ``probable" member \\
\enddata
\end{deluxetable*}


\clearpage
\onecolumngrid

\startlongtable
\begingroup
\setlength{\tabcolsep}{2pt}
\renewcommand{\arraystretch}{1.12}
\setlength{\emergencystretch}{2em}
\sloppy

\begin{deluxetable}{%
  r                                   
  p{0.16\textwidth}                   
  c                                   
  c                                   
  r                                   
  r                                   
  >{\centering\arraybackslash}p{0.08\textwidth}                     
  >{\raggedright\arraybackslash\hspace{0pt}\sloppy}p{0.15\textwidth}
  >{\raggedright\arraybackslash\hspace{0pt}\sloppy}p{0.36\textwidth}
}
\tabletypesize{\scriptsize}
\tablewidth{\textwidth}
\tablecolumns{9}
\tablecaption{Candidate Merger and Post-Merger Galaxies \label{tab:_types_mergers}}
\tablehead{
 \colhead{\#} & \colhead{ID [NGVSJ]} &  \colhead{NGVS} &  \colhead{VCC} &
 \colhead{$\alpha$(2000)} & \colhead{$\delta$(2000)} &
 \colhead{$g^\prime$} & \colhead{Type}  & \colhead{Notes} \\
 \colhead{} & \colhead{} & \colhead{} & \colhead{} &
 \colhead{(deg)} & \colhead{(deg)} & \colhead{(mag)} & \colhead{} & \colhead{}
}
\startdata
1 & 12:08:20.90+12:30:07.8 & 7 & - & 182.09 & 12.50 & 18.01 & EI(sp:\vbar{}du:\vbar{}b:\vbar{}sh:\vbar{}pm:)-I(c\vbar{}p\vbar{}g) & Possible EI galaxy with irregular structure and star formation. Post merger system?. Classified as ``certain" member \\
2 & 12:09:50.60+12:54:09.8 & 16 & - & 182.46 & 12.90 & 17.42 & E(n:\vbar{}pm:)-I(c:) & Unusual elongated LSB ETG, very likely the result of on-going merger event between two LSB ETG galaxies, each with a bright blue nucleus. A high-surface brightness blue filament connects the two nuclei. Possible post-merger remnant. Classified as ``certain" member \\
3 & 12:10:03.30+11:42:48.3 & 19 & - & 182.51 & 11.71 & 16.86 & ES(pm:)-I(c) & ES galaxy with SF in patches and filaments embedded in elliptical isophotes. Possible post-merger system. Classified as ``certain" member \\
4 & 12:11:45.72+13:01:27.4 & 49 & 37 & 182.94 & 13.02 & 17.74 & ES(n\vbar{}pm:)-Q & Nucleated ES galaxy with apparent disk. Possible flared or bow-tie morphology. Possible post merger remnant. Classified as ``certain" member. Cross ID: VCC37 \\
5 & 12:13:35.81+13:02:04.4 & 94 & 85 & 183.40 & 13.03 & 17.37 & EI(sh\vbar{}pm)-I(g\vbar{}e) & Structurally complex galaxy with active star formation. Galaxy contains a series of loops and shells. Peak of star formation offset from centroid of light. Post merger remnant. Classified as ``certain" member. Cross ID: VCC85 \\
6 & 12:15:35.69+13:37:11.0 & 171 & 155 & 183.90 & 13.62 & 15.04 & ES(di:\vbar{}b:\vbar{}sp:\vbar{}du\vbar{}pm:)-I(c\vbar{}d) & Complex ES galaxy with old stellar  population, but streams and possible intermediate-age bar, disk or lens. Possible post-merger remnant. Classified as ``certain" member. Cross ID: VCC155 \\
7 & 12:16:23.01+07:47:54.5 & 220 & 190 & 184.10 & 7.80 & 17.14 & EI(pm:)-I(g\vbar{}e) & Extended low-surface brightness EI galaxy with distributed star formation LTG with peculiar structure suggestive of a post-merger remnant. Irregular appearance with low level star formation and several compact clumps are found scattered near the centre. Classified as ``certain" member. Cross ID: VCC190 \\
8 & 12:17:09.91+07:11:29.3 & 268 & 222 & 184.29 & 7.19 & 11.99 & ES(n\vbar{}di\vbar{}sp:\vbar{}sh\vbar{}st\vbar{}du\vbar{}pm:)-I(g) & Nucleated ES galaxy seen nearly edge-on. Low level star formation in disk, with dust. Bow-shaped streams at both ends of the disk. Possible post merger remnant. Classified as ``certain" member. Cross ID: VCC222 \\
9 & 12:17:14.54+08:56:32.4 & 272 & 227 & 184.31 & 8.94 & 14.64 & ES(n\vbar{}sh\vbar{}st\vbar{}b\vbar{}pm:)-I(g) & ES galaxy with complex structure comprising a series of shells and streams. Possible post-merger remnant. Bright compact nucleus. Possible inner bar or disk. Classified as ``certain" member. Cross ID: VCC227 \\
10 & 12:18:15.35+13:44:56.8 & 330 & 281 & 184.56 & 13.754 & 15.39 & ES(sh\vbar{}du\vbar{}sp:\vbar{}pm:)-I(c) & ES galaxy with smoothly distributed older component and star formation and dust in the core. Dust may be distributed in an arc. Apparent shell to the NW. Possible post-merger remnant. Classified as ``certain" member. Cross ID: VCC281 \\
11 & 12:18:57.69+08:38:41.4 & 363 & - & 184.74 & 8.64 & 18.41 & E(sh:\vbar{}pm:)-Q & LSB ETG with irregular structure. Possible shells. Possible post-merger product. Classified as ``certain" member \\
12 & 12:19:05.14+13:58:51.9 & 371 & 322 & 184.77 & 13.98 & 15.37 & EI(sh:\vbar{}st:\vbar{}pm)-A(g) & Complex galaxy with shells, ripples and some central star formation. Possibly interacting with VCC319 (v=421 km/s), located 0.7' to the W, and sharing a common envelope. Also possibly associated with  VCC334 (v=-246 km/s), located 6.3' to the SE. VCC322 velocity is -195 km/s. Likely post-merger system. Classified as ``certain" member. Cross ID: VCC322 \\
13 & 12:19:15.56+06:17:38.0 & 388 & 331 & 184.81 & 6.29 & 15.43 & ES(n\vbar{}sp\vbar{}sh\vbar{}pm:)-I(c) & ES galaxy with offset nucleus extended star forming core or disk superimposed on older population with ripples, arms or shells. Possible post-merger remnant. Located 48\arcsec to the S of the similarly bright VCC332, although the two galaxies have significantly different velocities (2257 and 660 km/s).  A third galaxy, NGVS393, with unknown velocity, is located 110\arcsec to the NE. Classified as ``certain" member. Cross ID: VCC331 \\
14 & 12:19:16.39+15:23:46.3 & 389 & 335 & 184.82 & 15.40 & 17.32 & EI(sh\vbar{}pm:)-I(c) & Complex EI galaxy with a series of shells and lobes. Possible post-merger product. Cross ID: VCC360 is located 6' to the NE. Classified as ``certain" member. Cross ID: VCC335 \\
15 & 12:19:22.13+05:54:38.1 & 398 & 340 & 184.84 & 5.91 & 14.65 & E(sh\vbar{}st\vbar{}du\vbar{}pm:)-A(c) & Complex ETG or possibly EI galaxy with possible nucleus, blue and dusty inner regions and a shell to the S. Might be part of a small group that includes  VCC340 (v=1506 km/s), VCC329 (v=1618 km/s), located 5.0' to the NW, VCC315 (v=1594 km/s), located 12.2' to the NW, VCC341 (v=1838 km/s), located 11.2' to the N, VCC379 (v=1453 km/s), located 9.0' to the NE. Located close to a prominent stellar stream that emanates from VCC336. Possible tidal or post-merger features. Classified as ``certain" member. Presence of nucleus is impossible to assess due to the presence of dust or other contaminants. Cross ID: VCC340 \\
16 & 12:20:43.91+14:37:52.0 & 519 & 429 & 185.18 & 14.63 & 16.64 & EI(sh\vbar{}pm:)-I(c\vbar{}g\vbar{}p) & EI galaxy with active star formation in its core and "lobes" or clumps along its major axis. Shells in its inner regions. Possible post-merger remnant. Classified as ``certain" member. Cross ID: VCC429 \\
17 & 12:20:58.89+08:51:55.4 & 537 & 447 & 185.25 & 8.87 & 19.03 & E(pm:)-Q & Elongated LSB ETG with unusual peanut-like morphology. Possible post merger remnant. Classified as ``probable" member. Cross ID: VCC447 \\
18 & 12:21:03.68+07:04:39.4 & 543 & 450 & 185.27 & 7.08 & 15.38 & ES(di\vbar{}sh\vbar{}st:\vbar{}du\vbar{}pm:)-I(c) & ES galaxy with dust and very weak star formation offset from center. Series of outer faint shells or streams, suggestive of past merger or accretion. The galaxy is likely associated (and may be interacting) with VCC439, which is located 3.3' to the SW and has almost identical velocity (2326 km/s for VCC439 and 2358 km/s for VCC450) and VCC415 (v=2560 km/s), located 13.8' to the SW. Classified as ``certain" member. Cross ID: VCC450 \\
19 & 12:21:29.23+08:09:04.6 & 586 & 479 & 185.37 & 8.15 & 16.32 & EI(sh\vbar{}pm:)-I(c) & EI galaxy with active star formation in its core. Shell-like structure and S-shaped distortion at large radius. Possible post-merger system. Classified as ``certain" member. Cross ID: VCC479 \\
20 & 12:21:42.72+09:21:19.3 & 612 & 496 & 185.43 & 9.36 & 17.51 & ES(di\vbar{}st:\vbar{}sh:\vbar{}pm:)-Q & Complex ES galaxy with central structure and dumbbell appearance, including possible shells. Possible post-merger remnant. Classified as ``certain" member. Cross ID: VCC496 \\
21 & 12:21:47.10+05:59:20.8 & 620 & - & 185.45 & 5.99 & 16.86 & E(n\vbar{}sh:\vbar{}pm:)-I(c) & ETG with blue nucleus and core, and possible outer shells. Possible post-merger system. Classified as ``certain" member \\
22 & 12:22:19.98+15:40:46.6 & 680 & - & 185.58 & 15.68 & 18.75 & ES(di:\vbar{}pm:)-Q & Elongated ES galaxy with extended tails. Possible S-shaped morphology. Bow-tie morphology at large radii. The smaller and more compact NGVS687 is located 64\arcsec to the SE, although neither galaxy has a known velocity. Possible post-merger remnant. Classified as ``possible" member \\
23 & 12:22:21.67+10:36:06.4 & 683 & 546 & 185.59 & 10.60 & 15.65 & E(di:\vbar{}sh\vbar{}du:\vbar{}pm:)-I(g\vbar{}d) & Unusual ETG with prominent shells. Twisted isophotes, with galaxy having bow-shaped or bent appearance at large radius. Possible post-merger remnant. Classified as ``certain" member. Cross ID: VCC546 \\
24 & 12:22:28.32+12:53:55.6 & 696 & - & 185.62 & 12.90 & 18.35 & EI(sh\vbar{}st\vbar{}pm:)-I(c\vbar{}g) & Possible post-merger remnant with fans and shells, plus central SF. Classified as ``possible" member \\
25 & 12:23:21.46+07:35:30.4 & 833 & 633 & 185.84 & 7.59 & 17.10 & E(sh:\vbar{}pm:)-I(c:) & Elongated LSB ETG  with bow-tie or boxy structure at large radius. Possible post-merger remnant. Classified as ``certain" member. Cross ID: VCC633 \\
26 & 12:24:03.22+05:10:51.3 & 937 & 693 & 186.01 & 5.18 & 14.81 & EI(di:\vbar{}b\vbar{}sp:\vbar{}sh\vbar{}st\vbar{}du\vbar{}pm)-A(g\vbar{}e) & Morphologically complex galaxy with strong star formation but series of streams and shells that do not resemble organized spiral arms. Likely post-merger remnant. Classified as ``certain" member. Cross ID: VCC693 \\
27 & 12:24:07.55+06:36:27.4 & 949 & 699 & 186.03 & 6.61 & 13.74 &  EI(di\vbar{}sp:\vbar{}sh\vbar{}st:\vbar{}du\vbar{}pm:)-I(d) & EI galaxy object with series of rings/shells at large radius, and central irregular star forming component.  Possible post-merger remnant. Classified as ``certain" member. Cross ID: VCC699 \\
28 & 12:25:52.42+05:48:32.9 & 1204 & 848 & 186.47 & 5.81 & 14.77 & EI(b\vbar{}sh\vbar{}du\vbar{}pm)-A(g\vbar{}b\vbar{}e:) & EI galaxy and likely post merger system. The galaxy is roughly face-on and includes a bar  and a chaotic series of shells that extend to large radius. Classified as ``certain" member. Cross ID: VCC848 \\
29 & 12:27:11.68+09:25:13.8 & 1463 & 979 & 186.80 & 9.42 & 11.83 & ES(di\vbar{}b\vbar{}sp\vbar{}sh\vbar{}du\vbar{}pm:)-I(c\vbar{}f) & Morphologically complex ES galaxy with dust filaments and star formation in inner bar. Faint incomplete spiral structure and possible shells at large radius. Possible post merger remnant. Classified as ``certain" member. Cross ID: VCC979 \\
30 & 12:27:11.68+09:25:13.8 & 1524 & 1013 & 186.88 & 9.34 & 16.68 &  EI(sh\vbar{}pm:)-I(c\vbar{}p\vbar{}f\vbar{}g) & EI  galaxy with star formation throughout and filamentary/patchy formation in the core. Likely shell or 'shelf' to the NW. Possible post-merger remnant. Classified as ``certain" member. Cross ID: VCC1013\\
31 & 12:27:45.65+13:00:31.9 & 1570 & 1043 & 186.94 & 13.01 & 10.92 & EI(di:\vbar{}sh\vbar{}st\vbar{}du\vbar{}pm)-At(g\vbar{}f\vbar{}p\vbar{}e) & Catastrophically disturbed galaxy, classified as EI, which is possibly interacting with VCC1030 (NGVS1550, located 263\arcsec to the N), although the two galaxies have significantly different velocities (70 km/s for NGVS1570, and 793 km/s for NGVS1550). Two additional galaxies are located nearby: NGVS1567 (92\arcsec to the SW, with v=-142 km/s) and NGVS1587 (171\arcsec to the SE, unknown velocity). The galaxy is also located 23\arcsec to the SE of M86 (v=-233 km/s) and has similar velocity, suggesting it might be part of the M86 group. Dust and star formation, with very disturbed morphology. Classified as ``certain" member. Cross ID: VCC1043 \\
32 & 12:28:11.34+09:45:37.0 & 1638 & 1078 & 187.05 & 9.75 & 15.68 & ES(n\vbar{}di:\vbar{}sp\vbar{}sh\vbar{}st\vbar{}pm:)-I(c) & ES galaxy with slightly blue core,  a series of shells or clumps along the major axis, and offset nucleus. S-shaped isophotes at large radius, suggestive of a tidal encounter with  VCC1078, which is located 3.2' to the SE and has similar velocity (547 km/s for VCC1062 and 530 km/s for VCC1078). Possible post-merger remnant. Classified as ``certain" member. Cross ID: VCC1078 \\
33 & 12:28:33.75+08:38:24.6 & 1711 & 1114 & 187.14 & 8.64 & 14.41 & EI(n:\vbar{}sh\vbar{}du\vbar{}pm)-I(g\vbar{}p) & Elongated EI galaxy with shells and disturbed outer structure. Possible, maybe offset nucleus and off-centered clump of star formation in inner region. Boxy structure in the outer regions. Classified as ``certain" member. Cross ID: VCC1114 \\
34 & 12:28:41.01+11:07:58.4 & 1727 & 1121 & 187.17 & 11.13 & 16.81 & EI(sp:\vbar{}sh\vbar{}pm:)-I(g) & EI galaxy with star formation in shells and arcs. Possible weak spiral structure. Forms a very close pair with NGVS1732, which has similar morphology ad is located 12\arcsec to the SE. Very bright star projected close to center. Possible post-merger remnant. Classified as ``certain" member. Cross ID: VCC1121 \\
35 & 12:28:41.73+11:07:51.1 & 1732 & - & 187.17 & 11.13 & 16.72 & EI(pm)-I(g:) & Complex galaxy. Forms a very close pair with NGVS1727, which has similar morphology ad is located 12\arcsec to the NW. Interpretation is made difficult by the presence of a nearby star. Likely post-merger remnant. Classified as ``probable" member \\
36 & 12:29:28.02+08:44:59.2 & 1859 & 1190 & 187.37 & 8.75 & 11.75 & ES(di\vbar{}sp:\vbar{}du\vbar{}pm:)-I(c\vbar{}d) & Edge-on ES galaxy with weakly star forming disk and prominent dust lanes and patches. Dust gives the galaxy an ``X-shape" appearance in the inner regions. Classified as ``certain" member. Presence of nucleus is impossible to assess due to the presence of dust or other contaminants. Possible post-merger remnant. Cross ID: VCC1190 \\
37 & 12:30:57.11+11:29:00.7 & 2135 & 1326 & 187.74 & 11.48 & 12.96 & ES(n:\vbar{}di:\vbar{}b\vbar{}sh:\vbar{}st:\vbar{}du\vbar{}pm:)-I(c) & ES galaxy with complex inner structure highlighted by arcs, streams, dust, star formation, and possible nucleus. Possible shells at large radii. Possible post merger product. Classified as ``certain" member. Cross ID: VCC1326 \\
38 & 12:31:02.08+06:07:20.0 & 2149 & 1332 & 187.76 & 6.12 & 18.06 & E(sh\vbar{}pm:)-Q & LSB EI or ETG galaxy with asymmetric, irregular structure and shells. Possible SF. Possible merger product. Classified as ``certain" member. Cross ID: VCC1332 \\
39 & 12:31:27.11+09:44:00.5 & 2205 & 1361 & 187.86 & 9.73 & 17.19 & E(sh\vbar{}pm:)-I(c) & Peculiar ETG galaxy with old stellar populations and a series of lobes or clumps along the major axis. Possible post-merger remnant. Galaxy resembles Cross ID: VCC429. Classified as ``certain" member. Cross ID: VCC1361 \\
40 & 12:31:53.86+14:18:23.4 & 2280 & 1400 & 187.97 & 14.31 & 15.53 & E(n\vbar{}di:\vbar{}sh\vbar{}pm:)-I(c:) & Nucleated ETG. Galaxy has boxy, almost bow-tie or X-shaped appearance, with possible flaring or shells at large radius. Located 416\arcsec to the S of the bright LTG NGVS2297 (VCC1401), although the two galaxies have significantly different velocities (2283 vs 460 km/s). Possible post-merger remnant. Classified as ``certain" member. Cross ID: VCC1400 \\
41 & 12:32:05.63+11:49:03.6 & 2318 & 1411 & 188.02 & 11.82 & 15.37 & ES(n:\vbar{}sp:\vbar{}sh\vbar{}du\vbar{}f\vbar{}pm:)-I(c\vbar{}d:\vbar{}p) & ES  galaxy with significant star formation in its core and possible offset nucleus. Shells and filaments. May be associated with VCC1407, which is located 4.4' to the N and has similar velocity (906 km/s for VCC1411 and 1005 km/s for VCC1407), and VCC1426 (v=832 km/s), which is located 6.3' to the NE. Possible post-merger remnant. Classified as ``certain" member. Cross ID: VCC1411 \\
42 & 12:32:38.67+10:45:34.7 & 2402 & 1447 & 188.16 & 10.76 & 19.41 & E(st\vbar{}sh\vbar{}pm:)-I(c:) & Compact ETG with extended streams and prominent stellar shells and possible star formation in the core. Possible post-merger remnant. Classified as ``probable" member. Cross ID: VCC1447 \\
43 & 12:33:13.43+09:23:50.4 & 2493 & 1488 & 188.31 & 9.40 & 14.63 & E(sh\vbar{}di:\vbar{}pm:)-I(c) & ETG with possible blue lens/disk in core and a series of shells or ripples along the major axis. Isophotes become boxy at large radius. May be associated with VCC1509, which is located 5.8' to the NE and has similar velocity (808 km/s for VCC1509 and 1079 km/s for VCC1488). Possible post merger remnant. Classified as ``certain" member. Cross ID: VCC1488 \\
44 & 12:33:24.70+08:41:26.7 & 2519 & 1501 & 188.35 & 8.69 & 15.45 & ES(di:\vbar{}b:\vbar{}sh\vbar{}st\vbar{}pm:)-I(c) & ES galaxy with central star formation and numerous shells and streams. A larger LTG, NGVS2534 (VCC1508)  is located 153\arcsec to the SE and has similar velocity (1219 vs 1403 km/s). Possible merger remnant. Classified as ``certain" member. Cross ID: VCC1501 \\
45 & 12:34:31.29+09:37:23.4 & 2652 & 1567 & 188.63 & 9.62 & 14.42 & ES(n\vbar{}di:\vbar{}b:\vbar{}sh:\vbar{}pm:)-Q & Nucleated ETG. Galaxy is very boxy (bar-like) in the inner regions and becomes peanut-shaped (bone- or bow-tie shaped) at large radius. Possible post-merger remnant. Classified as ``certain" member. Cross ID: VCC1567 \\
46 & 12:34:31.81+13:30:13.3 & 2654 & 1569 & 188.63 & 13.50 & 15.38 & EI(di:\vbar{}b:\vbar{}sp:\vbar{}sh:\vbar{}pm:)-I(c\vbar{}d\vbar{}g) & EI with star formation, mostly in the inner regions. Possible weak spiral structure. Possible post-merger product. Classified as ``certain" member. Cross ID: VCC1569 \\
47 & 12:34:42.49+14:13:23.2 & 2676 & 1582 & 188.68 & 14.22 & 16.77 & EI(sh\vbar{}pm:)-I(g) & EI galaxy with widespread, low level star formation. Galaxy shows a generally disturbed morphology, with a series of faint loops and shells. Possible post merger remnant. Classified as ``certain" member. Cross ID: VCC1582 \\
48 & 12:35:24.15+06:30:04.3 & 2753 & 1611 & 188.85 & 6.50 & 18.38 & ES(di\vbar{}pm:)-Q & ES galaxy with boxy isophotes and strong disk flaring. Bow-tie morphology. Possible post-merger remnant. Classified as ``probable" member. Cross ID: VCC1611 \\
49 & 12:36:20.09+13:17:36.9 & 2850 & - & 189.08 & 13.29 & 19.58 & E(n:\vbar{}pm:)-Q & Very elongated, extremely LSB ETG with faint envelope and possible nucleus. Isophotal flaring at large radii. Possible post merger remnant. Classified as ``possible" member \\
50 & 12:36:30.13+13:32:33.2 & 2871 & 1668 & 189.13 & 13.54 & 17.73 & ES(b:\vbar{}di:\vbar{}sh:\vbar{}pm:)-Q & Unusual ES galaxy with bar-like, or bow-tie, morphology, and possible shells that define abrupt edge to galaxy. Possible post-merger remnant. Classified as ``certain" member. Cross ID: VCC1668 \\
51 & 12:36:34.93+08:03:16.7 & 2884 & 1675 & 189.15 & 8.05 & 14.24 & ES(n\vbar{}di:\vbar{}b\vbar{}sp\vbar{}du\vbar{}pm:)-I(c\vbar{}p\vbar{}n) & Nucleated ES galaxy with weak spiral pattern, and small, star forming bar. Possible post merger remnant. Classified as ``certain" member. Cross ID: VCC1675 \\
52 & 12:36:39.42+11:06:06.4 & 2894 & 1684 & 189.16 & 11.10 & 14.97 & EI(di:\vbar{}sh\vbar{}du\vbar{}pm:)-I(c) & EI galaxy with star formation in the core, aligned along a think disk. Peanut shaped isophotes at large radii. Located 189\arcsec to the S of NGVS2887 (VCC1681), whose velocity is unknown. Possible post-merger remnant. Classified as ``certain" member. Cross ID: VCC1684 \\
53 & 12:38:06.92+10:09:53.5 & 3028 & 1744 & 189.53 & 10.16 & 16.63 & EI(n:\vbar{}sh\vbar{}pm:)-I(c\vbar{}p) & EI galaxy with active star formation and possible offset nucleus. Peak of star formation occurs in a small patch that is slightly displaced from photocenter. A series of shells are visible along the major axis. May be associated with VCC1743, which is located 4.9' to the S and has similar velocity (1273 km/s for VCC1743 and 1139 km/s for VCC1744), although there is no sign of an interaction. Possible post-merger remnant. Classified as ``certain" member. Cross ID: VCC1744 \\
54 & 12:38:17.90+13:06:35.9 & 3046 & 1757 & 189.53 & 13.11 & 13.33 & ES(n:\vbar{}d\vbar{}sp\vbar{}sh\vbar{}du\vbar{}pm:)-I(c) & Face-on ES galaxy with faint residual spiral pattern and clear shell-like structures at large radius. Several point sources project near the photocenter, one of which might be an offset nucleus. Possible post-merger remnant. Classified as ``certain" member. Cross ID: VCC1757 \\
55 & 12:39:04.08+13:21:49.4 & 3094 & 1778 & 189.77 & 13.36 & 14.03 &  EI(n\vbar{}sh\vbar{}du\vbar{}di:\vbar{}pm)-I(c) & Structurally complex EI galaxy with faint nucleus, small dust disk, ongoing star formation and outer shells that fan out along major axis. Likely post-merger remnant. Classified as ``certain" member. Cross ID: VCC1778 \\
56 & 12:39:14.07+15:37:36.1 & 3103 & 1784 & 189.81 & 15.63 & 16.53 & EI(n:\vbar{}b:\vbar{}sp:\vbar{}sh\vbar{}pm:)-I(c:) & Disturbed EI galaxy with low level star formation. Twisted isophotes, with prominent S-shaped distortion or possible spiral arms at large radius. Possible offset nucleus and even a misaligned bar. Possible post merger remnant?. Classified as ``certain" member. Cross ID: VCC1784 \\
57 & 12:40:19.56+14:42:54.0 & 3187 & 1836 & 190.08 & 14.71 & 14.31 & ES(di:\vbar{}b\vbar{}sp:\vbar{}r:\vbar{}du:\vbar{}pm:)-I(g) & ES galaxy with complex structure including bar, possible ring, and spiral arms. Possible post-merger remnant. Classified as ``certain" member. Cross ID: VCC1836 \\
58 & 12:41:54.28+13:46:22.4 & 3285 & 1897 & 190.48 & 13.77 & 14.03 & E(n\vbar{}sh\vbar{}pm:)-Q & Nucleated, extremely boxy ETG with shells along major axis. Unsharp masked image shows entire galaxy to have a bow-tie appearance. Possible post merger remnant. Classified as ``certain" member. Cross ID: VCC1897 \\
59 & 12:42:03.23+12:28:48.1 & 3300 & 1905 & 190.51 & 12.48 & 17.49 & E(sh\vbar{}pm:)-Q & Very LSB ETG with shells or arcs. Possible merger remnant. Classified as ``certain" member. Cross ID: VCC1905 \\
60 & 12:42:17.76+10:01:47.4 & 3311 & - & 190.57 & 10.03 & 19.93 & ES(pm:)-I(c) & Possible ES galaxy with dumbbell lobes and central core. Possible post-merger remnant. Classified as ``possible" member \\
61 & 12:42:26.51+11:44:25.2 & 3321 & 1921 & 190.61 & 11.74 & 15.39 & ES(n\vbar{}di\vbar{}pm:)-I(c:) & Nucleated ES galaxy. Galaxy appears disk-like in its center, but isophotes become very boxy (almost peanut- or bow-tie shaped) at large radius. Located 489\arcsec to the NE of NGVS3296 (VCC1903, M59),  although its velocity is larger (1139 km/s vs 467 km/s for M59). May be associated with VCC1951, which is located 9.1' to the SE and has similar velocity (1138 km/s for VCC1921 and 1106 km/s for VCC1951). Possible post-merger remnant. Classified as ``certain" member. Cross ID: VCC1921 \\
62 & 12:44:45.45+12:21:03.4 & 3456 & 2006 & 191.19 & 12.35 & 13.60 & ES(sh:\vbar{}du\vbar{}pm:)-I(g\vbar{}d) & ES galaxy with boxy structure, ongoing star formation, dust and possible weak shells along major axis.  Possible post-merger remnant. Classified as ``certain" member. Cross ID: VCC2006 \\
63 & 12:45:50.97+15:29:29.2 & 3519 & - & 191.46 & 15.49 & 19.68 & E(n\vbar{}pm:)-Q & Compact, elongated, very LSB ETG with bright nucleus and outward flaring. Possible bow-tie morphology. Located 52\arcsec to the NW of NGVS3521, although the velocity of either galaxy is unknown.  Possible post merger remnant. Classified as ``probable" member \\
64 & 12:46:15.32+10:12:19.8 & 3531 & 2037 & 191.56 & 10.21 & 15.30 & EI(du\vbar{}di:\vbar{}sh\vbar{}pm:)-I(c) & Elongated EI galaxy with active star formation in its core. Unusual series of shells or lobes along its major axis at both ends. 3528 (VCC2037) is located 190\arcsec to the SW and has similar velocity (1139 km/s for VCC2037 and 1511 kms for VCC2034). Possible post merger remnant. Another galaxy, NGVS3530 is located 154\arcsec to the N. Classified as ``certain" member. Cross ID: VCC2037 \\
65 & 12:48:15.11+10:58:60.0 & 3597 & 2066 & 192.06 & 10.98 & 11.66 & ES(n:\vbar{}di\vbar{}sh:\vbar{}st:\vbar{}du\vbar{}pm:)-I(c) & Possibly nucleated ES galaxy with irregular star forming component in a diffuse, highly structured ETG-like component. Network of shells and ripples. The galaxy is likely interacting with the smaller irregular galaxy  NGVS3589 (VCC2062), which is located 216\arcsec  to the W and has similar velocity (1160 vs 1140 km/s). Possible post merger remnant. Classified as ``certain" member. Cross ID: VCC2066 \\
66 & 12:49:39.54+12:37:31.5 & 3634 & - & 192.41 & 12.63 & 14.97 & ES(n\vbar{}di\vbar{}sp:\vbar{}du:\vbar{}sh:\vbar{}pm:)-I(d:) & Nucleated, likely ES galaxy seen edge-on, with possible low-level star formation and shells. Possible post merger remnant. Classified as ``certain" member \\
\enddata
\end{deluxetable}
\endgroup

\twocolumngrid
\clearpage

\end{document}